\documentclass[tradiabstract]{aa}
\usepackage{graphicx}
\usepackage{txfonts}
\usepackage{amsmath}
\usepackage{amsfonts}
\usepackage{amssymb}
\usepackage{booktabs}
\usepackage{setspace}
\usepackage[colorlinks,citecolor=blue]{hyperref}
\usepackage{blindtext}
\usepackage[english]{babel}
\usepackage{natbib}
\usepackage{tabularx,multirow,booktabs,blindtext}
\usepackage{subfigure}
\usepackage{lipsum}
\usepackage{adjustbox}
\usepackage{longtable}
\usepackage{rotating} 
\usepackage{lscape}
\usepackage{footnote}
\usepackage{threeparttable}
\usepackage{multicol}
\usepackage{graphicx}
\usepackage{caption}
\usepackage{tablefootnote}
\usepackage[dvipsnames]{xcolor}
\usepackage{ulem}

\def\gsim{\;\lower4pt\hbox{${\buildrel\displaystyle >\over\sim}$}\,}
\def\lsim{\;\lower4pt\hbox{${\buildrel\displaystyle <\over\sim}$}\,}

\begin{document}

\title{Transitions in magnetic behavior at the substellar boundary}


\author{E. Magaudda \inst{1} \and B. Stelzer \inst{1,2} \and R. A. Osten \inst{3,4} \and J. S. Pineda \inst{5} \and St. Raetz\inst{1} \and M. McKay\inst{3,6}}


\institute{
 Institut f\"ur Astronomie \& Astrophysik, Eberhard Karls Universit\"at T\"ubingen, Sand 1, 72076 T\"ubingen, Germany\label{inst1} 
\and
 INAF - Osservatorio Astronomico di Palermo, Piazza del Parlamento 1, 90134 Palermo, Italy\label{inst2} 
\and
Space Telescope Science Institute, 3700 San Martin Drive, Baltimore, MD, USA \label{inst3}
\and
Center for Astrophysical Sciences, Johns Hopkins University, 3701 San Martin Drive, Baltimore, MD USA \label{inst4}
\and
Laboratory for Atmospheric and Space Physics, University of Colorado Boulder, Boulder, CO, USA\label{inst5}
\and
Department of Astronomy, University of Washington, Seattle, WA USA\label{inst6}
}

\titlerunning{Transitions in magnetic behavior at the substellar boundary}

\date{}

\abstract{
We aim at advancing our understanding of magnetic activity and the underlying dynamo mechanism at the end of the main-sequence.
To this end, we have embarked in collecting simultaneous X-ray and radio observations for a sample of M7..L0 dwarfs in the solar neighborhood using {\it XMM-Newton} jointly with the Jansky Very Large Array (JVLA) and the Australia Telescope Compact Array (ATCA). We supplemented the data from these dedicated campaigns with X-ray data from the all-sky surveys of the ROentgen Survey with an Imaging Telescope Array (eROSITA) on board the Russian Spektrum-Roentgen-Gamma mission (SRG). Moreover, we complement this multi-wavelength data set with rotation periods that we measure from light curves acquired with the Transiting Exoplanet Survey Satellite (TESS). We limit the sample to objects with rotation period $<1$\,d, focusing on the study of a transition in magnetic behavior suggested by a drastic change in the radio detection rate at $v \sin{i} \approx 38\,\rm km/s$, corresponding to $P_{\rm rot} \approx 0.2$\,d for a typical UCD radius of $R_{\star} = 0.15\,R_\odot$. Finally, to enlarge the target list we have compiled archival X-ray and radio data for UCDs from the literature, and we have analysed the above-mentioned ancillary eROSITA and TESS observations for these objects analogous to the targets from our dedicated X-ray/radio campaigns.
We compiled the most up-to-date radio/X-ray luminosity ($L_{\rm R,\nu}-L_{\rm x}$) relation for $26$ UCDs with rotation periods ($P_{\rm rot}$) lower than $1$\,d, finding that rapid rotators lie the furthest away from the ``G\"udel-Benz'' relation previously studied for earlier-type stars. Radio bursts are mainly (although not exclusively) experienced by very fast UCDs ($P_{\rm rot}\leq0.2$\,d), while X-ray flares are seen by objects spanning the whole range of rotation. 
Finally, we examined the $L_{\rm x}/L_{\rm bol}$ vs $P_{\rm rot}$ relation, where our sample of UCDs spans a large range of activity level, i.e. $\log(L_{\rm x}/L_{\rm bol}) = -5.5$ to $\log(L_{\rm x}/L_{\rm bol}) = -3$. Although they are all fast rotating, evidently X-ray activity decouples from that of normal dynamos. In fact, we found no evident relation between the X-ray emission and rotation, reinforcing previous speculations on a bimodal dynamo across late-type dwarfs.
One radio-detected object, 2MJ0838, has a rotation period consistent with the range of auroral bursting sources; while it displays moderately circularly polarized emission, there is no temporal variation in the polarized flux. 
A radio flare from this object is interpreted as gyrosynchrotron emission, and it displays X-ray and optical flares. 
Among the $10$ UCDs observed with the dedicated X-ray/radio campaigns, we found a slowly rotating apparent auroral emitter (2MJ0752), that is also one of the X-ray brightest radio-detected UCDs.
We speculate that this UCD is experiencing a transition of its magnetic behavior since it produces signatures expected from higher mass M~dwarfs along with emerging evidence of auroral emission.
}


\maketitle

\section{Introduction}\label{sect:intro}

Stellar magnetic activity is a fundamental ingredient in defining the observed spectral flux of low-mass stars. Magnetism powers chromospheres and coronae, which exhibit emissions from X-ray to radio wavebands \citep[see review by][]{Linsky2017ARA&A..55..159L}, but also generates variability through, for example, photospheric spots, active regions, and flaring reconnection events \citep[see within][]{Solanki2006RPPh...69..563S}. Observations of these emissions serve as critical diagnostics of the dynamic plasma of the stellar atmosphere, and must be physically linked with the internally generated magnetic dynamo \citep[e.g.,][]{Browning2008ApJ...676.1262B}. This connection is well evinced by multi-wavelength results demonstrating rotation-activity-age correlations \citep[e.g.,][]{Wright2018, Newton2017ApJ...834...85N,Magaudda2020, Pineda2021ApJ...911..111P}, tying the activity to long-term angular momentum evolution. Despite concerted effort studying stellar activity, many aspects of the underlying physics remain uncertain (e.g., dynamo mechanism, winds, coronal heating, etc.), especially in low-mass stars which often show stronger magnetic fields \citep[e.g.,][]{Shulyak2017NatAs...1E.184S} and more energetic eruptions \citep[e.g.,][]{Gunther2020AJ....159...60G} relative to what has been observed from the Sun. 

Understanding these effects of the stellar magnetic field are more important now than ever before because of the proliferation of detected exoplanetary systems around low-mass stars. These stars host many of the extrasolar planets most favorable for atmospheric characterization \citep[e.g.,][]{Morley2017ApJ...850..121M}, and interpreting those results requires understanding the high-energy stellar emissions, and their evolutionary history \citep[e.g.,][]{Johnstone2021A&A...649A..96J}. Of particular interest are very-low mass stars at the end of the main sequence which are capable of hosting multiple temperate Earth-sized planets \citep[e.g., TRAPPIST-1;][]{Gillon2017Natur.542..456G}, and can display enhanced magnetic activity for several billions of years \citep[e.g.,][]{West08.1}. 

However, relative to warmer stars, the nature of the stellar activity must change in this mass range, and toward lower mass brown dwarfs, as effective temperatures cool and the atmospheres become more neutral, decoupling the atmospheric dynamics from the magnetic field \citep[e.g.,][]{Mohanty03.1, RodriguezBarrera2015MNRAS.454.3977R}. This regime of ultracool dwarfs (UCDs; spectral types $\ge$M7) displays clear deviations from the standard coronal paradigm defined by Solar activity \citep[see within][]{Pineda2017}, best demonstrated by their remarkable radio emissions. 

For a wide range of stars (main-sequence flare stars, RS CVn binaries, weak-line T Tauri stars, and Algol binaries) there is a tight relation between the luminosity in the radio and in  X-rays, underlining the close connection between non-thermal gyrosynchrotron radio emission  and thermal X-rays \citep{Guedel1993ApJ...415..236G, Guedel93.2}. This `G\"udel-Benz' relation (hereafter GB-relation), 
however, breaks down in the UCD regime. Several UCDs with sensitive X-ray and radio measurements are overluminous in radio with respect to the GB-relation by factors of $>$ 1000 \citep[e.g.,][]{Williams14.0, Pineda2017}. Super-imposed on this quiescent, quasi-steady (synchrotron) emission, tracing radiation belt structures \citep{Kao2023Natur.619..272K,Climent2023Sci...381.1120C}, are highly circularly polarized periodically pulsed bursts \citep[e.g.,][]{Burgasser05.1, Hallinan06.1, Hallinan08.1, Route2012ApJ747L..22R,Williams2015ApJ799..192W,Kao2016,Vedantham2020ApJ...903L..33V,Rose2023ApJ951L..43R}. This `bursting' radio component has been attributed to the electron 
cyclotron maser instability \citep[ECMI; ][]{Hallinan08.1}, a phenomenon seen predominately in solar system giant planets that is tied to auroral electron precipitation \citep[e.g.,][]{Zarka2007P&SS...55..598Z}.

Conversely, the short-term variability of typical UCDs in the X-ray band exhibits the typical signatures of flares that are likely caused by magnetic reconnection events \citep[e.g.,][]{Stelzer2006, Berger08.1, Williams2014}. Long-term optical photometric monitoring of this population shows the capacity for these stars to also show white-light flares \citep[e.g.,][]{Paudel2018ApJ...858...55P, Murray2022MNRAS.513.2615M}. These behaviors suggest that much UCD magnetic activity remains analogous to that of higher-mass GKM stars.

The existing radio and X-ray data thus present a heterogeneous picture for UCDs. Indeed, the accumulation of data over the past decade have only reinforced the suggestion by \citet{Stelzer2012} of a dichotomy between radio-bursting / X-ray faint and radio-quiet / X-ray flaring UCDs, although the boundary may not be rigid. Still, the nature of the onset of this behavior near the end of the stellar main sequence is unclear. The emergence of UCD ECMI radio emission is most likely shaped by magnetic field structure/strength and the dwarf rotation rate \citep[see within][]{Stelzer2012, Pineda2017,Pineda18.0}. 

The confirmation of resolved quiescent radio emission in the UCD LSR~J1835+3259 from energetic electrons trapped in large-scale magnetic loops reinforces the idea of large-scale dipolar topologies being a critical ingredient for the radio-bright UCDs \citep{Kao2023Natur.619..272K,Climent2023Sci...381.1120C}. Theoretical treatments of the engines powering the radio-bright UCDs strongly depend on the object's 
rotation rate  \citep[e.g.,][]{Nichols2012ApJ...760...59N, Turnpenney2017MNRAS.470.4274T, Saur2021A&A...655A..75S}. Moreover, \citet{Pineda2017} used spectroscopic rotation measurements to reveal a sharp rise in the detection fraction of UCD radio emission: above $v \sin{i} \approx 35$\,km s$^{-1}$ (rotation period, $P_{\rm rot} \sim 3.5$~hr) the UCD detection fraction at radio wavelengths rises steeply, 4-5 times the overall rate of unbiased surveys \citep[$\sim$$10\%$; e.g.][]{Lynch2016MNRAS.457.1224L}. 
However, the unknown inclination angle makes it difficult to ascertain the precise role of rotation in observed UCD radio emissions. 

The \textit{Transiting Exoplanet Survey Satellite} \citep[TESS; ][]{Ricker2015JATIS...1a4003R}, through its broadband optical monitoring all-sky survey enables examination of photometric rotation periods across a broad sample of stars, including nearby UCDs. With TESS, we are now in the position to put reliable constraints on all relevant parameters: rotation, X-ray and radio emission, and clearly test whether the dichotomy of UCD X-ray/radio behavior emerges strictly across the rotation period boundary suggested by \cite{Pineda2017}. 

In this article, we build a mini-survey of GHz radio observations and X-ray data for a carefully-selected sample of UCDs accessible to TESS. Our multi-wavelength analysis is designed to examine UCDs on both sides of this rotation boundary ($\sim$3.5~hr), and assess the nature of their magnetically driven emissions. This sample is presented in Sect.~\ref{sect:sample}, with its radio, X-ray and optical data analysis in Sect.~\ref{sect:analysis}. In Sects.~\ref{sect:radio_int_obj} we present interesting radio emissions for specific objects and the coronal/magnetospheric models, respectively. We finally discuss our sample across all three wavebands adopted throughout this paper (Sect.~\ref{sect:discussion}) and propose our conclusions in Sect.~\ref{sect:conclusion}.

\addtolength{\tabcolsep}{-4pt}{
\begin{table}
    \begin{center}
        \begin{threeparttable}[b]
            \caption{Overview of the UCDs from our X-ray/radio campaigns: 
            cols.1\&2 show the 2MASS\,ID and the alternative name from Simbad, and the last two columns inform about the instruments used for the X-ray and radio observations.} 
            \label{tab:ucds_par}
                \begin{tabular}{cccc}
                    \midrule[0.5mm]     
                        \multicolumn{1}{c}{2MASS ID}
                       &\multicolumn{1}{c}{Name} 
                       &\multicolumn{1}{c}{X-ray\tnote{1}}
                       &\multicolumn{1}{c}{Radio\tnote{2}}\\

                         \multicolumn{1}{c}{}  
                        &\multicolumn{1}{c}{}  
                        &\multicolumn{1}{c}{}  
                        &\multicolumn{1}{c}{}\\
                        
                        \midrule[0.5mm]
    
    
     
                J07522390$+$1612 & LP\,423-31 & X/eR & VLA\\ 
                J03510004$-$0052 & LP\,593-68 & X/eR & VLA\\
                J17571539$+$7042 & LP\,44-162 & X & VLA\\ 
                J04402325$-$0530 & LP\,655-48 & X/eR & Ant13\\
                J03061159$-$3647 & SSSPM\,J0306-3648 & X/eR & VLA\\
                
                
                J08380224$-$5855 & SCR\,J0838-5855 & X/eR & ATCA\\
                J04351612$-$1606 & LP\,775-31 & X/eR & Ant13\\
                J06521977$-$2534 & DENIS\,J065219.7-253450 & X & JVLA \\
                J02150802$-$3040 & LP\,885-35 & X/eR & Ant13\\
                J10551532$-$7356 & WISEA J10551571-735611.3 & eR & ATCA\\
 
    \bottomrule[0.5mm] 
  
    \end{tabular}
     	\begin{tablenotes}
 				\item[1]  X: XMM-Newton; eR: eROSITA
 				\item[2] VLA: Very Large Array; JVLA: Jansky Very Large Array; ATCA: Australia Telescope Compact Array; Ant13:\citet{Antonova2013}.
 			\end{tablenotes}
\end{threeparttable}
\end{center}
\end{table}
}

\section{Sample}\label{sect:sample}


\label{sect:XMMsample}

The focus of this work is to investigate the  activity of UCDs across the boundary in terms of rotation rate, $v \sin{i} \sim 35$\,km/s, at which previous data has suggested a drastic change of the radio properties \cite{Pineda2017}. For a typical late-M dwarf 
this corresponds to a rotation period $P_{\rm rot} \sim 0.15$\,d ($\approx 3.5$\,hr), 
justified by most radio bursters being seen equator-on. Our focus is on quantifying the multi-wavelength behavior of UCDs in the areas just above and below this threshold.

To this end, we have obtained deep X-ray and radio data for $10$ nearby UCDs. The targets were selected with the criterion that $P_{\rm rot}$ be $< 1$\,d, drawing from a larger list of UCDs for which we had preliminarily determined the rotation period from TESS light curves.
Considering the decay of activity in the L dwarf regime which pushes current instruments beyond their sensitivity limits, for our campaigns we focused on UCDs with spectral types from M7 to L0.  Objects with small distances were preferred to guarantee the highest achievable sensitivity for the activity measurements. The target list and the X-ray and radio instruments used for each of them is provided in Table~\ref{tab:ucds_par}. 
The data were obtained in three dedicated X-ray/radio campaigns carried out in 2012-2013, 2014-2015 and 2020-2021.

The X-ray observations were carried out with the X-ray Multi-Mirror mission ({\it XMM-Newton}) and the radio data were obtained from the Jansky Very Large Array (JVLA) and the Australia Telescope Compact Array (ATCA). For four UCDs X-ray and radio data were acquired simultaneous or nearly simultaneous, see Tables~\ref{tab:obslog_xmm} \& \ref{tab:obslog_radio}. To enlarge the data base we have searched the literature for UCDs observed in both the X-ray and the radio band, and we found $16$ objects.

\begin{table*}
\begin{center}
\begin{threeparttable}[b]
\caption{Stellar parameters and {\it Gaia}-DR3 data of the UCDs studied in this work, see text in Sect.~\ref{sect:sample}. }
    \label{tab:stepar}
    \begin{tabular}{lccccccccccccc}
      \midrule[0.5mm]
      \multicolumn{1}{l}{2MASS\,ID}
      &\multicolumn{1}{c}{SpT}
      &\multicolumn{1}{c}{$J$}
     &\multicolumn{1}{c}{$T_{\rm eff}$}
      &\multicolumn{1}{c}{$\log L_{\rm bol}$}
      &\multicolumn{1}{c}{$R_\star$}
      &\multicolumn{1}{c}{{\it Gaia}-DR3 IDs}
      &\multicolumn{1}{c}{$d$}
      &\multicolumn{1}{c}{$PM_{\rm \alpha}$}
      &\multicolumn{1}{c}{$PM_{\rm \delta}$}
      &\multicolumn{1}{c}{$G$}
      &\multicolumn{1}{c}{$G_{\rm bp}$}
      &\multicolumn{1}{c}{$G_{\rm rp}$} \\

      \multicolumn{1}{l}{}
      &\multicolumn{1}{l}{}
      &\multicolumn{1}{l}{[mag]}
     &\multicolumn{1}{c}{[K]}
      &\multicolumn{1}{c}{[$L_{\odot}$]}
      &\multicolumn{1}{c}{[$R_\odot$]}
      &\multicolumn{1}{c}{}
      &\multicolumn{1}{c}{[pc]}
      &\multicolumn{1}{c}{[mas/yr]}
      &\multicolumn{1}{c}{[mas/yr]}
      &\multicolumn{1}{c}{[mag]}
      &\multicolumn{1}{c}{[mag]}
      &\multicolumn{1}{c}{[mag]} \\

      \midrule[0.5mm]
      
 \multicolumn{12}{c}{New UCD sample}\\
\midrule[0.1mm]\\
2MJ0752 & M7.0 & 10.9 & 2683 & -2.67 & 0.20$\pm$0.02 & 666988221840703232 & 18.88$\pm$0.01 & 183.1 & -350.3 & 14.75 & 17.22 & 13.33  \\ 
2MJ0351 & M7.5 & 11.3 & 2611 & -3.02 & 0.14$\pm$0.01 & 3257243312560240000 & 14.67$\pm$0.01 & 10.9 & -469.9 & 15.31 & 18.28 & 13.84 \\ 
2MJ1757 & M7.5 & 11.5 & 2611 & -2.93 & 0.17$\pm$0.01 & 1638180413086979840 & 19.13$\pm$0.08 & 6.8 & 326.6 & 15.65 & 18.92 & 14.14 \\ 
2MJ0440 & M7.0 & 10.7 & 2683 & -3.20 & 0.11$\pm$0.01 & 3200303384927512960 & \phantom{0}9.74$\pm$0.01 & 334.5 & 127.9 & 14.90 & 18.05 & 13.40 \\
2MJ0306 & M8.5 & 11.7 & 2467 & -3.32 & 0.12$\pm$0.01 & 5047423236725995136 & 13.26$\pm$0.01 & -172.1 & -669.2 & 15.99 & 19.53 & 14.46 \\ 


2MJ0838 & M6.0 & 10.3 & 2830 & -2.98 & 0.13$\pm$0.01 & 5302788969815543936 & 11.11$\pm$0.03 & -56.9 & -313.6 & 14.52 & 17.33 & 12.87 \\ 
2MJ0435 & M7.0 & 10.4 & 2683 & -3.04 & 0.14$\pm$0.01 & 3171631420210205056 & 10.60$\pm$0.02 & 160.1 & 315.1 & 14.6 & 17.92 & 13.09 \\ 
2MJ0652 & L0.0 & 12.7 & 2248 & -3.57 & 0.10$\pm$0.01 & 2920995300823950720 & 16.06$\pm$0.02 & -235.5 & 87.9 & 17.43 & 20.64 & 15.87 \\ 
2MJ0215 & M7.5 & 11.6 &2611 & -3.24 & 0.12$\pm$0.01 & 4971892010576979840 & 14.05$\pm$0.01 & 768.6 & -360.0 & 15.8 & 19.13 & 14.32 \\ 
2MJ1055 & M7.0 & 10.6 & 2683 & -2.93 & 0.16$\pm$0.01 & 5225863906515477376 & 12.90$\pm$0.08 & 163.1 & -216.3 & 14.72 & 17.84 & 13.22 \\ 

\midrule[0.1mm]
\multicolumn{12}{c}{Literature sample}\\
\midrule[0.1mm]

 2MJ0036 & L3.5 & 12.4 & 3286 & -3.97 & 0.03$\pm$0.01 & 2794735086363871360 & 8.74$\pm$0.01 & 901.5 & 124.3 & 17.5 & 20.44 & 15.91 \\
 2MJ0523 & L2.5 & 13.0 & 3554 &-3.78 & 0.03$\pm$0.01 & 2985035874544160384 & 12.73$\pm$0.02 & 107.5 & 161.47 & 18.03 & 21.27 & 16.48 \\
 2MJ0602 & L1.0 & 12.3 & 4150 & -3.73 & 0.03$\pm$0.01 & 3457493517036545280 & 11.66$\pm$0.02 & 157.6 & -506.4 & 17.23 & 20.8 & 15.66 \\
 2MWJ1507 & L5.0 & 12.8 & 2990 & -4.26 & 0.03$\pm$0.01 & 6306068659857135232 & 7.41$\pm$0.01 & -151.8 & -896.0 & 17.95 & 21.17 & 16.34 \\
 BRI0021-0214 & M9.5 & 11.5 & 2314 & -3.44 & 0.12$\pm$0.01 & 2541756977144595712 & 12.45$\pm$0.03 & -74.6 & 140.2 & 16.58 & 20.02 & 15.03 \\
 DENIS J1048 & M9.0 & 9.5 & 2349 & -3.53 & 0.10$\pm$0.01 & 5393446658454453632 & 4.05$\pm$0.01 & -1179.3 & -988.1 & 14.01 & 17.62 & 12.46\\
 Gl 569 B & M9.1 & 11.1 & 2380 & -3.17 & 0.15$\pm$0.01 & 1187851653287128576 & 9.94$\pm$0.01 & 279.1 & -117.9 & 9.12 & 10.31 & 8.03 \\
 Kelu 1 & L2.0 & 13.4 & 3722 & -3.6 & 0.04$\pm$0.01 & 6187779556809793024 & 20.39$\pm$0.3 & -305.4 & -27.3 & 18.74 & 21.16 & 16.79 \\
 LHS2065 & M9.0 & 11.2 & 2394 & -3.51 & 0.10$\pm$0.01 & 5761985432616501376 & 8.66$\pm$0.01 & -516.6 & -199.6 & 15.89 & 19.11 & 14.35 \\
 LP 349-25 & M7.9 & 10.6 & 2553 & -2.78 & 0.21$\pm$0.02 & 2799992744809482112 & 14.13$\pm$0.09 & -490.3 & -843.5 & 14.96 & 18.31 & 13.38 \\
 LP 412-31 & M7.9 & 11.8 & 2546 & -3.26 & 0.12$\pm$0.01 & 0056252256123908096 & 14.65$\pm$0.02 & 392.7 & -186.6 & 16.1 & 19.22 & 14.58 \\
 LP 944-20 & M9.5 & 10.7 & 2321 & -3.84 & 0.07$\pm$0.01 & 4860376345833699840 & 6.43$\pm$0.01 & 309.0 & 269.0 & 15.44 & 18.94 & 13.89 \\
 LSR J1835 & M8.4 & 9.5 & 2474 & -3.4 & 0.11$\pm$0.01 & 2091177593123254016 & 5.69$\pm$0.01 & -72.6 & -755.1 & 14.85 & 18.46 & 13.31 \\
 TVLM513 & M8.5 & 11.8 & 2467 &-3.65 & 0.08$\pm$0.01 & 1262763648230973440 & 10.73$\pm$0.02 & -43.1 & -65.1 & 16.51 & 20.13 & 14.95 \\
 vB 10 & M8.0 & 9.9 & 2539 & -3.39 & 0.10$\pm$0.01 & 4293315765165489536 & 5.92$\pm$0.01 & -598.8 & -1366.1 & 14.3 & 17.75 & 12.77 \\
 vB 8 & M6.9 & 9.8 & 2690 & -3.19 & 0.12$\pm$0.01 & 4339417394313320192 & 6.49$\pm$0.01 & -813.0 & -870.6 & 13.81 & 17.05 & 12.32 \\
\bottomrule[0.5mm]
\end{tabular}
\end{threeparttable}
 \end{center}
\end{table*}

\begin{figure}
    \centering
    \includegraphics[width=0.5\textwidth]{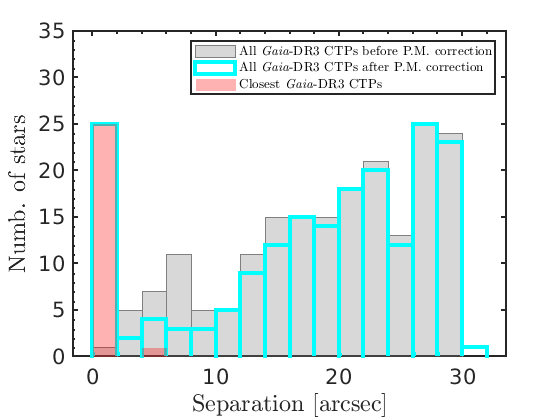}
    \caption{Separation between the {\it Gaia}-DR3 (J2016) and the  J2000 positions for the $10$ UCDs from our dedicated X-ray/radio campaigns and the $16$ UCDs from the literature sample. All $191$ {\it Gaia}-DR3 counterparts within $30^{\prime\prime}$ are shown in gray. In turquoise there are the same objects but after the application of their P.M. correction. The red histogram represents the subsample of closest {\it Gaia}-DR3 CTPs to all UCDs after the P.M. correction.}
    \label{fig:rev_match}
\end{figure}

Our nearby targets have high proper motion, and for the match of the objects' optical position with X-ray and radio images their apparent space motion needs to be taken into account. We, thus, have to propagate the optical positions to the X-ray/radio observing date. To this end, we first search for the {\it Gaia} counterparts (CTPs) of our targets. We started with identifying \underline{all} {\it Gaia} objects in their vicinity and then narrowed down the search as described in the following. First, we matched the J2000 coordinates of our UCD sample with the coordinates given in the {\it Gaia}-DR3 catalog \citep[epoch J2016,][]{GaiaDR32022} within a radius of $30^{\prime\prime}$. Within this radius we found $54$ and $137$ {\it Gaia}-DR3 CTPs for the $10$ UCDs from our dedicated X-ray/radio observations and the literature sample, respectively, shown in gray in Fig.~\ref{fig:rev_match}. We then performed the proper motion\footnote{The uncertainties on the proper motions range from $0.03$\,mas/yr to $0.3$\,mas/yr and are neglected here.} (P.M.) correction propagating the J2016 coordinates of all these potential {\it Gaia}-DR3 CTPs backwards to the J2000 epoch. Then we calculated again the separation between the position of the UCDs in our input catalog and the position of the {\it Gaia} sources. In Fig.~\ref{fig:rev_match} we show how the separation between the {\it Gaia}-DR3 CTPs and the input coordinates of the UCD sample has changed after applying the P.M. correction (before and after the P.M. correction in gray and in turquoise, respectively). 
Each UCD has now its closest {\it Gaia}-DR3 CTP within a radius of $1^{\prime\prime}$, except for one target from the literature with separation $\leq 6^{\prime\prime}$.
Given this excellent match, we adopted for each UCD this closest {\it Gaia}-DR3 CTP after the P.M. correction. We confirmed our associations checking the {\it Gaia}-DR3 {\tt source\_id} on Simbad\footnote{\url{https://simbad.u-strasbg.fr/simbad/sim-fid}}, from where we retrieved also the spectral types (SpTs). 

We also made use of the All-Sky survey data of the {\it ROentgen Survey with an Imaging Telescope Array} \citep[eROSITA,][]{Predehl2021} on board the Russian Spektrum-Roentgen-Gamma mission \citep[SRG,][]{Sunyaev2021}.
Due to the data sharing policy our consortium eROSITA\_DE has access to the western galactic half of the sky that comprises all but one of the UCDs in our sample and $7$ out of $16$ of the literature sample. Thus, we performed a match within $30^{\prime\prime}$ between the P.M. corrected coordinates of our samples with the boresight position of the merged eRASS:4 catalog (all\_s4\_SourceCat3B\_221031\_poscorr\_mpe\_clean.fits) provided by the German consortium, that combines the data from all $4$ eROSITA All-sky surveys. The details about how the official catalogs are compiled will be published with the first data release of eROSITA within the work by \citet{Merloni23.0} on the first all-sky survey (eRASS1). Eight UCDs of our sample are detected in the eRASS:4 catalog, specifically the eRASS proved to be useful for one object, for which no observational time was allocated with {\it XMM-Newton} during our campaigns. Among the objects of the literature sample only $2$ are associated with an eROSITA source listed in eRASS:4.

For both samples we derived the bolometric luminosity ($L_{\rm bol}$) and the effective temperature ($T_{\rm eff}$) that we used to calculate the radius ($R_{\star}$) through the Stefan-Boltzmann law. First, we computed $T_{\rm eff}$ and the bolometric correction in the $J$ band \citep[$BC_{\rm J}$, $J$ band magnitude from the Two Micron All-Sky Survey~--~2\,MASS,][]{Cutri2003} with the polynomial fits as a function of spectral type from \cite{Filippazzo2015}. Then, we calculated the absolute $J$ band magnitude ($M_{\rm J}$) with the {\it Gaia}-DR3 \citep[][]{GaiaDR32022} distances listed in Table~\ref{tab:stepar}.
Finally, we used the distances combined with $BC_{\rm J}$ to estimate $L_{\rm bol}$. 
The uncertainties for $L_{\rm bol}$ and $T_{\rm eff}$ are taken from \citet{Filippazzo2015} and they are $0.133$\,dex and $113$\,K, respectively. We computed those of the radius by performing the error propagation.

The two samples count $26$ UCDs in total. Their {\it Gaia}-DR3 parameters and stellar parameters are reported in Table~\ref{tab:stepar}. In particular, in col.1 we show a reduced 2MASS name when only the epoch and the first three digits of the right ascension are displayed\footnote{The source designation for objects in 2MASS Catalogs represents the epoch, right ascension and declination of the source during the observation: Jhhmmss[.]ss$\pm$ddmmss[.]s \citep{Cutri2003}.}. Spectral types are in col.2, $J$ band magnitude and effective temperature in cols.3\&4, bolometric luminosity and radii in cols.5\&6. From cols.7 to 13 we report the {\it Gaia}-DR3 {\sc source\_id} with distance, proper motion and photometry. 

\begin{figure*}
\begin{center}
\parbox{18cm}
{
\parbox{6cm}{\includegraphics[width=0.33\textwidth]{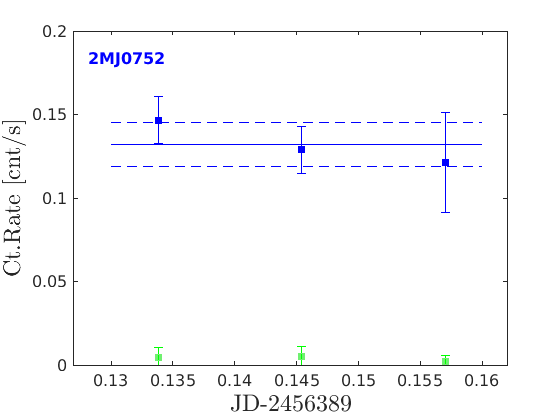}}
\parbox{6cm}{\includegraphics[width=0.33\textwidth]{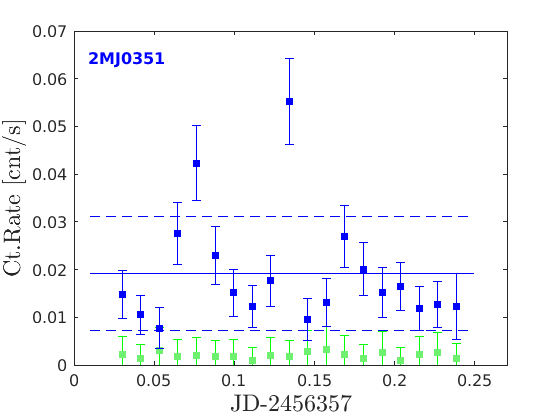}}
\parbox{6cm}{\includegraphics[width=0.33\textwidth]{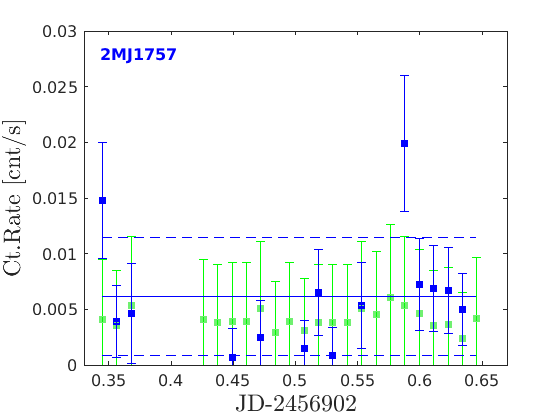}}
}

\parbox{18cm}
{
\parbox{6cm}{\includegraphics[width=0.33\textwidth]{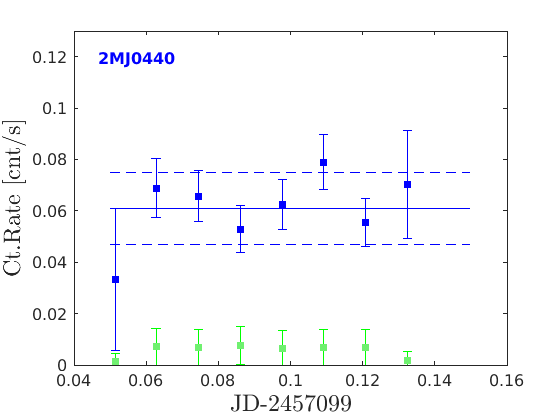}}
\parbox{6cm}{\includegraphics[width=0.33\textwidth]{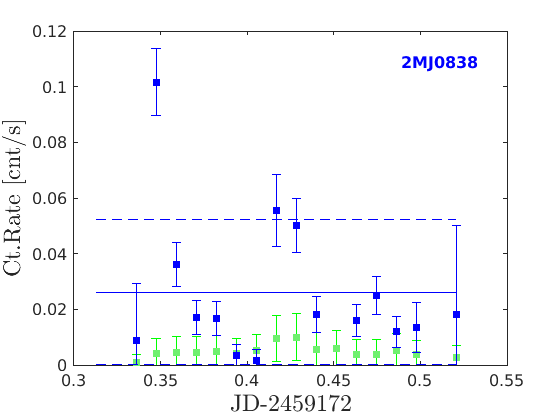}}
\parbox{6cm}{\includegraphics[width=0.33\textwidth]{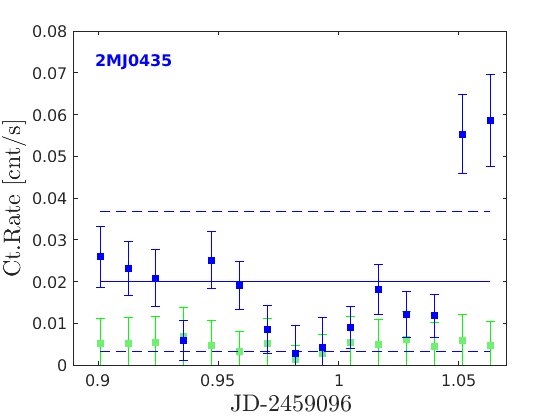}}
}
\caption{Background-subtracted {\it XMM-Newton} EPIC/pn
light curves in the energy band of $0.2-2.0$\,keV of the detected UCDs (blue) and the light curves of the background (green) both with $1$\,ks binsize. The mean count rate and its standard deviation are shown as horizontal solid and dashed bars.} 

\label{fig:lc}
\end{center}
\end{figure*}


\begin{figure*}
    \begin{center} 
     
    \parbox{18cm}
    {
    \parbox{6cm}{\includegraphics[width=0.33\textwidth]{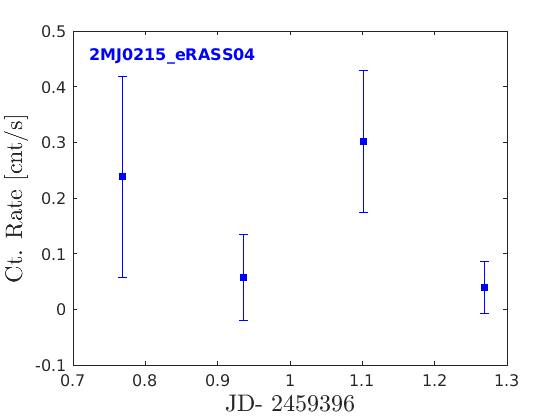}}
    \parbox{6cm}{\includegraphics[width=0.33\textwidth]{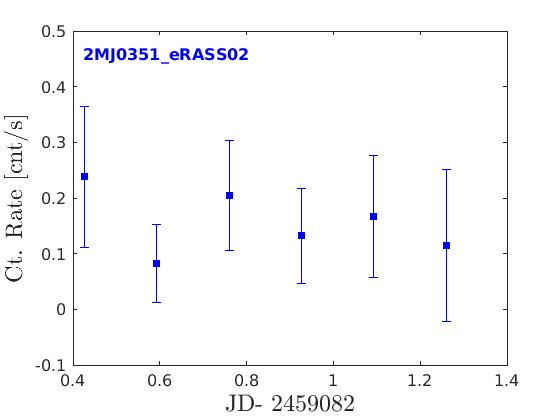}}
    \parbox{6cm}{\includegraphics[width=0.33\textwidth]{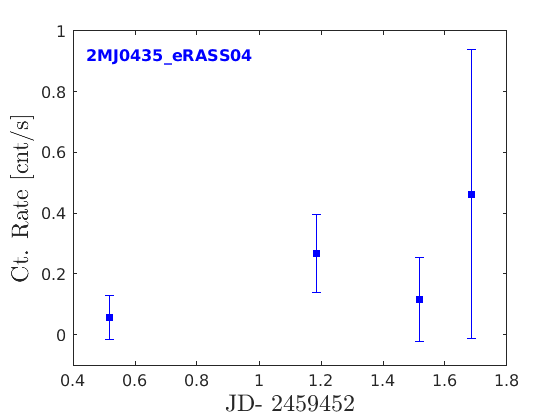}}
    }
    
    \parbox{18cm}
    {
    \parbox{6cm}{\includegraphics[width=0.33\textwidth]{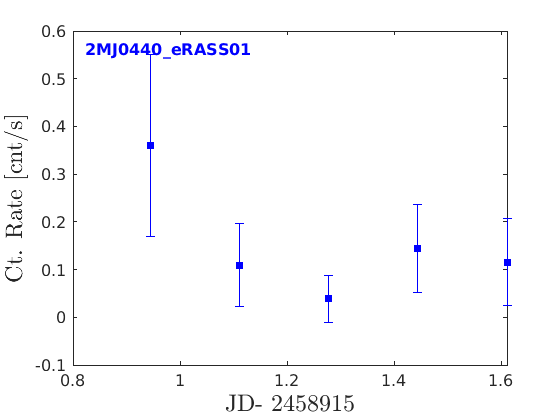}}
    \parbox{6cm}{\includegraphics[width=0.33\textwidth]{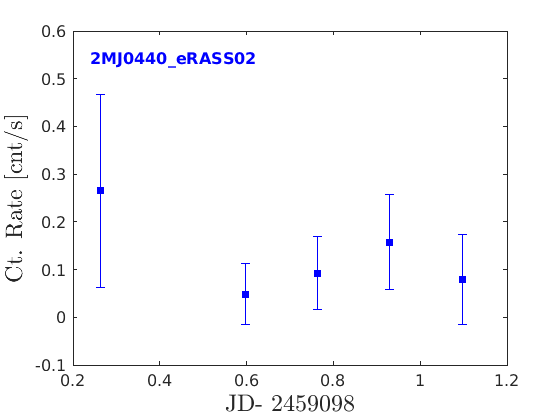}}
    \parbox{6cm}{\includegraphics[width=0.33\textwidth]{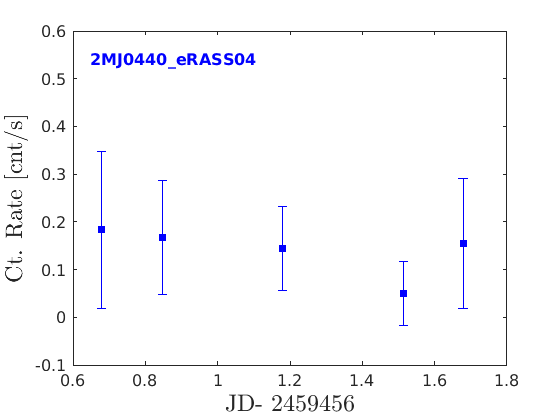}}
    }
    
    \parbox{18cm}
    {
    \parbox{6cm}{\includegraphics[width=0.33\textwidth]{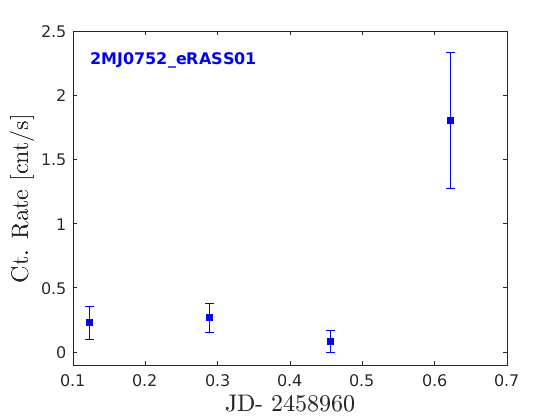}}
    \parbox{6cm}{\includegraphics[width=0.33\textwidth]{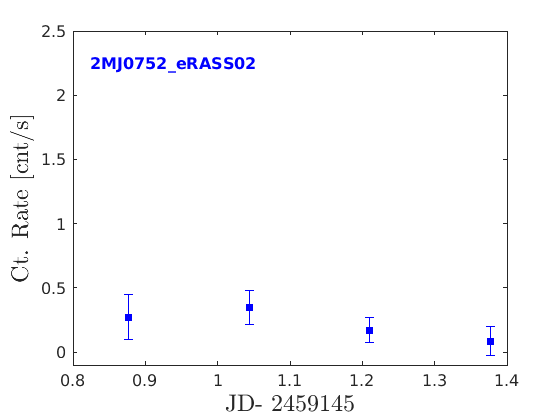}}
    \parbox{6cm}{\includegraphics[width=0.33\textwidth]{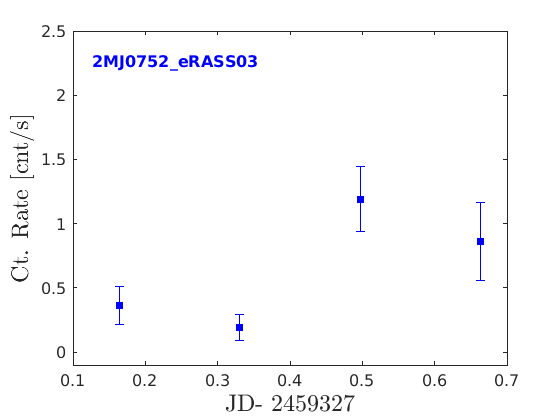}}
    }
    
    \parbox{18cm}
    {
    \parbox{6cm}{\includegraphics[width=0.33\textwidth]{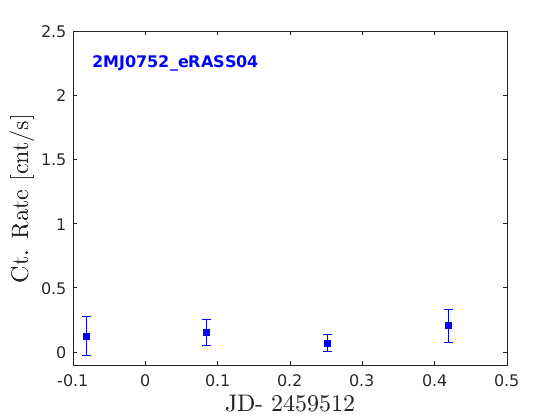}}
    \parbox{6cm}{\includegraphics[width=0.33\textwidth]{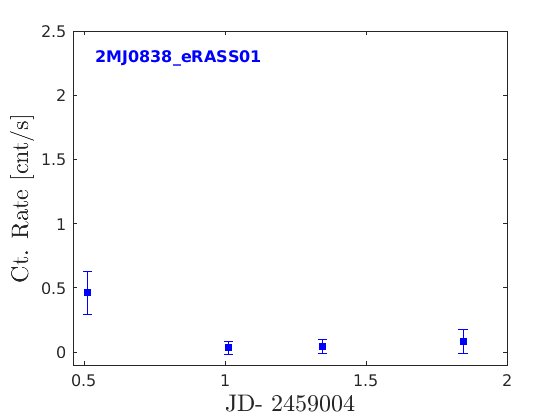}}
    \parbox{6cm}{\includegraphics[width=0.33\textwidth]{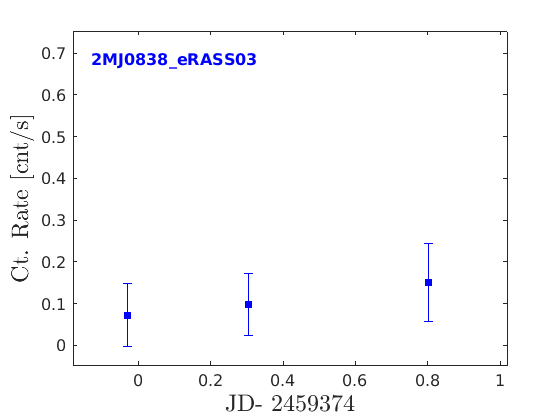}}
   
    }
     
    \parbox{18cm}
    {
    \parbox{6cm}{\includegraphics[width=0.33\textwidth]{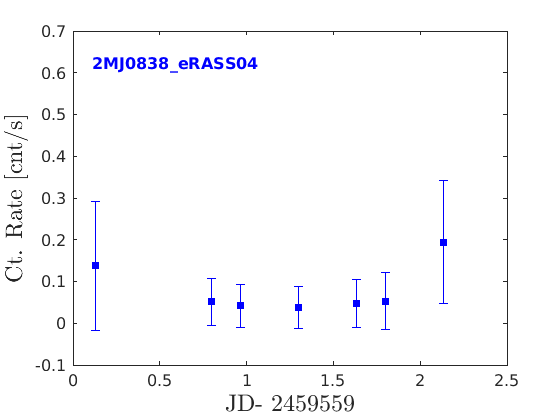}}
    \parbox{6cm}{\includegraphics[width=0.33\textwidth]{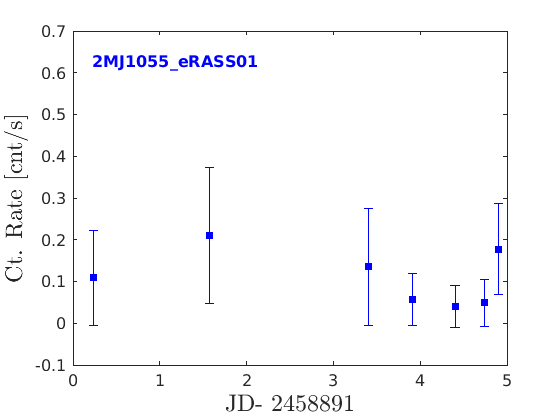}}
    \parbox{6cm}{\includegraphics[width=0.33\textwidth]{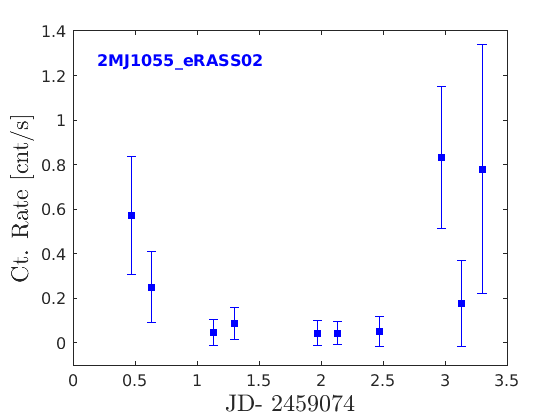}}
    }
    \caption{Background subtracted eROSITA light curves of the two samples presented in this work, the new and the literature UCDs in the $0.2-2.0$\,keV energy band.}
    \label{fig:er_LCs}
    \end{center}
\end{figure*}

\begin{figure*}
    \begin{center} 
    \ContinuedFloat
    \parbox{18cm}
    {
    \parbox{6cm}{\includegraphics[width=0.33\textwidth]{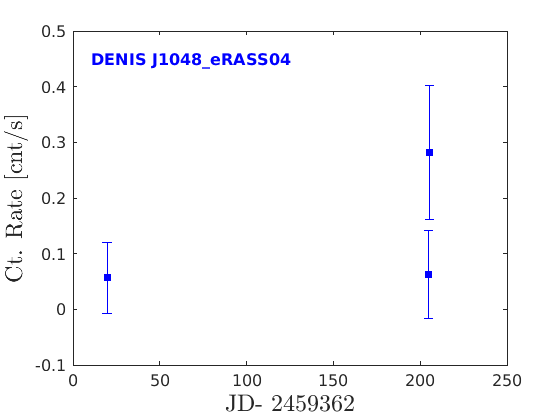}}
    \parbox{6cm}{\includegraphics[width=0.33\textwidth]{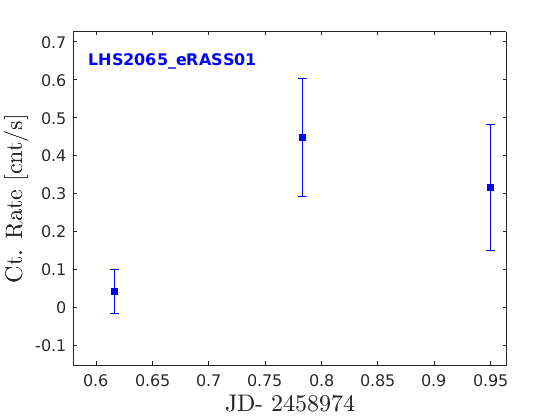}}
    \parbox{6cm}{\includegraphics[width=0.33\textwidth]{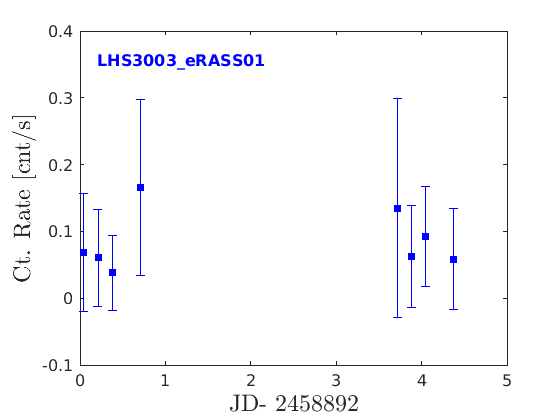}}
    }
    \caption{\it continued}
\label{fig:er_LCs2}
    \end{center}
\end{figure*}

\section{Data analysis}\label{sect:analysis}

We describe here separately the X-ray, radio and optical photometric data analysis steps.


\subsection{X-ray data analysis}
\subsubsection{XMM-Newton}
\label{subsubsect:analysis_xmm}
\begin{table*}
\begin{center}
\begin{threeparttable}[b]
\caption{{\it XMM-Newton} observation log: target name (col. 1), observation ID (col.2), date of observation (col.3),  offset between proper motion corrected expected position and X-ray source (col.4), net source counts and detection maximum likelihood  (cols.5 and 6), count rate (col.7), X-ray and fractional X-ray luminosity (cols.~8 and 9). The last two columns refer to the energy band of $0.2-2.0$\,keV.}
    \label{tab:obslog_xmm}
    \begin{tabular}{lcccccccc}
      \midrule[0.5mm]
      \multicolumn{1}{l}{2MASS\,ID} &
      \multicolumn{1}{c}{ObsID} &
      \multicolumn{1}{c}{Obs.Date} &
      \multicolumn{1}{c}{Offset} &
      \multicolumn{1}{c}{SCTS} &
      \multicolumn{1}{c}{Det.ML} &
      \multicolumn{1}{c}{Rate} &
      \multicolumn{1}{c}{$\log(L_{\rm x})$} &
      \multicolumn{1}{c}{$\log(L_{\rm x}/L_{\rm bol})$} \\
      
      \multicolumn{1}{l}{} &
      \multicolumn{1}{c}{} &
      \multicolumn{1}{c}{} &
      \multicolumn{1}{c}{[arcsec]} &
      \multicolumn{1}{c}{} &
      \multicolumn{1}{c}{} &
      \multicolumn{1}{c}{[$\times 10^{-3}$\,cnt/s]} &
      \multicolumn{1}{c}{[erg/s]} &
      \multicolumn{1}{c}{} \\
      \midrule[0.5mm]
      2MJ0752 & 0690720101 & 2013-04-06 & 7.01 & 243.1$\pm$17.1 & 690 & 130.8$\pm$9.2 & 27.87$\pm$ 0.03 & -2.99\\
      2MJ0351 & 0690720501 & 2013-03-05 & 5.63 & 310.4$\pm$21.1 & 455 & 19.1$\pm$1.3 & 26.78$\pm$0.03 & -3.72\\
      2MJ1757 & 0743900101 & 2014-09-01 & 6.06 & 53.9$\pm$10.9 & 25 & 2.8$\pm$0.5 & 27.34$\pm$ 0.08 & -3.25\\
      2MJ0440 & 0743900201 & 2015-03-17 & 6.03 & 333.5$\pm$20.6 & 706 & 61.8$\pm$3.8 & 26.99$\pm$ 0.03 & -3.32\\
      2MJ0306 & 0743900301 & 2015-01-18 & $\cdot\cdot$ & $\cdot\cdot$ & $\cdot\cdot$ & $<$17.3 & $<$26.76 & $<$-3.44\\
      
      
      2MJ0838 & 0860200101 & 2020-11-18 & 6.13 & 272.9$\pm$19.5 & 361 & 26.9$\pm$1.9 & 26.70$\pm$0.03 & -3.84\\
      2MJ0435 & 0860200201 & 2020-09-04 & 7.47 & 212.8$\pm$17.1 & 138 & 42.2$\pm$3.4 & 26.57$\pm$0.04 & -3.91\\
      2MJ0652 & 0860200301 & 2020-09-22 & $\cdot\cdot$ & $\cdot\cdot$ & $\cdot\cdot$ & $<$0.3 & $<$25.90 & $<$-4.04 \\
      2MJ0215 & 0860200401 & 2020-06-22 & $\cdot\cdot$ & $\cdot\cdot$ & $\cdot\cdot$ & $<$2.1 & $<$26.40 & $<$-3.88 \\
      
      \bottomrule[0.5mm]\end{tabular}
\end{threeparttable}
\end{center}
\end{table*}

For nine UCDs marked with `X' in Table~\ref{tab:ucds_par} we present 
new observations with {\it XMM-Newton} obtained during AO\,11, AO\,13, and AO\,19 with exposure times ranging from $16$ to $45$\,ksec.
We analyzed only the observations from the most sensitive instrument, the EPIC/pn imaging camera, using the {\it XMM-Newton} Science Analysis System (SAS)\footnote{SAS Data Analysis Threads: {\tt \url{https://www.cosmos.esa.int/web/xmm-newton/sas-threads}}.} pipeline, version 20.0.0. We started the data processing with {\tt epproc} which produces the EPIC/pn photon events list. 
 We extracted the high energy light curve (LC) of the whole detector ($10.0<E<12.0$\,keV) to determine the count rate above which the particle background is too high. The corresponding time intervals are removed for the further analysis. The remaining observing time forms the good time intervals (GTIs). The threshold count rate in the whole-detector light curve that defines the GTI was chosen individually for each observation based on visual inspection, and it varies from $0.4$\,cnt/s to $1.5$\,cnt/s. We then filtered the event lists  applying the GTIs, pixel pattern ({\tt 0 $\leq$ pattern $\leq$ 12}), and the \#{\tt xmmea\_ep} filter.

We performed the source detection in five energy bands: $0.2-0.5$\,keV, $0.5-1.0$\,keV, $1.0-2.0$\,keV, $2.0-4.5$\,keV, and $4.5-12.0$\,keV using SAS tools. To identify the X-ray sources associated with our sample we first propagated the {\it Gaia}-DR3 coordinates using the {\it Gaia}-DR3 proper motions to the epoch of the specific {\it XMM-Newton} observation and then we matched these expected positions of the UCDs with the X-ray coordinates from the output list of the source detection process allowing for a maximum distance of $15^{\prime\prime}$. 
This way we establish that $6$ out of $9$ UCDs were detected with EPIC/pn and that their emission is mainly within $2$\,keV, with only few and zero counts in the $2.0-4.5$\,keV and $4.5-12.0$\,keV energy bands, respectively. The remaining three targets out of the $9$ observed are undetected and they are identifiable in Table~\ref{tab:obslog_xmm} where upper limits are displayed. In this table we present all relevant X-ray parameters for the {\it XMM-Newton} observations and the results of the source detection process, namely target name (col.1), observation ID and date (cols.2 \& 3), offset between the optical and X-ray coordinates (col.4), source counts (col.5), detection likelihood (col.6), net count rate (col.7), all provided by the detection algorithm for the broad energy band ($0.2-12.0$\,keV). 
For the three undetected targets we calculated their upper limit count rates from the broad band sensitivity map of the EPIC/pn observations that we extracted with the SAS tool {\tt esensmap}.

To enable a comparison with the literature, the energy band adopted hereafter is the combination of the softest bands we used for the source detection: $0.2-0.5$\,keV, $0.5-1.0$\,keV, and $1.0-2.0$\,keV.  Thus, the X-ray luminosity and the fractional X-ray luminosity, given in cols.8\,\&\,9 of Table~\ref{tab:obslog_xmm}, refer to $0.2-2.0$\,keV (see Sect.~\ref{subsubsect:analysis_xrays_lum} for the calculation of these parameters).

For the spectral and temporal analysis we retained only pixels with {\tt flag} $= 0$. We defined a circular photon extraction region with a radius of $35^{\prime\prime}$ centered on the EPIC/pn source position. The background was extracted from an adjacent source-free circular region on the same CCD chip with a radius twice that of the source region. We grouped the spectral channels such that each bin comprises a minimum of $10-20$\,photons depending on the brightness of the source.
The response matrix and ancillary response for the spectral analysis were created using the appropriate SAS tools, {\tt rmfgen} and {\tt arfgen}. 

As mentioned above, the X-ray emission of the detected UCDs is mainly within $2$\,keV, with less than $5$\,\% of the source counts in the harder energy band of $2.0-12.0$\,keV. 
This justifies our choice of the $0.2-2.0$\,keV band for the extraction of the X-ray light curves.
The EPIC/pn light curves for source and background were generated with the SAS tool {\tt evselect}, and the background subtraction was applied using the {\sc epiclccorr} task, which also takes care of corrections for instrumental effects such as vignetting. The corrected light curves of our sample are displayed in Fig.~\ref{fig:lc} for a $1$\,ks bin size and with the background emission overlaid in green. Events that are not within the GTIs, have been removed during the filtering process. This may lead to a reduced number of bins in the light curve. For instance, the light curve of LP\,423-31 (see left top panel in Fig.~\ref{fig:lc}) does not show any ``good'' event  during more than half of its observation.

Visual inspection of the light curves reveals significant variability for four UCDs: 2MASS\,J03510004-0052 (2MJ0351), 2MASS\,J17571539+7042 (2MJ1757), 2MASS\,J08380224-5855 (2MJ0838), and 2MASS\,J04351612-1606 (2MJ0435). In Fig.~\ref{fig:lc} we present the average count rate of the corrected source light curve and its standard deviation with solid and dashed blue lines, respectively. The bins located above the standard deviation upper bound can be considered to represent possible flaring activity. The four UCDs with bins that have a count rate above the standard deviation are, indeed, the same ones that we identified as variable from visual inspection.


\begin{table*}
\begin{center}
\begin{threeparttable}[b]
\caption{Results from the source detection of eROSITA data: target name (col. 1), survey number and X-ray coordinates with their uncertainties (col.2 to 5),  net count rate and detection maximum likelihood in the broad band $0.2-10.0$\,keV (cols.6\&7),  offset between proper motion corrected expected position and eROSITA X-ray source (col.8), X-ray and fractional X-ray luminosity (cols.~9 and 10). The last two columns refer to the energy band of $0.2-2.0$\,keV.}
    \label{tab:obslog_er}
    \begin{tabular}{lccccccccc}
      \midrule[0.5mm]
    \multicolumn{1}{c}{2MASS ID} &
    \multicolumn{1}{c}{eRASS} &
    \multicolumn{1}{c}{RA} &
    \multicolumn{1}{c}{DEC} &
    \multicolumn{1}{c}{RADEC\_ERR} &
    \multicolumn{1}{c}{Rate} &
    \multicolumn{1}{c}{Det.ML} &
    \multicolumn{1}{c}{$\rm Sep_{\rm opt,X}$} &
    \multicolumn{1}{c}{$\log L_{\rm x}$} &
    \multicolumn{1}{c}{$\log (L_{\rm x}/L_{\rm bol})$} \\
      
    \multicolumn{1}{l}{} &
    \multicolumn{1}{c}{} &
    \multicolumn{1}{c}{[deg]} &
    \multicolumn{1}{c}{[deg]} &
    \multicolumn{1}{c}{[arcsec]} &
    \multicolumn{1}{c}{[$\times 10^{-3}$\,cnt/s]} &
    \multicolumn{1}{c}{} &
    \multicolumn{1}{c}{[arcsec]} &
    \multicolumn{1}{c}{[erg/s]} &
    \multicolumn{1}{c}{} \\
    \midrule[0.5mm]
    
 2MJ0752 & 1 & 118.101629 & 16.202432 & 3.64 & 340.15 $\pm$ 68.44 & 61.67 & 3.13 & 28.06$\pm$0.10 & -2.79  \\
 2MJ0752 & 2 & 118.100579 & 16.202269 & 3.11 & 209.96 $\pm$ 53.05 & 49.23 & 0.84 & 27.85$\pm$0.12 & -3.00 \\
 2MJ0752 & 3 & 118.102579 & 16.202739 & 2.65 & 451.07 $\pm$ 71.40 & 125.98 & 6.54 & 28.19$\pm$0.08 & -2.67 \\
 2MJ0752 & 4 & 118.100588 & 16.202026 & 6.31 & 136.29 $\pm$ 66.28 & 9.01 & 0.86 & 27.66$\pm$0.21 & -3.19 \\
 
 2MJ0351 & 1 & $\cdot\cdot$ & $\cdot\cdot$ & $\cdot\cdot$ & $<$84.60 & $\cdot\cdot$ & $\cdot\cdot$ & $<$27.24 & $<$-3.27 \\
 2MJ0351 & 2 & 57.751248 & -0.881977 & 4.50 & 115.89 $\pm$ 34.89 & 33.89 & 3.88 & 27.37$\pm$0.13 & -3.13\\
 2MJ0351 & 3 & $\cdot\cdot$ & $\cdot\cdot$ & $\cdot\cdot$ & $<$52.80 & $\cdot\cdot$ & $\cdot\cdot$ & $<$27.03 & $<$-3.47 \\
 2MJ0351 & 4 & $\cdot\cdot$ & $\cdot\cdot$ & $\cdot\cdot$ & $<$36.40 & $\cdot\cdot$ & $\cdot\cdot$ & $<$26.87 & $<$-3.63 \\
 
 2MJ0440 & 1 & 70.096931 & -5.501754 & 3.96 & 81.68 $\pm$ 25.11 & 22.83 & 6.96 & 26.87$\pm$0.14 & -3.45 \\
 2MJ0440 & 2 & 70.099682 & -5.502688 & 5.87 & 102.56 $\pm$ 47.18 & 13.67 & 5.03 & 26.97$\pm$0.20 & -3.55 \\
 2MJ0440 & 3 & $\cdot\cdot$ & $\cdot\cdot$ & $\cdot\cdot$ & $<$66.50 & $\cdot\cdot$ & $\cdot\cdot$ & $<$26.78 & $<$-3.54 \\
 2MJ0440 & 4 & 70.100214 & -5.501299 & 2.19 & 258.81 $\pm$ 42.59 & 128.07 & 4.38 & 27.37$\pm$0.08 & -2.95 \\
 
 2MJ0838 & 1 & 129.509037 & -58.936883 & 5.71 & 51.72 $\pm$ 21.96 & 12.17 & 8.07 & 26.78$\pm$0.19 & -3.77 \\
 2MJ0838 & 2 & $\cdot\cdot$ & $\cdot\cdot$ & $\cdot\cdot$ & $<$72.06 & $\cdot\cdot$ & $\cdot\cdot$ & $<$26.92 & $<$-3.62 \\
 2MJ0838 & 3 & 129.507423 & -58.935126 & 5.64 & 26.98 $\pm$ 11.55 & 8.42 & 2.19 & 26.50$\pm$ 0.20 & -4.05 \\
 2MJ0838 & 4 & 129.511589 & -58.930834 & 4.84 & 35.01 $\pm$ 12.30 & 9.10 & 15.49 & 26.61$\pm$0.18 & -3.94 \\

 2MJ0435 & 1 & $\cdot\cdot$ & $\cdot\cdot$ & $\cdot\cdot$ & $<$60.80 & $\cdot\cdot$ & $\cdot\cdot$ & $<$26.81 & $<$-3.67 \\
 2MJ0435 & 2 & $\cdot\cdot$ & $\cdot\cdot$ & $\cdot\cdot$ & $<$36.70 & $\cdot\cdot$ & $\cdot\cdot$ & $<$26.60 & $<$-3.90 \\
 2MJ0435 & 3 & $\cdot\cdot$ & $\cdot\cdot$ & $\cdot\cdot$ & $<$47.90 & $\cdot\cdot$ & $\cdot\cdot$ & $<$26.71 & $<$-3.77 \\
 2MJ0435 & 4 & 68.818034 & -16.115683 & & 85.30 $\pm$ 0.0 & 11.07 & 6.15 & 26.96$\pm$0.04 & -3.53 \\
 
 2MJ0215 & 1 & $\cdot\cdot$ & $\cdot\cdot$ & $\cdot\cdot$ & $<$16.64 & $\cdot\cdot$ & $\cdot\cdot$ & $<$26.49 & $<$-3.78 \\
 2MJ0215 & 2 & $\cdot\cdot$ & $\cdot\cdot$ & $\cdot\cdot$ & $<$5.30 & $\cdot\cdot$ & $\cdot\cdot$ & $<$25.99 & $<$-4.28 \\
 2MJ0215 & 3 & $\cdot\cdot$ & $\cdot\cdot$ & $\cdot\cdot$ & $<$63.80 & $\cdot\cdot$ & $\cdot\cdot$ & $<$27.08 & $<$-3.20 \\
 2MJ0215 & 4 & 33.791500 & -30.671689 & 6.67 & 68.35 $\pm$ 21.69 & 15.98 & 11.97 & 27.11$\pm$0.14 & -3.17 \\
 
 2MJ1055 & 1 & 163.829431 & -73.936249 & 7.64 & 27.99 $\pm$ 8.75 & 11.44 & 12.65 & 26.6$\pm$0.14 & -3.95 \\
 2MJ1055 & 2 & 163.817294 & -73.936770 & 4.38 & 85.32 $\pm$ 21.61 & 33.50 & 1.20 & 27.13$\pm$0.12 & -3.46 \\ 
 2MJ1055 & 3 & $\cdot\cdot$ & $\cdot\cdot$ & $\cdot\cdot$ & $<$48.40 & $\cdot\cdot$ & $\cdot\cdot$ & $<$26.88 & $<$-3.71 \\
 2MJ1055 & 4 & $\cdot\cdot$ & $\cdot\cdot$ & $\cdot\cdot$ & $<$22.00 & $\cdot\cdot$ & $\cdot\cdot$ & $<$26.54 & $<$-4.05 \\

 \midrule[0.1mm]
 \multicolumn{10}{c}{Literature sample}\\
 \midrule[0.1mm]
 DENIS J1048 & 1 & $\cdot\cdot$ & $\cdot\cdot$ & $\cdot\cdot$ & $<$54.24 & $\cdot\cdot$ & $\cdot\cdot$ & $<$25.92 & $<$-4.12  \\
 DENIS J1048 & 2 & $\cdot\cdot$ & $\cdot\cdot$ & $\cdot\cdot$ & $<$27.52 & $\cdot\cdot$ & $\cdot\cdot$ & $<$25.63 & $<$-4.42 \\
 DENIS J1048 & 3 & $\cdot\cdot$ & $\cdot\cdot$ & $\cdot\cdot$ & $<$26.46 & $\cdot\cdot$ & $\cdot\cdot$ & $<$25.62 & $<$-4.44 \\
 DENIS J1048 & 4 & 162.052061 & -39.939468 & 5.33 & 48.72$\pm$23.86 & 7.16 & 6.54 & 25.88$\pm$0.22 & -4.12 \\
 LHS2065 & 1 & 133.398799 & -3.494837 & 4.13 & 152.08$\pm$43.39 & 31.40 & 6.38 & 27.04$\pm$0.13 & -2.98 \\
 LHS2065 & 2 & $\cdot\cdot$ & $\cdot\cdot$ & $\cdot\cdot$ & $<$68.83 & $\cdot\cdot$ & $\cdot\cdot$ & $<$26.70 & $<$-3.37 \\ 
 LHS2065 & 3 & $\cdot\cdot$ & $\cdot\cdot$ & $\cdot\cdot$ & $<$28.47 & $\cdot\cdot$ & $\cdot\cdot$ & $<$26.31 & $<$-3.76 \\ 
 LHS2065 & 4 & $\cdot\cdot$ & $\cdot\cdot$ & $\cdot\cdot$ & $<$68.65 & $\cdot\cdot$ & $\cdot\cdot$ & $<$26.70 & $<$-3.37 \\ 
        \bottomrule[0.5mm]\end{tabular}
\end{threeparttable}
\end{center}
\end{table*}

\subsubsection{eROSITA}
\label{subsubsect:analysis_erosita}

eROSITA data was analyzed for both the UCDs from our new dedicated X-ray/radio campaigns and the literature sample defined in Sect.~\ref{sect:sample}. Hereby, we made use of the preliminary version of the catalog provided by the eROSITA\_DE consortium (all\_s4\_SourceCat3B\_221031\_poscorr\_mpe\_clean.fits) that combines the X-ray data from all $4$ eROSITA All-sky surveys (eRASS:4). We propagated the {\it Gaia}-DR3 coordinates of all UCDs for their proper motion (P.M.) to the mean epoch of eRASS:4 catalog, 15 Dec 2020, and we then performed a match with the boresight corrected coordinates (RA\_CORR, DEC\_CORR) of eRASS:4 within $30^{\prime\prime}$. We found potential eROSITA matches
for $2$ UCDs from the literature and $8$ of the UCDs from our new X-ray/radio sample.

To confirm our association we performed a reverse match as in \citet{Magaudda2022}. Starting from the eRASS:4 coordinates we looked for all possible {\it Gaia}-DR3 CTPs within $30\,^{\prime\prime}$. We found $43$ and $11$ possible {\it Gaia}-DR3 CTPs for our new UCD and the literature samples, respectively. 
After propagating the positions of all possible {\it Gaia} CTPs to the mean eRASS:4 epoch, we selected the closest to each target of our sample we imposed that the closest eROSITA CTP must satisfy the condition that the optical and X-ray separation be smaller than three times the uncertainty of the eRASS:4 coordinates ($Sep_{\rm X-ray,opt} < 3 \times \rm{RADEC\_ERR}$). Finally, the match provided an eROSITA detection for all $8$ new UCDs and $2$ targets from the literature, that is all matches achieved before the reverse match
were confirmed. 
We aim to compare the results from the analysis of {\it XMM-Newton} data to those from eROSITA in the same energy bands. Since eROSITA bands used for eRASS:4 are not the same as those of {\it XMM-Newton} that we adopted for this work, we performed our own source detection on the eROSITA data, and we extracted the eRASS spectra and light curves for the {\it XMM-Newton} energy bands. 

From the DETUID column of eRASS:4 we found in which eROSITA sky maps our targets are located.  Then we downloaded the event files of the sky maps of our sample for each of the four surveys.
We performed the source detection  using the eSASSusers\_211214 software release \citep{Brunner22.0} in the same energy bands adopted for {\it XMM-Newton} data extraction, except for the hardest band that is slightly narrower ($4.5-10$\,keV) because eROSITA has very little effective area above $10.0$\,keV.
After creating the exposure map, detection mask and background map within the adopted energy bands, we compiled the list of the detected sources for each of the $4$ surveys using the eSASS pipeline {\tt ermldet} for which we used a threshold detection maximum likelihood of $6.0$. We found that $7$ of the new UCDs are detected in at least one survey, that is one target less than according to the official eRASS:4 catalog. We recovered the $2$ UCDs of the literature sample that are also present in eRASS:4. The missing UCD detection can be explained by different detection likelihood (DET\_LIKE) thresholds. The value adopted for the compilation of the eRASS:4 catalog ($\rm DET\_LIKE=5$)  is smaller than ours and the missing object (2MJ0306) shows $<20$\,counts in eRASS:4. 

With eROSITA data we added one more X-ray detection to our UCD sample that is associated to 2MASS\,J10551532-7356 (2MJ1055), for which no time was allocated with {\it XMM-Newton}.  Moreover, we enabled a long-term variability analysis, for instance 2MASS\,J02150802-3040 (2MJ0215) is an upper limit with {\it XMM-Newton} but appears as a detection during the last eROSITA survey, eRASS4.

The UCDs without detection in any of the 4 eRASS all have a detection with {\it XMM-Newton}. The eRASS upper limits are shallower than the {\it XMM-Newton} data and would, therefore, not provide additional information. Therefore, we decided to ignore them and calculate the upper limits only for those targets that are detected during at least one eROSITA survey. The X-ray parameters of the eROSITA sources detected during one or more eRASSs are given in Table~\ref{tab:obslog_er} with the results of the upper limit calculation from the extraction of the aperture photometry with the {\tt apetool} pipeline.
Specifically, in Table~\ref{tab:obslog_er} we provide the acronym for the 2MASS name of the object and the identifying number of the eROSITA survey
(col.1\&2), the X-ray coordinates with their uncertainty (cols.3$-$5), the offset between the P.M. corrected optical position and the X-ray coordinates (col.6), the $0.2-10.0$\,keV count rate obtained from the source detection procedure (col.7), and the detection maximum likelihood in the same energy band (col.8). The last two columns show the $L_{\rm x}$ and $L_{\rm x}/L_{\rm bol}$ values computed for the energy band of $0.2-2.0$\,keV, as for the {\it XMM-Newton} data (see Sect.~\ref{subsubsect:analysis_xrays_lum}). 

Finally, we extracted the spectrum and light curve for each of the eROSITA detections using the {\tt srctool} pipeline with the option {\tt AUTO} for the choice of the source and background region sizes. The eROSITA light curves and spectra are shown in Figs.~\ref{fig:er_LCs} \& \ref{fig:xspec}. 
None of the literature targets show a bright eROSITA detection, thus only their light curves are shown in Fig.~\ref{fig:er_LCs}. The bin size of the light curves is $14.4$\,ks corresponding to one eRODay, i.e. the time that eROSITA takes to return to the same sky position during its scanning motion \citep[][]{Predehl2021}.
This means there is one bin in the light curve for each of eROSITA's visits of the source.
 

\subsubsection{X-ray spectra}\label{subsubsect:analysis_xrays_xspec}

We fitted all EPIC/pn and eROSITA spectra with net counts $>$200 and net counts $>$30, respectively. 
These thresholds were set  after several tests to find the best-fit parameters. 
They are met by $6$ UCds, $5$ detected with {\it XMM-Newton} and one from eRASS where it exceeds the above threshold in two out of 4 surveys.

We made use of XSPEC version 12.13 \citep{Arnaud1996} using a one- or two-temperature thermal {\tt APEC} model. Given that nearly all observed counts are concentrated below $2.0$\,keV (see Sect.~\ref{subsubsect:analysis_xmm}) we fit the {\tt apec} model to the spectrum in the  
$0.2-2.0$\,keV energy band. 
Interstellar absorption is negligible for these nearby objects. Given the low statistics of the spectra we fixed the global abundance ($Z$) to $0.3\,Z_{\odot}$, a typical value for stellar coronae \citep{Maggio2007}, and we adopted the solar abundance vector set from \citet{Asplund2009}.  The free parameters are thus, the temperatures ($kT$) and emission measures ($EM$). We calculated the $1\,\sigma$ uncertainties of the fit parameters using the {\sc error} command in XSPEC. We used the best-fit parameters to compute for each spectrum the emission measure weighted mean coronal temperature ($<kT_{\rm corona}>$) defined as
\begin{equation}
  <kT_{\rm corona}> = \frac{\sum(\rm EM_{\rm n} \cdot kT_{\rm n})}{\sum{\rm EM_{\rm n}}}
\end{equation}
where $kT_{\rm n}$ \& $EM_{\rm n}$ with $n=1,2$ are the two temperatures and two emission measures of the best-fitting model.

\begin{figure*}
\begin{center}
\parbox{18cm}
{
\parbox{6cm}{\includegraphics[width=0.33\textwidth]{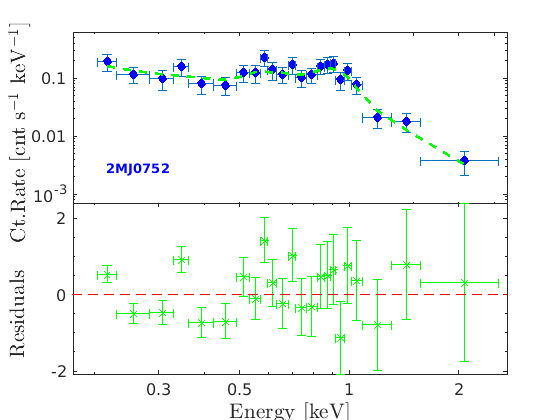}}
\parbox{6cm}{\includegraphics[width=0.33\textwidth]{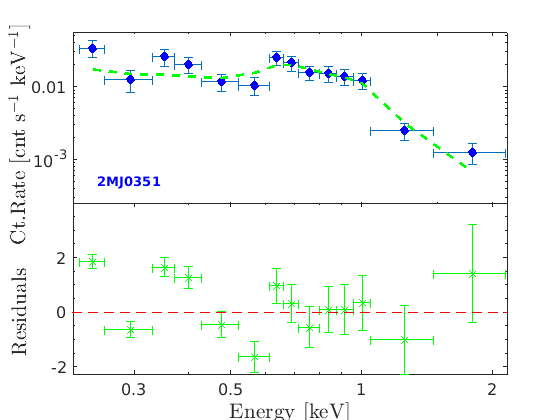}}
\parbox{6cm}{\includegraphics[width=0.33\textwidth]{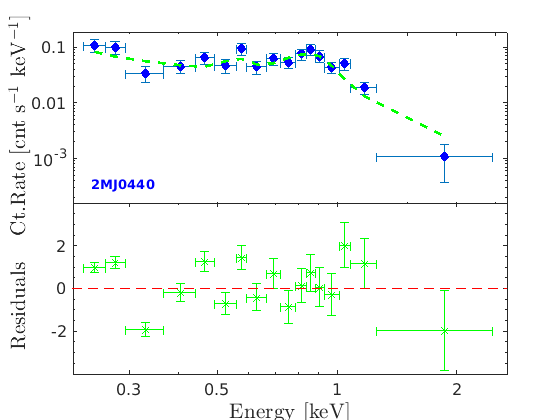}}
}

\parbox{18cm}
{
\parbox{6cm}{\includegraphics[width=0.33\textwidth]{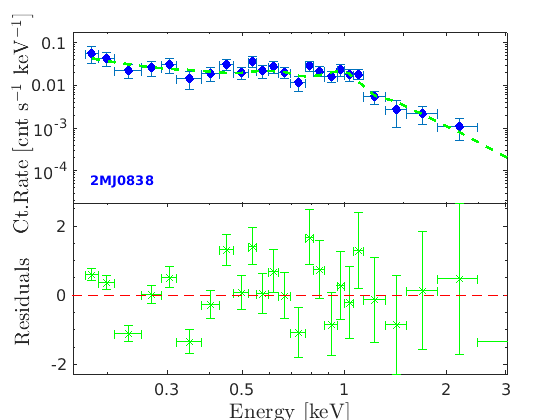}}
\parbox{6cm}{\includegraphics[width=0.33\textwidth]{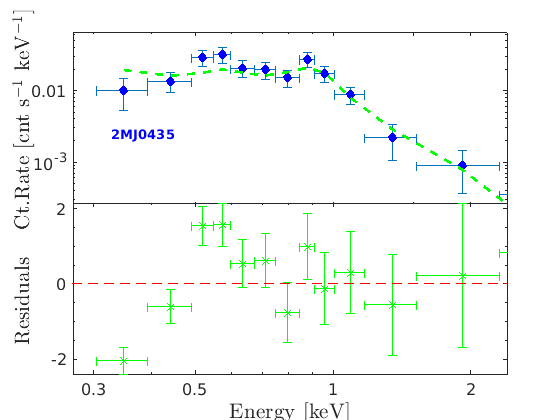}}
\parbox{6cm}{\includegraphics[width=0.33\textwidth]{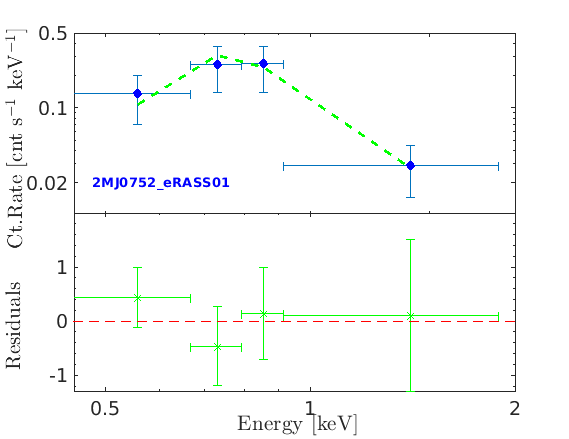}}
}

\parbox{18cm}
{
\parbox{6cm}{\includegraphics[width=0.33\textwidth]{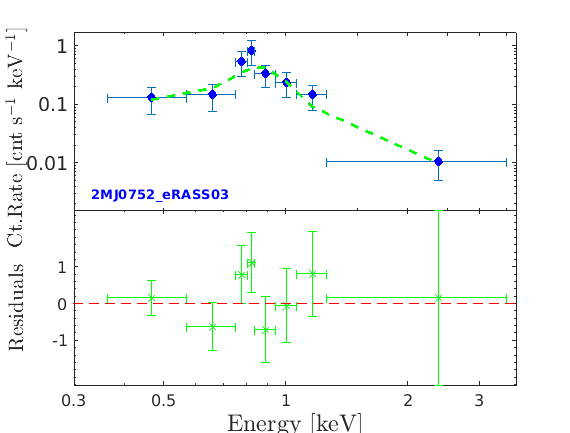}}
\parbox{12cm}{\caption{X-ray spectra of UCDs associated with {\it XMM-Newton} (the first five panels) and eROSITA (the last two panels) sources that have more than $200$ and $30$ net counts, respectively. The spectra are plotted together with the best-fitting {\tt apec} model and the residuals (both shown in green); see text in Sect.~\ref{subsubsect:analysis_xrays_xspec} and Table~\ref{tab:xspec}.}
\label{fig:xspec}}
}
\end{center}
\end{figure*}

The results of the best-fitting models are summarized for all UCDs in Table~\ref{tab:xspec}, where we show the values for the two $kT$ and $EM$ components with their $1\,\sigma$ uncertainties (cols.~2 to~5), the reduced $\chi$ square ($\chi^{2}_{\rm red}$) and the degrees of freedom (d.o.f.) in cols.~6 \& 7 and the average coronal temperature ($<kT_{\rm corona}>$) in the last column.
The seven spectra, their best-fit models and the relative residuals between data and model are displayed in Fig.~\ref{fig:xspec}.
\addtolength{\tabcolsep}{5pt}

\begin{table*}
    \begin{center}
    \caption{Best-fit X-ray spectral parameters for the five UCDs with net counts $>200$ detected with {\it XMM-Newton} and for the only object that shows two eROSITA detections with net counts $>30$. The best-fit parameters refer to the $0.2-2.0$\,keV energy band.}
    \label{tab:xspec}
        \begin{tabular}{lccccccc}
        \midrule[0.5mm]
      \multicolumn{1}{l}{Other\_Name} &
      \multicolumn{1}{c}{$kT_{\rm 1}$} &
      \multicolumn{1}{c}{$\log (EM_{\rm 1})$} &
      \multicolumn{1}{c}{$kT_{\rm 2}$} &
      \multicolumn{1}{c}{$\log(EM_{\rm 2})$} &
      \multicolumn{1}{c}{$\chi^{2}_{red}$} &
      \multicolumn{1}{c}{d.o.f.} &
      \multicolumn{1}{c}{$<kT_{\rm corona}>$} \\
      
      \multicolumn{1}{l}{} &
      \multicolumn{1}{c}{[keV]} &
      \multicolumn{1}{c}{[$\rm cm^{-3}$]} &
      \multicolumn{1}{c}{[keV]} &
      \multicolumn{1}{c}{[$\rm cm^{-3}$]} &
      \multicolumn{1}{c}{} &
      \multicolumn{1}{c}{} &
      \multicolumn{1}{c}{[keV]}\\
      \midrule[0.5mm]
      \multicolumn{8}{c}{{\it XMM-Newton}: Sect.~\ref{subsubsect:analysis_xmm}} \\
      \midrule[0.1mm]

        2MJ0752 & 0.22 $\pm$ 0.09 & 50.58 $\pm$ 0.20 & 0.86 $\pm$ 0.15 & 50.67 $\pm$ 0.12 & 0.65 & 19 & 0.57 $\pm$ 0.09 \\ 
        2MJ0351 & 0.28 $\pm$ 0.05 & 49.57 $\pm$ 0.11 & 1.08 $\pm$ 0.29 & 49.33 $\pm$ 0.16 & 1.25 & 25 & 0.58 $\pm$ 0.11 \\ 
        2MJ0440 & 0.15 $\pm$ 0.06 & 49.78 $\pm$ 0.24 & 0.75 $\pm$ 0.09 & 49.79 $\pm$ 0.08 & 1.94 & 13 & 0.46 $\pm$ 0.06 \\ 
        2MJ0838 & 0.22 $\pm$ 0.07 & 49.40 $\pm$ 0.18 & 1.12 $\pm$ 0.18 & 49.58 $\pm$ 0.11 & 0.85 & 21 & 0.77 $\pm$ 0.11 \\ 
        2MJ0435 & 0.17 $\pm$ 0.10 & 49.38 $\pm$ 0.18 & 0.84 $\pm$ 0.25 & 49.37 $\pm$ 0.12 & 1.48 & 19 & 0.50 $\pm$ 0.13 \\ 
      \midrule[0.1mm]
     \multicolumn{8}{c}{ eROSITA: Sect.~\ref{subsubsect:analysis_erosita}} \\
      \midrule[0.1mm]
        2MJ0752 (eRASS01) & 0.48$\pm$0.28 & 51.06$\pm$0.28 & $\cdot\cdot$ & $\cdot\cdot$ & 0.22 & 2 & 0.48$\pm$0.28 \\
        2MJ0752 (eRASS03)  & 0.81$\pm$0.22 & 51.18$\pm$0.13 & $\cdot\cdot$ & $\cdot\cdot$ & 0.58 & 6 & 0.81$\pm$0.22 \\
        \bottomrule[0.5mm]
        \end{tabular}
    \end{center}
\end{table*}

\begin{figure}
\centering
    \includegraphics[width=0.5\textwidth]{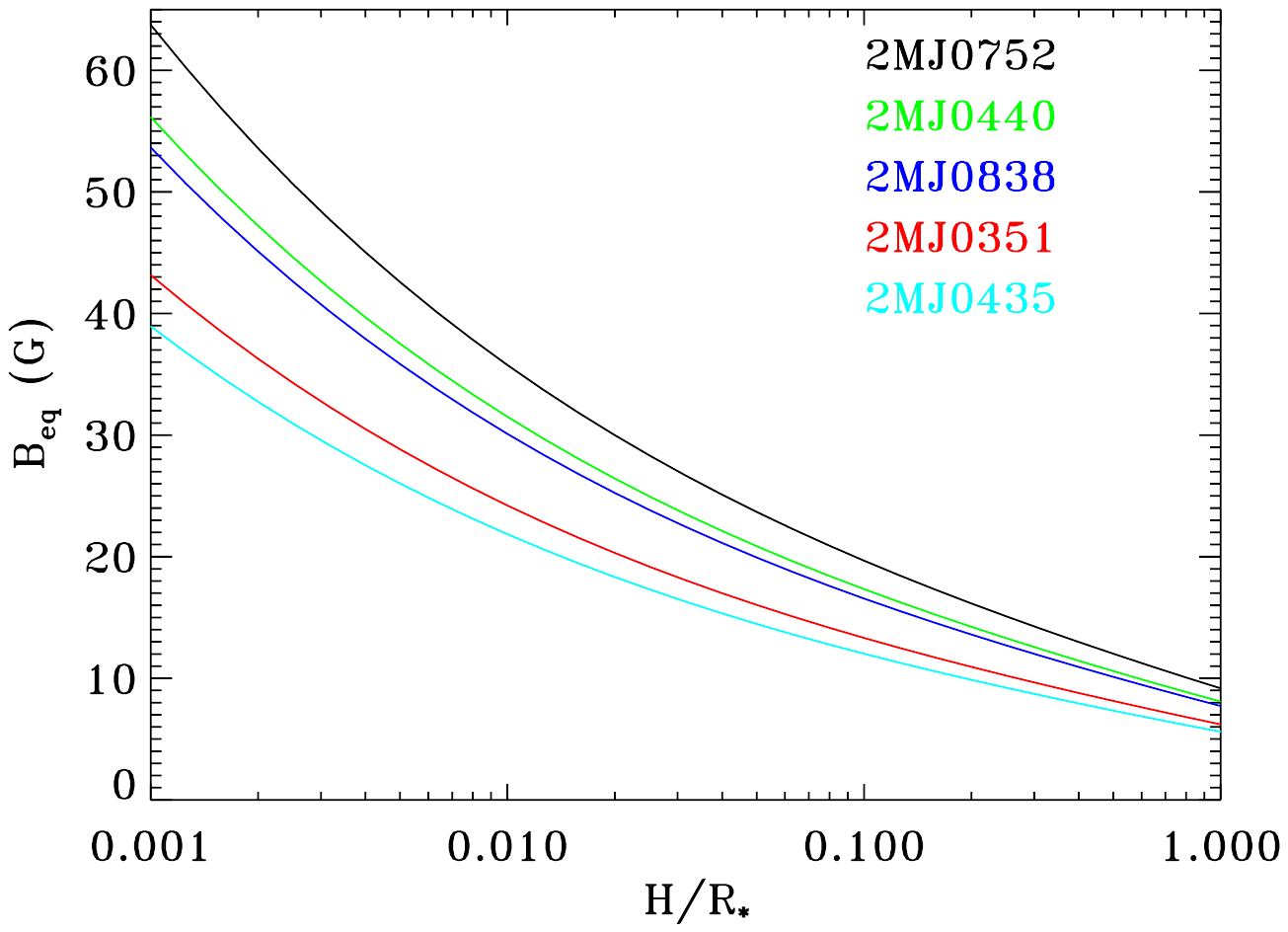}
    \caption{Plot of equipartition magnetic field strength for objects with spectral fits from {\it XMM-Newton} as a function of assumed thickness of a spherically symmetric corona, using coronal temperatures and emission measures from Table~\ref{tab:xspec} and radius measurements from Table~\ref{tab:stepar}.
    \label{fig:xray_eq}}
\end{figure}

We use the X-ray emission properties derived from spectral fits to the five objects with more than 200\,counts in the {\it XMM-Newton} data
(Table~\ref{tab:xspec}) to 
construct constraints on the size scale of the thermal coronal emission. 
We start with 
the pressure equilibrium equation, \\
\begin{equation}
    B_{\rm eq}^{2}/8\pi = n_{e}k_{B}T 
\end{equation}
where $B_{\rm eq}$ is the equipartition magnetic field strength that balances thermal coronal pressure represented by electron density $n_{e}$, and coronal temperature $k_{B}T$, where $k_{B}$ is Boltzmann's constant.
The electron  density is obtained from the volume emission measure (VEM) according to
\begin{equation}
    VEM = n_{e}^{2} V 
\end{equation} 
where $V$ is the emitting volume.
Combining Eqs.~2 and~3 and assuming
that the corona is a spherical shell with height $H$ we can derive the equipartition magnetic field strength as a function of coronal height according to
\begin{equation}
    \frac{VEM (8\pi k_{B}T)^{2}}{B_{\rm eq}^{4}} = V=4\pi/3(R_{\star}+H)^{3} - 4\pi/3R_{\star}^{3}.
\end{equation}
Inspection of Table~\ref{tab:xspec} shows that, while there is a significant difference between the two X-ray temperatures fitted, the volume emission measures of each component are roughly equal within the error bars. Therefore, because the hotter temperature plasma will provide a stronger constraint on the confining magnetic field, we use the temperature and emission measure values of the
hotter component. 
The results are shown in Fig.~\ref{fig:xray_eq} and further discussed in Sect.~\ref{sect:results_multil}.

\subsubsection{X-ray luminosities}\label{subsubsect:analysis_xrays_lum}

For the brightest detections for which we performed the spectral analysis, we computed the X-ray flux ($f_{\rm x}$) with the {\texttt flux} routine of XSPEC. For the fainter sources, as well as for the upper limits, we converted the count rate into flux with a conversion factor based on the results of our spectral analysis and in the $0.2-2.0$\,keV energy band. Specifically, we calculated the CF of each bright target as the ratio of the XSPEC flux and the count rate from the source detection. Then we calculated the mean value,

\begin{equation}
    <CF> = mean\left(\frac{f_{\rm x}}{Ct.Rate}\right)
\end{equation}

We adopted this approach for both instruments (EPIC/pn and eROSITA) separately, so that the instrumental properties are implicitly  considered. The final CFs are $\rm CF_{XMM}=2.71\times10^{-12}\pm3.44\times10^{-12}\,\rm erg/cm^2/cts$ and $\rm CF_{eR}=8.01\times10^{-13}\pm7.33\times10^{-14}\,\rm erg/cm^2/cts$. 
We then computed X-ray luminosities ($L_{\rm x}$) and X-ray to bolometric luminosity ratios ($L_{\rm x}/L_{\rm bol}$) for the $0.2-2.0$\,keV energy band using the \textit{Gaia}-DR3 distances and bolometric luminosities from Table~\ref{tab:stepar}. 
The X-ray luminosity and $L_{\rm x}/L_{\rm bol}$ ratios as well as their upper limits for the undetected UCDs are listed in Tables~\ref{tab:obslog_xmm} \& \ref{tab:obslog_er} for {\it XMM-Newton} and eROSITA, respectively.

To be able to compare the literature sample with that of our sample of UCDs we converted the published X-ray luminosities to the energy band chosen for this work. To this end, we calculated a CF for each energy band reported in the literature that is different from our $0.2-2.0$\,keV band. We used the online count rate simulator WebPIMMS,\footnote{Count-rate simulator PIMMS: {\tt \url{http://heasarc.gsfc.nasa.gov/cgi-bin/Tools/w3pimms/w3pimms.pl}}} assuming an unabsorbed $1T$-{\tt apec} model with temperature of $0.61$\,keV and coronal abundance of $0.2\,\rm Z_{\odot}$. Among the values implemented in WebPIMMS we chose $0.61$\,keV because it is the closest to the measured mean value, computed with the emission measure weighted temperatures of the six UCDs with spectral fits (see Sect.~\ref{subsubsect:analysis_xrays_xspec} and Table~\ref{tab:xspec} for the spectral analysis). 
The updated X-ray literature data are shown in Table~\ref{tab:lit_sample_xray_radio} with the computed CFs, the original values, and their references.
In this table we report the results of observations taken while the target was in a quiescent state (upper limits are included). In cols.~8\&11 we indicate if a specific object has data from the literature of enhanced activity (X-ray flares and/or radio bursts). When the count rate is not provided in the literature we directly converted the literature flux from the original energy band into the $0.2-2.0$\,keV band, so no CF is provided in this case. Finally, we show the published quiescent and bursting radio luminosities ($L_{\rm rad}$) and their references in the last four columns of the same table. 

\addtolength{\tabcolsep}{-6pt}
\begin{table*}
\begin{center}
\begin{threeparttable}[b]
\caption{X-ray and Radio data of the literature sample retrieved as explained in Sect.~\ref{sect:XMMsample}. We show the X-ray instrument, the X-ray flux and its reference (from col. 2 to 4), the conversion factor used to calculate the flux in the $0.2-2.0$\,keV energy band used in this work (cols.5\&6), the X-ray luminosity (col.7), with the reference of detected X-ray flares (col.8) and the quiescent and bursting radio luminosities with their reference (from col.9 to 12).}
\label{tab:lit_sample_xray_radio}
\begin{tabular}{lccccccccccc}
\toprule[0.5mm]

  \multicolumn{1}{l}{Name} &
  \multicolumn{1}{c}{Instr\tnote{1}} &
  \multicolumn{1}{c}{$f_{\rm x,Lit}$} &
  \multicolumn{1}{c}{$\rm Ref_{xray}$\tnote{2}} &
  \multicolumn{1}{c}{$\rm CF_{0.2-2.0}$} &
  \multicolumn{1}{c}{$f_{\rm x,0.2-2.0}$} &
  \multicolumn{1}{c}{$\log L_{\rm x}$} &
  \multicolumn{1}{c}{X-ray\tnote{2}} &
  \multicolumn{1}{c}{$\log L_{\rm R,\nu}$} &
  \multicolumn{1}{c}{$\rm Ref_{ radio}$\tnote{2}} &
  \multicolumn{1}{c}{$\log L_{\rm R,\nu}^{\rm burst}$} &
  \multicolumn{1}{c}{Radio\tnote{2}} \\ 
  
  \multicolumn{1}{c}{} &
  \multicolumn{1}{c}{} &
  \multicolumn{1}{c}{[$\rm \times 10^{-14} erg/cm^2/s$]} &
  \multicolumn{1}{c}{} &
  \multicolumn{1}{c}{[$\rm \times 10^{-11} erg/cm^2/cnt]$} &
  \multicolumn{1}{c}{[$\rm \times 10^{-14} erg/cm^2/s$]} &
  \multicolumn{1}{c}{[erg/s]} &
  \multicolumn{1}{c}{flares} &
  \multicolumn{1}{c}{[erg/s/Hz]} &
  \multicolumn{1}{c}{} &
  \multicolumn{1}{c}{[erg/s/Hz]} &
  \multicolumn{1}{c}{bursts} \\

  \midrule[0.5mm]


  2MJ0036 &  C &  $<0.10^{0.2-8.0}$ & B05 & 3.54 & $<$0.79 & $<$25.86 & $\cdot\cdot$ & 13.4 & Bur05 & 13.8 & B02\\
  2MJ0523 & C &  $<0.08^{0.2-2.0}$ & B10 & 3.84 & $<$0.90 & $<$26.24 & $\cdot\cdot$ & $<$13.0 & Ant07 & 13.7 & B06\\
  2MJ0602 & C & $<0.08^{0.2-2.0}$ & B10 & 3.84 & $<$0.90 & $<$26.17 & $\cdot\cdot$ & $<$12.8 & B10 & $\cdot\cdot$ & $\cdot\cdot$\\
  2MWJ1507 & C & $<0.10^{0.2-2.0}$ & B05 & 3.84 & $<$0.85 & $<$25.75 & $\cdot\cdot$ & $<$12.6 & B02 & $\cdot\cdot$ & $\cdot\cdot$ \\
  BRI0021-0214 & C &  $<0.08^{0.2-2.0}$ & B10 & 3.84 & $<$0.90 & $<$26.22 & $\cdot\cdot$ & 12.6 & B02 & 13.8 & B02\\
  DENIS J1048 & X & 0.66$^{0.6-10.0}$  & St12 & 0.18 & 0.21 & 24.62 & $\cdot\cdot$ & 12.3 & Rav11 & 14.8 & Bur05\\
  Gl 569 B & C &  0.55$^{0.3-1.2}$ & St04 & 8.8 & 10.56 & 27.08 & St04 & $<$12.5 & B06 & $\cdot\cdot$ & $\cdot\cdot$\\
  Kelu 1 & C &  0.07$^{0.1-10.0}$ & Au07 & $\cdot\cdot$ & 0.06 & 25.45 & $\cdot\cdot$ & $<$12.8 & Au07 & $\cdot\cdot$\\
  LHS2065 & X & 2.92$^{0.3-8.0}$ & R08 & $\cdot\cdot$ & 6.46 & 26.76 & SL02 & $<$12.8 & B02 & $\cdot\cdot$ & $\cdot\cdot$\\
  LP 349-25 & C & 0.32$^{0.2-2.0}$ & W14 & $\cdot\cdot$ & 0.32 & 25.88 & W14 & 14.0 & PB07 & $\cdot\cdot$ & $\cdot\cdot$\\
  LP 412-31 & X & 6.23$^{0.5-10.0}$  & St06 & 0.17 & 188.26 & 27.25 & St06 & $<$12.8 & B10 & $\cdot\cdot$ & $\cdot\cdot$\\
  LP 944-20 & C & $<0.03^{0.1-4.0}$ & Ru00 & 3.54 & $<$0.42 & $<$25.32 & Ru00 & 12.3 & B06 & 13.9 & B06\\
  LSR J1835 & C &  $<0.08^{0.2-2.0}$ & B08 & 3.84 & $<$0.95 & $<$25.56 & $\cdot\cdot$ & 13.3 & B06 & 14.0 & Hal08\\
  TVLM513 & C & 0.63$^{0.3-2.0}$ & B08 & 4.11 & 1.11 & 26.18 & $\cdot\cdot$ & 13.4 & B02 & 14.9 & B08\\
  vB 10 & R & 0.42$^{0.1-2.4}$ & F00 & 2.22 & 0.40 & 25.22 & F03 & $<$12.6 & Kri99 & $\cdot\cdot$ & $\cdot\cdot$\\
  vB 8 & R & 15.77$^{0.1-2.4}$ & F93 & 0.54 & 16.71 & 26.92 & T90 & $<$12.1 & Kri99 & $\cdot\cdot$ & $\cdot\cdot$\\
  
  
\bottomrule[0.5mm]
\end{tabular}
	\begin{tablenotes}
				\item[1]  XMM-Newton: X; ROSAT: R; Chandra: C
				\item[2]  Ant07: \cite{Antonova07.1}; Au07: \cite{Audard2007}; B02: \cite{Berger02.1}; B05: \cite{Berger05.1}; B06: \cite{Berger06.1};  B08: \cite{Berger08.1}; B10: \cite{Berger2010}; Bur05: \cite{Burgasser05.1}; F93: \cite{Fleming1993}; 
				F95: \cite{Fleming1995}; F00: \cite{Fleming2000}; F03: \cite{Fleming2003}; Hal08: \cite{Hallinan08.1}; Kri99: \cite{Kri99}; PB07: \cite{PhanBao07.1}; R08: \cite{Robrade2008}; Rav11: \cite{Ravi11.0}; Ru00: \cite{Rutledge2000}; St04: \cite{Stelzer2004}; 
				St06: \cite{Stelzer2006};  St12: \cite{Stelzer2012}; SL02: \cite{SL2002}; T90: \cite{Tagliaferri1990}; W14: \cite{Williams2014}
			\end{tablenotes}
\end{threeparttable}
\end{center}
\end{table*}

\subsection{Radio data analysis}
\label{subsec:analysis_radio}
Observations were obtained as part of three NRAO/JVLA projects and one ATCA project; observation specifics are listed
in Table~\ref{tab:obslog_radio}. 
Note that we add in the radio flux densities for 3 of our targets which were reported previously in the literature. 
Table~\ref{tbl:radio_freq} lists the specific radio 
frequency ranges used in pre-upgrade VLA observations (pertinent to the literature references used in this paper), and the JVLA and ATCA observations presented. 
Table~\ref{tab:ucds_par}
Our literature references are \citet{Antonova2013} for 2M0215 and \citet{Berger06.1} for 2M0440 and 2M0435.
\citet{Berger06.1} does not quantify what the quoted upper limits are, and we are assuming that they are 3 sigma. 
We have recalculated the 3$\sigma$ upper limits on the radio luminosities from the literature using {\it Gaia}-DR3  distances from Table~\ref{tab:stepar}. 
These updated values are listed in \ref{tab:radio_limits}.

\subsubsection{JVLA data}
\label{subsubsect:jvla}
The flux calibrator (col.~7 of Table~\ref{tab:obslog_radio}) was used to establish both 
overall gain solutions as well as determine bandpass calibrations, while the phase 
calibrator (column 8 of Table~\ref{tab:obslog_radio}) was used to correct for phase variations. 
Data editing and further calibration steps were performed using standard techniques in 
the {\it Common Astronomy Software Applications (CASA)} package \citep{McMullin07.0}. 
The calibrated visibility data for each target were split off from 
the main dataset after successful calibration and imaged. 
The individual epochs of observations in January 2015 were calibrated and imaged separately,
then the visibility datasets were concatenated and imaged as a whole to increase the sensitivity. 
A similar approach was used for the two epochs of observations in September 2020.
The cell size
for imaging the D configuration X-band datasets
was 1.5 arcsec, while for the CnB configuration X-band datasets the cell size was 0.34
arcsec. 
The cell size for the B configuration C-band datasets was 0.5 arcsec.
The clean algorithm used to create images employed natural weighting with a 
threshold of 10 $\mu$Jy to discover faint radio sources. 

We derived the expected position of our targets at the time of each JVLA observation, 
taking into account the P.M. provided by {\it Gaia}\,DR3 (see Table~\ref{tab:stepar}).
Once images had been created for each of the four fields, we searched for any radio sources 
in the vicinity of the expected target positions. 
For the six epochs of 2MJ0306 and two of 2MJ0652, we searched 
the image for each epoch as well as that made from combining all epochs.

\begin{figure}
\begin{center}
\includegraphics[scale=0.45]{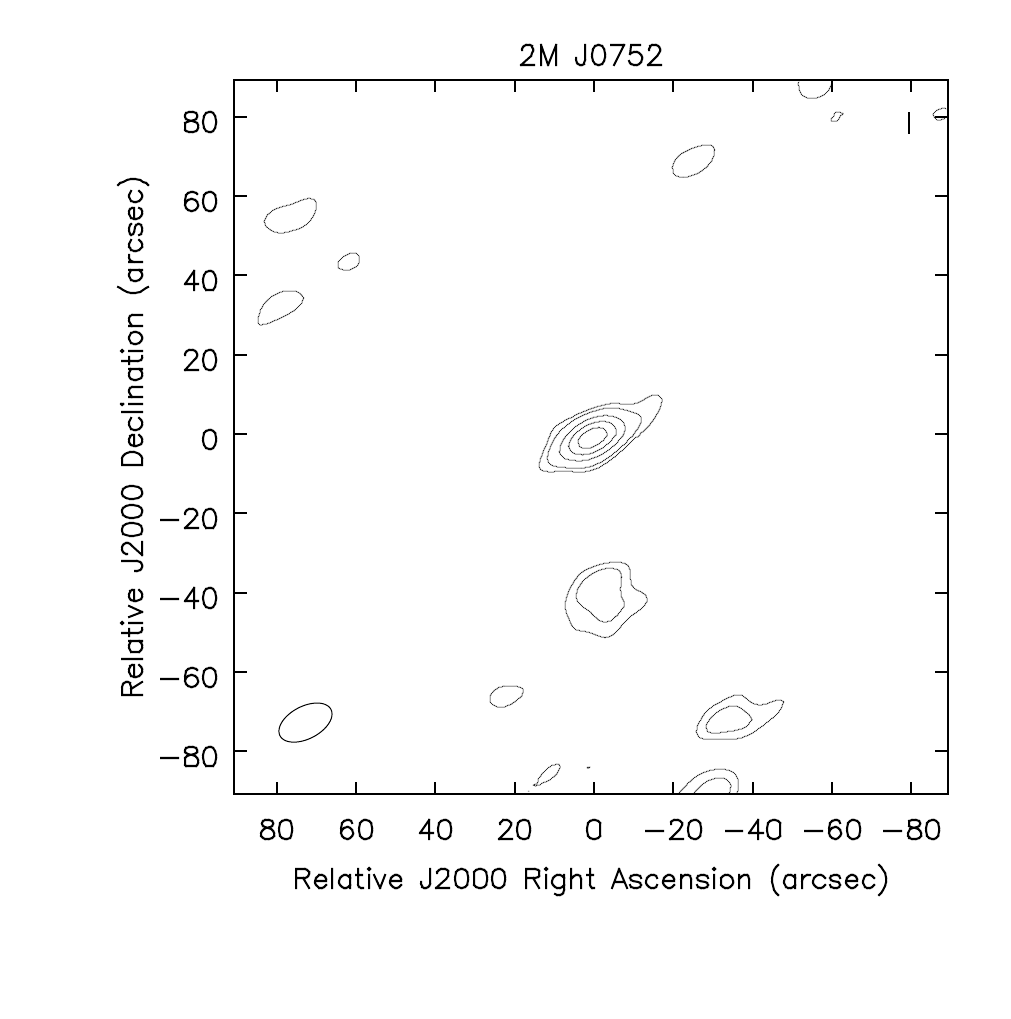}
\caption{Contour image of the JVLA field around the target 2MJ0752, which has a source only 0.27$^{\prime\prime}$ from the expected position. The size is 
$3^{\prime} \times 3^{\prime}$ and 
contour levels are $3$, $5$, $10$, $15$, $20$, $30$, $50$, and $100$ times
the image rms, which is listed in Table~\ref{tab:radio_limits}. 
\label{fig:radio_zoomfields}}
\end{center}
\end{figure}

The only field with a radio source near the expected target position is 2MJ0752, with a 
radio counterpart offset from the expected position by only $0.27^{\prime\prime}$
(Figure~\ref{fig:radio_zoomfields}). 
After deconvolving from the clean beam it appears to be a point source with a peak flux density
of 78.9$\pm$3.4 $\mu$Jy/beam in total intensity. 
The same field imaged in Stokes V also shows a source at the
same position, with a 
peak flux density of $-59.8\pm$2.6 $\mu$Jy/beam, or circular polarization fraction
$V/I = -76\pm$5\%.
Late-type stellar radio sources are typically the only type to experience high degrees of 
circular polarization.
We consider the association of this highly circularly polarized radio point source with the 
brown dwarf robust, considering the beam size of $14.45^{\prime\prime} \times 8.2^{\prime\prime}$ and the
significance of the detection. 
For all other targets the 
radio sources in the images are offset by $>30^{\prime\prime}$ from the expected target position,
i.e. they are undetected.  
The fields around the other four sources
are displayed in the Appendix (Figures~\ref{fig:appendix_JVLA} 
and ~\ref{fig:appendix_2MJ0652}). 
Table~\ref{tab:radio_limits} lists the 3$\sigma$ upper limits 
and the detected radio flux  with rms for all of the program sources, and Table~\ref{tbl:fluxdetails} describes the average, maximum and minimum values of total and circularly polarized flux.

\subsubsection{ATCA data }

Reduction and calibration of ATCA data proceeded as noted above in Sect.~\ref{subsubsect:jvla}, using Miriad software \citep{MIRIADref}, with amplitude and phase calibration using sources identified in cols.~7 and~8, respectively, of Table~\ref{tab:obslog_radio}. 
The cell size for imaging C band observations was 0.8 arcsec, while for X band observations it was 0.4 arcsec.

\begin{figure*}
    \begin{center}
    \parbox{18cm}{
     \parbox{9cm}
     {\includegraphics[scale=0.3]{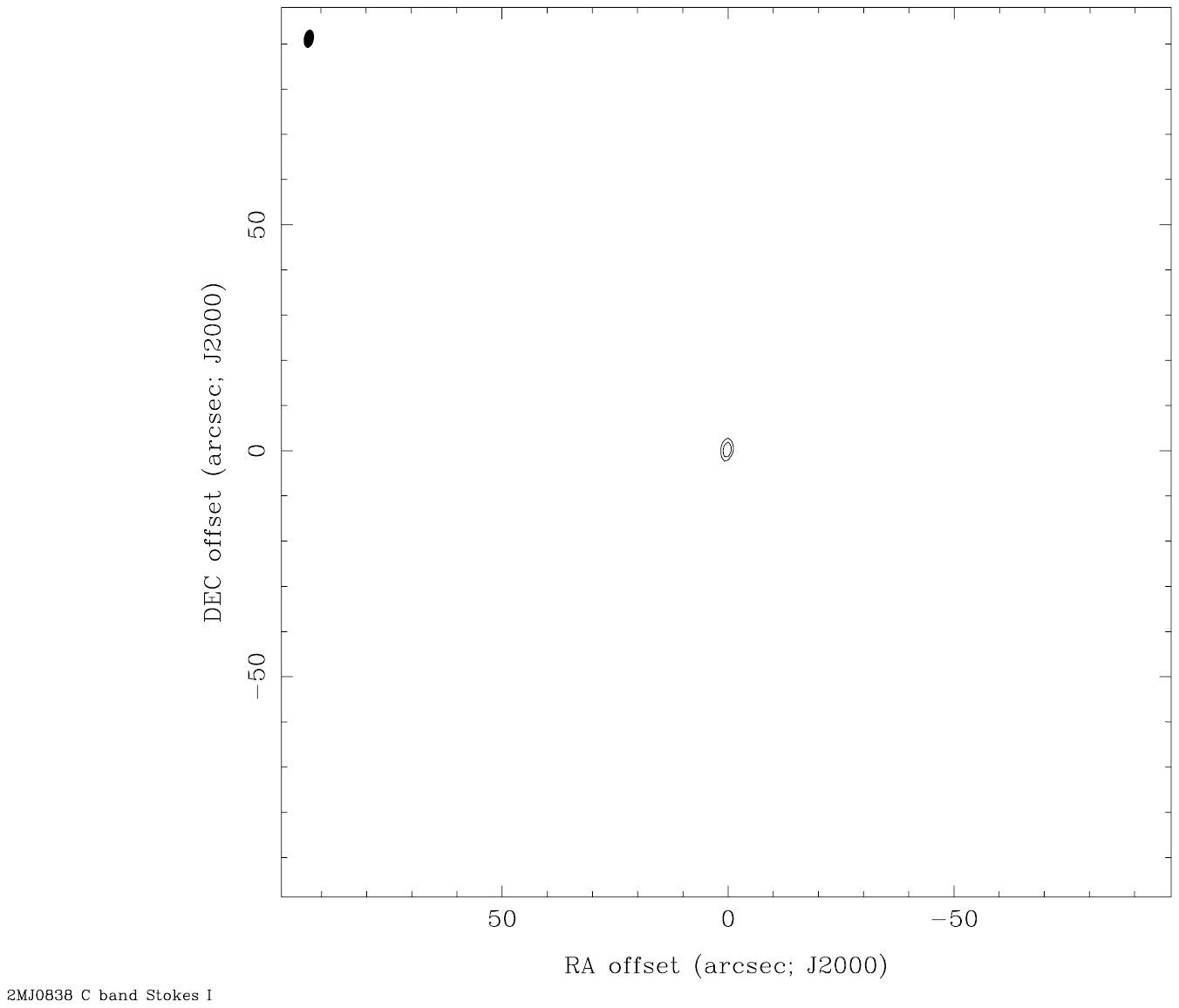}}
     \parbox{9cm}
     {\includegraphics[scale=0.3]{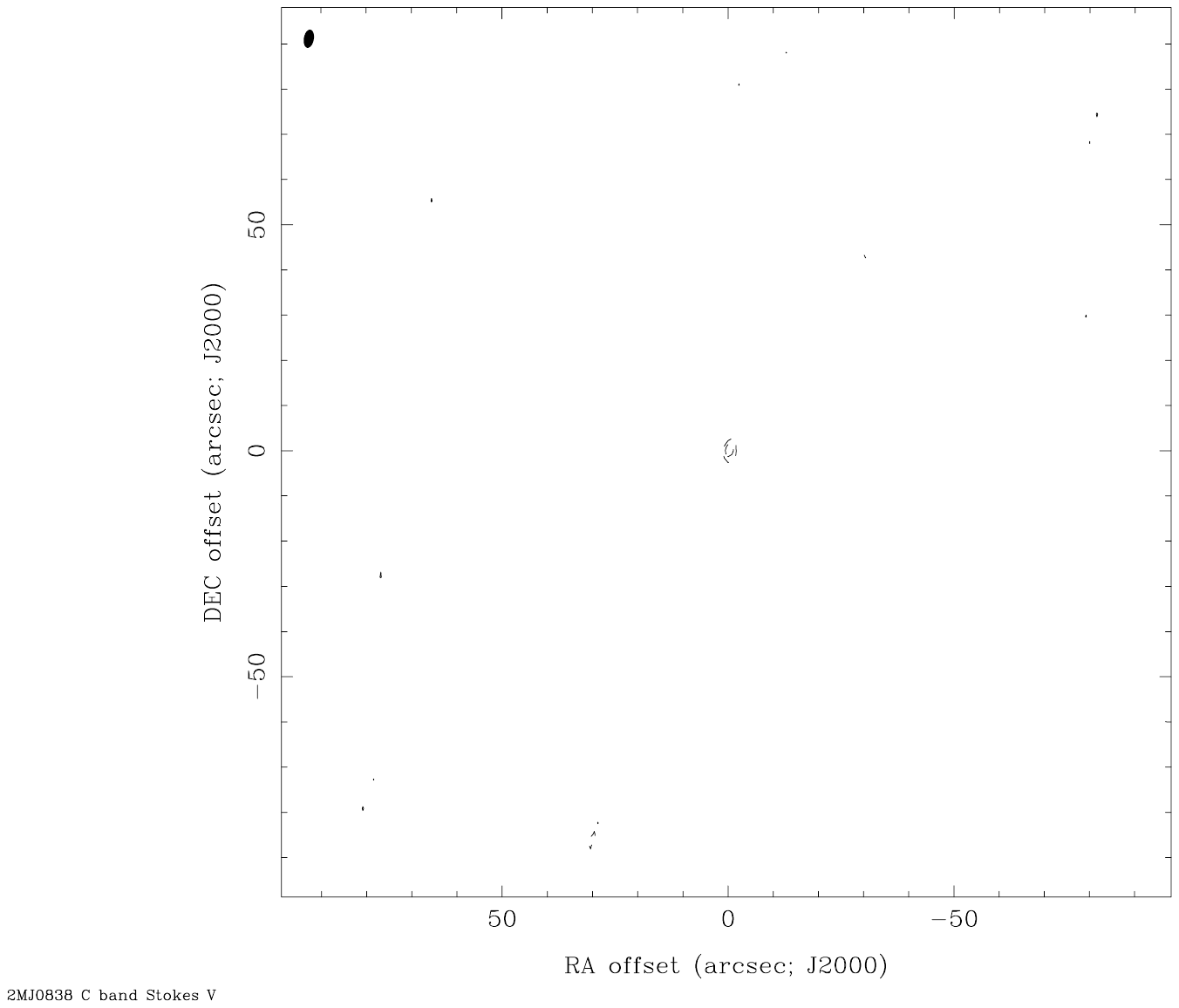}}
     }
    \caption{Total intensity and Stokes V maps of the three arcminute region around the  target 2MJ0838 observed with ATCA. Contours are shown at 3, 5, and 10 times the rms values listed in Table~\ref{tab:radio_limits}. }
    \label{fig:atca_maps}
    \end{center}
\end{figure*}

We performed similar calculations as described 
above in Sect.\ref{subsubsect:jvla}
to determine the expected position of the targets at the epoch of the observation. The only
object with a confident detection was 2MJ0838, with a point
source only 0.3$^{\prime\prime}$ away from the expected position at both X and C bands. 
Maps of the region around the expected position for this object in Stokes I and V are shown in Fig.~\ref{fig:atca_maps}, while those for the undetected objects are illustrated in Fig.~\ref{fig:appendix_ATCA}.
Table~\ref{tab:radio_limits} lists the 3$\sigma$ upper limits and detected radio luminosities; the only object with a detection is 2MJ0838. Table~\ref{tbl:fluxdetails} lists the average values of the flux densities for total intensity and circularly polarized emission averaged across the observation.

\begin{table*}[]
\caption{Radio Frequency Coverage}
    \centering
    \begin{tabular}{cc}
    \midrule[0.5mm]
    Frequency Code    & Frequency Range (GHz) \\
     \midrule[0.2mm]
AC &  4.476-6.525\\
AX & 7.976-10.025\\
VC & 3.976-8.024\\
VX & 7.991-10.015 \\
LVC & 2$\times$50 MHz at 4.9 GHz \\
LVX & 2$\times$50 MHz at 8.46 GHz\\
\bottomrule[0.5mm]
\multicolumn{2}{l}{AC=ATCA C-band, AX=ATCA X-band} \\
\multicolumn{2}{l}{VC=JVLA C-band, VX=JVLA X-band} \\
\multicolumn{2}{l}{LVC=Legacy VLA C-band, LVX=Legacy VLA X-band} \\
    \end{tabular}
    \label{tbl:radio_freq}
\end{table*}

\begin{table*}[hbtp]
\begin{center}
\caption{Observing log for radio data.} 
\label{tab:obslog_radio}
\begin{tabular}{lclllcccc}
\midrule[0.5mm]
Target & Project & Obs. Date & Obs. Start & Obs. End & Conf. & 
Flux Calib. & Phase Calib. & Frequency  \\ 
\midrule[0.2mm]
2MJ0752 & 13A-360    & 2013 April 6  & 19:54:25 &  23:53:46 & D   & 3C147 & J0738+1742 &VX\\
2MJ0351 & " & 2013 March 5  & 00:10:49 &  03:10:20 & D   & 3C147 & J0339-0146 & VX\\
2MJ1757 & "       & 2013 April 27 & 05:49:08 &  11:47:57 & D   & 3C286 & J1716+6836 &VX\\
2MJ0306 & 14B-158 & 2015 Jan. 13  & 00:58:08 &  01:56:31 & CnB & 3C138 & J0334-4008 & VX\\
2MJ0306 & "       & 2015 Jan. 14  & 02:07:35 &  03:07:24 & "   & " & " & "\\
2MJ0306 & "       & 2015 Jan. 14  & 03:07:27 &  04:07:15 & "   & " & " & "\\
2MJ0306 & "       & 2015 Jan. 15  & 01:33:45 &  02:33:33 & "   & " & " & "\\
2MJ0306 & "       & 2015 Jan. 15  & 02:33:37 &  03:33:24 & "   & " & " & "\\
2MJ0306 & "       & 2015 Jan. 19  & 00:48:03 &  04:17:28 & "   & " & " & "\\
2MJ0838    &C3388 & 2021 Feb. 19   &   06:30:05          &    18:28:35      &6D & 1934-638 & J0835-5953 & AX,AC \\
2MJ0652 & SJ6020 & 2020 Sept. 15 &12:35:22 &17:34:34 &B & 3C48 & J0648-3044 & VC \\
2MJ0652 & " & 2020 Sept. 24 & 11:46:27& 16:45:39& B & 3C48 & J0648-3044 & VC \\
2MJ1055 & C3388 & 2021 Feb. 20 & 00:39:35  &  12:59:05&6D & 1934-638 & J1057-797 & AX\&AC \\
\bottomrule[0.5mm]
\end{tabular}
\end{center}
\end{table*}

\begin{table} [htbp]
\begin{center}
\caption{Radio flux densities (with $1\,\sigma$ error) or $3\,\sigma$ upper limit for non-detections.}
\label{tab:radio_limits} 
\begin{tabular}{cccc}
\midrule[0.5mm]
Target & Freq. & rms & $L_{\rm R,\nu}$ \\
      &GHz & [$\mu$Jy] & [erg s$^{-1}$ Hz$^{-1}$] \\
\hline
2MJ0752 & VX&3.3 & 1.05$\times$10$^{13}\pm$4.5$\times$10$^{11}$ \\
2MJ0351 & VX&8.2 & $<$6.4$\times$10$^{12}$ \\
2MJ1757 & VX & 5.4 & $<$7.1$\times$10$^{12}$ \\
2MJ0440 &LVX & 13$^{1}$ &$<$4.5$\times$10$^{12}$\\
2MJ0306 & VX & 2.6 & $<$1.2$\times$10$^{12}$ \\
2MJ0838 & AC & 63 & 7.1$\times$10$^{13}\pm$0.94$\times$10$^{13}$ \\
"       & AX & 35 & 4.9$\times$10$^{13}\pm$0.52$\times$10$^{13}$ \\
2MJ0435 &LVX &16$^{1}$ & $<$6.5$\times$10$^{12}$ \\
2MJ0652 & VC   & 1.8 & $<$1.7$\times$10$^{12}$\\
2MJ0215 &LVC &25$^{2}$ & $<$1.8$\times$10$^{13}$ \\
2MJ1055 & AC & 36 & $<$2.4$\times$10$^{13}$ \\
"            & AX & 30 & $<$2.0$\times$10$^{13}$ \\

\bottomrule[0.5mm]
\multicolumn{4}{l}{$^{1}$ upper limits reported in Berger (2006) are presumed } \\
\multicolumn{4}{l}{to be 3$\sigma$; $^{2}$ Antonova et al. (2013)} \\
\multicolumn{4}{l}{$^{\dagger}$ separation of two UCDs is greater than the synthesized beam }\\
\multicolumn{4}{l}{width at both frequencies, so limit applies to each component }\\
\multicolumn{4}{l}{of the binary separately; see text for details.}
\end{tabular}
\end{center}
\end{table}

\begin{table}[]
    \centering
    \caption{Total and Circularly Polarized Intensities of Radio-Detected Objects}
    \label{tbl:fluxdetails}
    \begin{tabular}{llllll}
    \hline
    Object & Freq. & Stokes & Avg F$_{\nu}$ &Min. F$_{\nu}$ & Max. F$_{\nu}$ \\
            &      &        & ($\mu$Jy) & ($\mu$Jy)& ($\mu$Jy) \\
    \hline
        2MJ0752 & VX & I &  78.9$\pm$3.4 &38.4$\pm$8.9 & 147$\pm$17\\
      " & "& V  & $-$58.9$\pm$2.6&$-$28$\pm$9 &$-$130$\pm$14 \\
      2MJ0838 & AC & I &476$\pm$63 &231$\pm$68 & 1220$\pm$58\\
      " & " & V &$-$102$\pm$18 &$-$208$\pm$66 &$-$704$\pm$175 \\
      " & AX & I & 330$\pm$35 & 251$\pm$60&485$\pm$60 \\
      " & " & V & -39$\pm$13 &$-$178$\pm$59 & $-$214$\pm$51\\
      \hline
    \end{tabular}
\end{table}

\subsection{TESS data}\label{subsect:analysis_tess}

We analysed TESS light curves with the primary aim to obtain the rotation periods of the UCDs, and secondly to identify optical flares as additional activity diagnostics.
The results of the TESS analysis and TESS identification numbers (TIC) for all UCDs studied in this article, namely the $26$ objects from our dedicated X-ray/radio campaigns and the literature sample,  are given in Table~\ref{tab:tess}.

We uploaded {the target list of the UCDs to the {\it Barbara A. Mikulski Archive for Space Telescopes} (MAST) interface and found two-minute cadence light curves processed by the TESS collaboration for $17$ out of the $26$ targets. Most of them were observed in multiple sectors (see Table~\ref{tab:tess}), and we downloaded $69$ individual light curves from MAST.

For our analysis we used the \texttt{PDCSAP\_FLUX}, in which the pipeline processing  has removed some of the instrument systematics from the flux measurements. 
TESS assigns a quality flag to all measurements. In the first step, we removed all data points with a quality flag different from 0 except of `Impulsive outlier' (which could be real stellar flares) and `Cosmic ray in collateral data' \citep[see more detailed information in ][]{Magaudda2022}. We then normalized the individual sector light curves of each UCD by dividing all data points by the median flux. The period search was performed using the method described by \citet{Magaudda2022}. 
In short, we used three different standard time series analysis techniques, the generalized Lomb-Scargle periodogram \citep[\begin{scriptsize}GLS\end{scriptsize};][]{2009A&A...496..577Z}, the autocorrelation function (ACF), and sine fitting. We then phase-folded the light curves with the best period estimate from each method. By visual inspection we selected the methods that yielded the period that best represents the light curves. As final period we adopted the mean value of these methods. If light curves from more than one sector were available for a given UCD all good period values from the different methods in the different sectors were averaged (up to three values per sector). Uncertainties on the $P_{\rm rot}$ values are given as the standard deviation of those period values that were used to calculate our adopted period.
The typical results of our analysis are demonstrated on an example for one sector of the UCD TIC\,29890705 in Fig.~\ref{fig:tessexample}. 

This initial period search resulted in a significant period detection for seven of $26$ targets. However, by eye the ACF of $3$ of the remaining UCDs  without a period estimate (i.e. TIC\,70555405, TIC\,117733581 and TIC\,401945077) appears to display a clear signal. Therefore, we decided to bin the original light curves of all $19$ UCDs without a detected period with a factor of $15$ and we repeated the period search. This way, we found a reliable rotation period for further 5 UCDs, including the three mentioned above.

Therefore, as a final result we were able to detect a period for $12$ out of $26$ UCDs. The period for one of them,  TIC\,308243298, is, however, only marginally detected in the binned data. The object was observed in $4$ sectors but only two of them show a detection. 
Since this period is consistent with the period upper limit from the $v\sin{i}$ measurement, we consider it when we discuss the X-ray vs radio luminosity relation for UCDs with $P_{\rm rot}\leq 1$\,d (see Sect.~\ref{sect:discussion}).

For the $14$ UCDs without a period estimate from our method or without TESS two-minute cadence data we adopted the value listed by \cite{Crossfield2014}, who collected published measurements of photometric rotation periods for $58$ UCDs.  When no photometric period is available, we calculated its upper limit by using $v\sin{i}$ measurements from high-resolution spectroscopic data taken from the same reference and the computed radii listed in Table~\ref{tab:stepar}.

For those UCDs for which we found a valid photometric rotation period we calculated the inclination of the systems by using the radii listed in Table~\ref{tab:stepar} and the $v\sin{i}$ measurements. These data are taken from \citet{Crossfield2014} for all UCDs, but for one object (2MJ0752) for which no value is listed in their work thus we adopted the one from \citet{Hughes2021}. For $5$ UCDs $v\sin{i}$ is larger than the actual rotational velocity, thus we could not retrieve a reliable inclination. We flagged these cases with ``$v\sin{i}>v_{\rm rot}$'' in Table~\ref{tab:tess} where we summarize  the results of the analysis of our TESS data. In particular, cols.~2\&3 show the TESS ID of the UCD and the number of sectors in which it was observed, cols.4 and~5 list the photometric period and its reference, col.6 holds the literature $v\sin{i}$ value, and col.~7 the inclination. 

To detect flares we applied the method that we explained in detail in \citet{Stelzer2022} and \citet{Magaudda2022}. In short, we smoothed the light curves and removed all data points that deviate more than 2-$\sigma$ from the smoothed light curve. The treatment was repeated three times with decreasing size of the boxcar width that was used for the smoothing. The final smoothed light curve was then subtracted from the original light curve in order to remove the rotational signal. All groups of at least three consecutive data points that deviate more than 3-$\sigma$ from this `flat' version of the light curve were identified as potential flares. In a 5-step validation process false positive detections were removed from this list. The validation criteria are (1) the flare event must not occur right before or after a data point gap in the light curve, (2) the flux ratio between flare maximum and last flare point must be $\geq2$, (3) the flare maximum cannot be the last flare point, (4) the decay time has to be longer than the rise time, and (5) a fit conducted using the flare template defined by \citet{2014ApJ...797..122D} must fit the flare better than a linear fit. The UCDs with validated flares that fulfill all of the above mentioned criteria are marked in the last column of Table~\ref{tab:tess}.

A detailed analysis of the TESS flares is beyond the scope of this paper. Systematic studies of large samples of flares observed with TESS on UCDs have been published \citep{Petrucci23.0}. In this work we merely us the presence of optical flares as an additional activity indicator in the joint multi-wavelength analysis of our sample.

\begin{table*}
    \begin{center}
        \begin{threeparttable}[b]
            \caption{TESS observations of all UCDs studied in this article, $v\cdot\sin{i}$ measurements retrieved from the literature and the computed inclination.
            For $7$ literature targets we calculated the upper limits from the $v\sin{i}$ measurements given by \cite{Crossfield2014}.} 
            \label{tab:tess}
                \begin{tabular}{lccccccc}
                    \midrule[0.5mm]     
                       \multicolumn{1}{l}{Object} 
                        &\multicolumn{1}{c}{TIC\,ID}
                       &\multicolumn{1}{c}{$N_{\mathrm{TESS}}^{a}$}
                       &\multicolumn{1}{c}{$P_{\rm rot}$}
                       &\multicolumn{1}{c}{Ref$^{c}$}
                       &\multicolumn{1}{c}{Flares}&\multicolumn{1}{c}{$v~\sin{i}^{d}$}
                        &\multicolumn{1}{c}{$i$} \\

                         \multicolumn{1}{l}{} 
                         &\multicolumn{1}{l}{}  
                        &\multicolumn{1}{c}{} 
                        &\multicolumn{1}{c}{[d]} 
                        &\multicolumn{1}{c}{}
                        &\multicolumn{1}{c}{}
                        &\multicolumn{1}{c}{[km/s]}
                        &\multicolumn{1}{c}{[deg]}\\
                        
                        \midrule[0.5mm]
    \multicolumn{8}{c}{UCDs with new X-ray/radio campaigns} \\
    \midrule[0.2mm]                    
    2MJ0752 &  17773418 & 3 & 0.8840\,$\pm$\,0.0022 & (1) & \checkmark & 9.0$\pm$2.0 & 47.3$\pm$14.7 \\
    2MJ0351 & 401945077 & 1 & 0.7288\,$\pm$\,0.0002 & (1) & --- & 6.5$\pm$2.0 & 38.3$\pm$14.5\\   
    2MJ1757  & 219095664 & 26  & 0.1622\,$\pm$\,0.0002 & (1) & \checkmark & 34.6$\pm$2.6 & 41.5$\pm$5.8\\   
    2MJ0440 & 298907057 & 2 & 0.4975\,$\pm$\,0.0059 & (1) & \checkmark & 17.9$\pm$1.4 & $v\sin{i}>v_{\rm rot}$ \\ 
    2MJ0306 & 308243298 & 4 & 0.2947\,$\pm$\,0.0003$^{b}$ & (1) & --- & 20.4$\pm$2.0 & 83.9$\pm$72.0  \\   
    
    
    2MJ0838 & 44984200 & 9 & 0.1130\,$\pm$\,0.0001 & (1) & \checkmark & $\cdot\cdot$ & $\cdot\cdot$  \\
    2MJ0435 & 117733581 & 2 & 0.6255\,$\pm$\,0.0045 & (1)  & \checkmark & 12.1$\pm$3.0 & $v\sin{i}>v_{\rm rot}$ \\
    2MJ0652 & 53603145 & 3 & 0.2095\,$\pm$\,0.0002 & (1) & --- & $\cdot\cdot$ & $\cdot\cdot$ \\
    2MJ0215 & 70555405 & 2 & 0.7025\,$\pm$\,0.0016 & (1) & \checkmark & $\cdot\cdot$ & $\cdot\cdot$ \\ 
    2MJ1055 & 277539431 & 4 & 0.1901\,$\pm$\,0.0001 & (1) & \checkmark & $\cdot\cdot$ & $\cdot\cdot$ \\
    
    \midrule[0.2mm]                    
    \multicolumn{8}{c}{Literature UCD sample} \\
    \midrule[0.2mm]                    

    2MJ0036 & $\cdot\cdot$ & $\cdot\cdot$ & 0.12 & (2) & $\cdot\cdot$ & 40.0$\pm$2.0 & 62.2$\pm$15.0 \\
    2MJ0523 & 442929628 & 3 & $<$0.09 & (2) & --- & 18.0$\pm$2.0 & $\cdot\cdot$ \\
    2MJ0602 & $\cdot\cdot$ & $\cdot\cdot$ & $<$0.14 & (2) & $\cdot\cdot$ & 9.0$\pm$3.0 & $\cdot\cdot$ \\ 
    2MWJ1507 & $\cdot\cdot$ & $\cdot\cdot$ & $<$0.05 & (2) & $\cdot\cdot$ & 30.0$\pm$2.3 & $\cdot\cdot$ \\
    BRI0021-0214 & 244167275 & 2 & 0.17 & (2) & --- & 34.2$\pm$1.6 & 76.0$\pm$25.0 \\
    DENIS J1048 & 107012050 & 3 & 0.2540\,$\pm$\,0.0004 & (1) & \checkmark & 18.0$\pm$2.0 & 65.0$\pm$17.9  \\
    Gl 569 B & 258105174 & 1 & $<$0.40 & (2) & \checkmark & 19.3$\pm$2.0 & $\cdot\cdot$ \\
    Kelu 1 & 229255770 & 1 & 0.07 & (2) & --- & 60.9$\pm$1.9 & 41.0$\pm$6.0 \\
    LHS2065 & $\cdot\cdot$ & $\cdot\cdot$ & $<$0.45 & (2) & $\cdot\cdot$ & 11.3$\pm$1.1 & $\cdot\cdot$ \\
    LP 349-25 & $\cdot\cdot$ & $\cdot\cdot$ & 0.07 & (2) & $\cdot\cdot$ & 55.0$\pm$2.0 & 23.9$\pm$2.4 \\
    LP 412-31 & $\cdot\cdot$ & $\cdot\cdot$ & 0.61 & (2) & $\cdot\cdot$ & 14.8$\pm$2.0 & $v\sin{i}>v_{\rm rot}$ \\
    LP 944-20 & 143029977 & 2 & 0.1598\,$\pm$\,0.0003 & (1) & --- & 28.9$\pm$0.8 & $v\sin{i}>v_{\rm rot}$ \\
    LSR J1835 & $\cdot\cdot$ & $\cdot\cdot$ & 0.11 & (2) & $\cdot\cdot$  & 43.9$\pm$2.2 & 71.1$\pm$17.5 \\
    TVLM513 & 311188315 & 1 & 0.08 & (2) & --- & 59.2$\pm$2.0 & $v\sin{i}>v_{\rm rot}$ \\
    vB 10 & $\cdot\cdot$ & $\cdot\cdot$ & $<$0.80 & (2) & $\cdot\cdot$ & 6.5$\pm$2.0 & $\cdot\cdot$ \\
    vB 8 & $\cdot\cdot$ & $\cdot\cdot$ & $<$0.66 & (2) & $\cdot\cdot$ & 9.0$\pm$2.0 & $\cdot\cdot$ \\ 
    \bottomrule[0.5mm] 
  \end{tabular}
  \begin{tablenotes}
        \item[$^{a}$] Number of sectors in which TESS observed the UCD
	\item[$^{b}$] Marginal period detection (see text in Sect.~\ref{subsect:analysis_tess})
        \item[$^{c}$]  References: (1) this work, (2) \cite{Crossfield2014}.
        \item[$^{d}$] $v\sin{i}$ measurements 
        from \cite{Crossfield2014}, except for the value for 2MJ0752 that comes from \cite{Hughes2021}.
  \end{tablenotes}
  \end{threeparttable}
\end{center}
\end{table*}

\section{Objects with particular interesting radio behavior}
\label{sect:radio_int_obj}

\subsection{2MJ0752}\label{subsect:results_2m0752}

The source 2MJ0752 was previously reported as undetected by \citet{Berger06.1} with an 
upper limit of $39\,\mu$Jy. Because of the high degree of circular polarization in the 
time- and frequency-averaged data, we explored
subsets to determine whether there was any evidence of time- or frequency-dependent 
variability, as an indication of the bursting behavior observed in some radio-detected
UCDs.

We initially explored the time domain only, creating a light curve by integrating over all 
spectral windows, and dividing each scan in half, to create a total of 38 time bins. 
The top left panel of Fig.~\ref{fig:2m0752_radio} displays the time variation of total 
intensity and circularly polarized flux. 
Upper limits in total intensity are depicted as downward facing arrows. 
There are two local peaks in the light curve of total intensity, each having an enhancement 
of 1.7-1.8 times the time-integrated flux density, and separated by about 1.7 hours. 
These however do not show up as local peaks in circularly polarized flux.

In the bottom left panel of Fig.~\ref{fig:2m0752_radio} we show the percent circular polarization, computed only 
for time bins in which there was a detection in both total intensity and circularly 
polarized flux. 
The average value of percent circular polarization, taken from the time-averaged values 
of Stokes V and Stokes I flux, is shown as a black horizontal line.
The high degree of circularly polarized flux obtained from
the integrated maps does not reveal any significant variation over the course of the observation.

We next explored dividing the data in both time and frequency to explore spectro-temporal variations.
Our initial investigation used the original 38 time bins, but subdividing the data into a
few frequency bins did not reveal any significant variations.  We converged on four
spectral spans and six time intervals.  These light curves are displayed in the right panels of Fig.~\ref{fig:2m0752_radio}.
These again did not display evidence for significant temporal or frequency-dependent variations.

The object's low rotational velocity of  $v\sin i =9$\, {\rm km/s}  and rotation period of $21.1$\,h measured on the TESS light curves 
reveal that we are seeing the UCD at the relatively high inclination angle of $i$ $\approx $50$^{\circ}$ (see Table~\ref{tab:tess}). The emission is highly circularly polarized ($-76\pm5\%$, see Sect.~\ref{subsubsect:jvla}), which is very suggestive of bursting behavior. While the total intensity did show some variations, there was no apparent change in the amount of circular polarization. Comparison with a previous epoch of observation $\sim$ 9 years earlier, with an upper limit below the detected level of emission, suggests that the detection spanned a long-lived burst.

The light curve of 2MJ0752 across the entire band shown in Fig.~\ref{fig:2m0752_radio} reveals two apparent double peaks, which at face value could be periodic bursts
corresponding to a short rotation period ($<2$\,h). However, with a known rotation period $\sim$21\,h, perhaps this observation instead reveals the leading and trailing edges of an auroral loss cone as it traverses our line of sight. The time between peaks of about 1.7\,h would approximately constrain the full-width opening angle of the loss cone. And
the width of each of burst, which is roughly 0.4-0.5\,h, would
approximately constrain the loss-cone width. 
Compared to previous radio detections of periodic bursting behavior that was ascribed to auroral loss cones 
\citep{Hallinan06.1,Yu2011}, 2MJ0752 has a relatively long rotation period.
Additional observations are needed to confirm that it is in fact producing periodic bursts, and to constrain the timing of the bursts at a range of frequencies.
Detailed modelling of frequency- and time-dependence of the emission, along the lines of \citet{Leto2016}, will likewise enable constraints on the physical properties of the emitting region.



\begin{figure*}
\begin{center}
\parbox{18cm}{
\parbox{9cm}{
\includegraphics[scale=0.3]{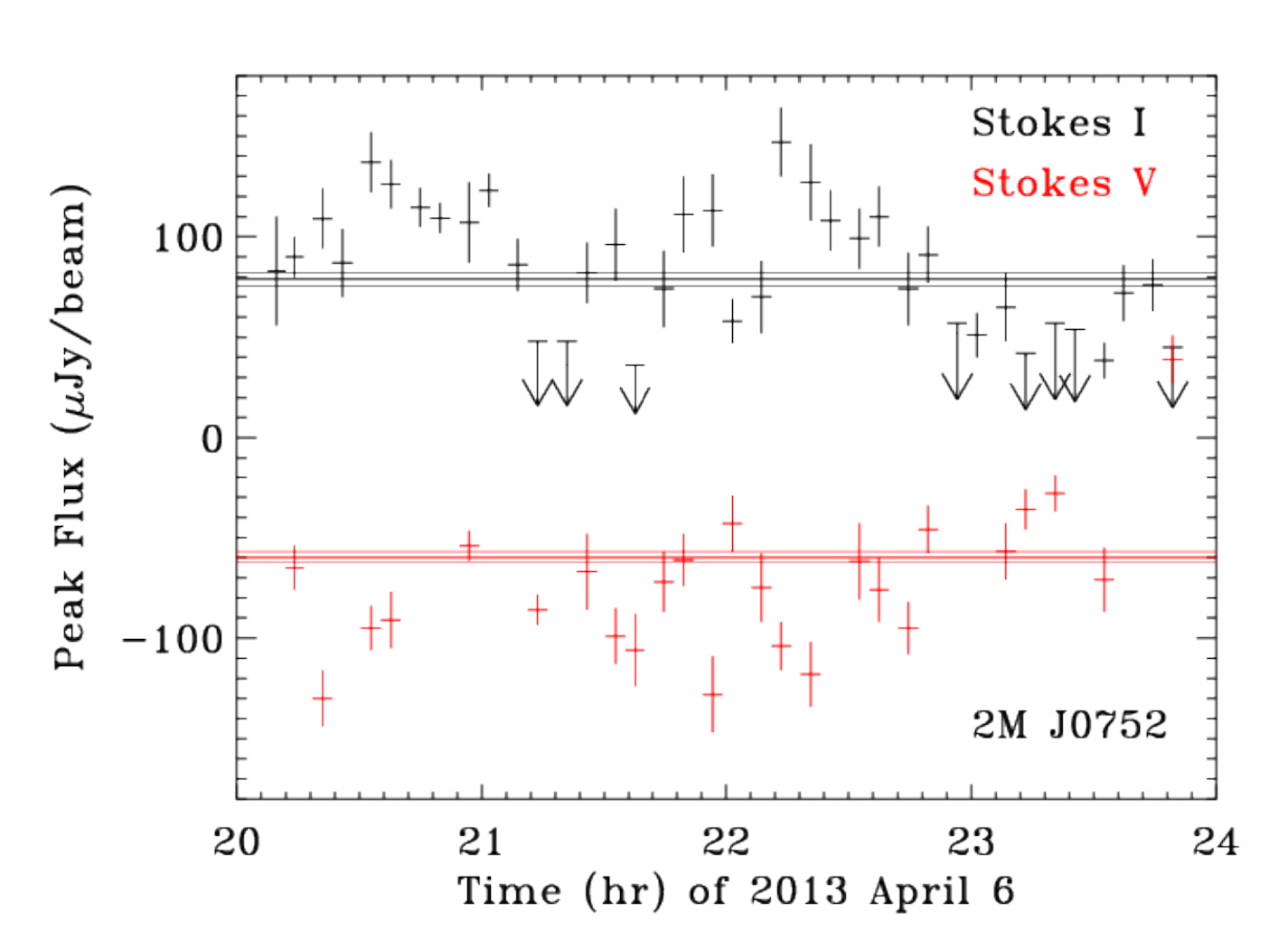}
}
\parbox{9cm}{
\includegraphics[scale=0.3]{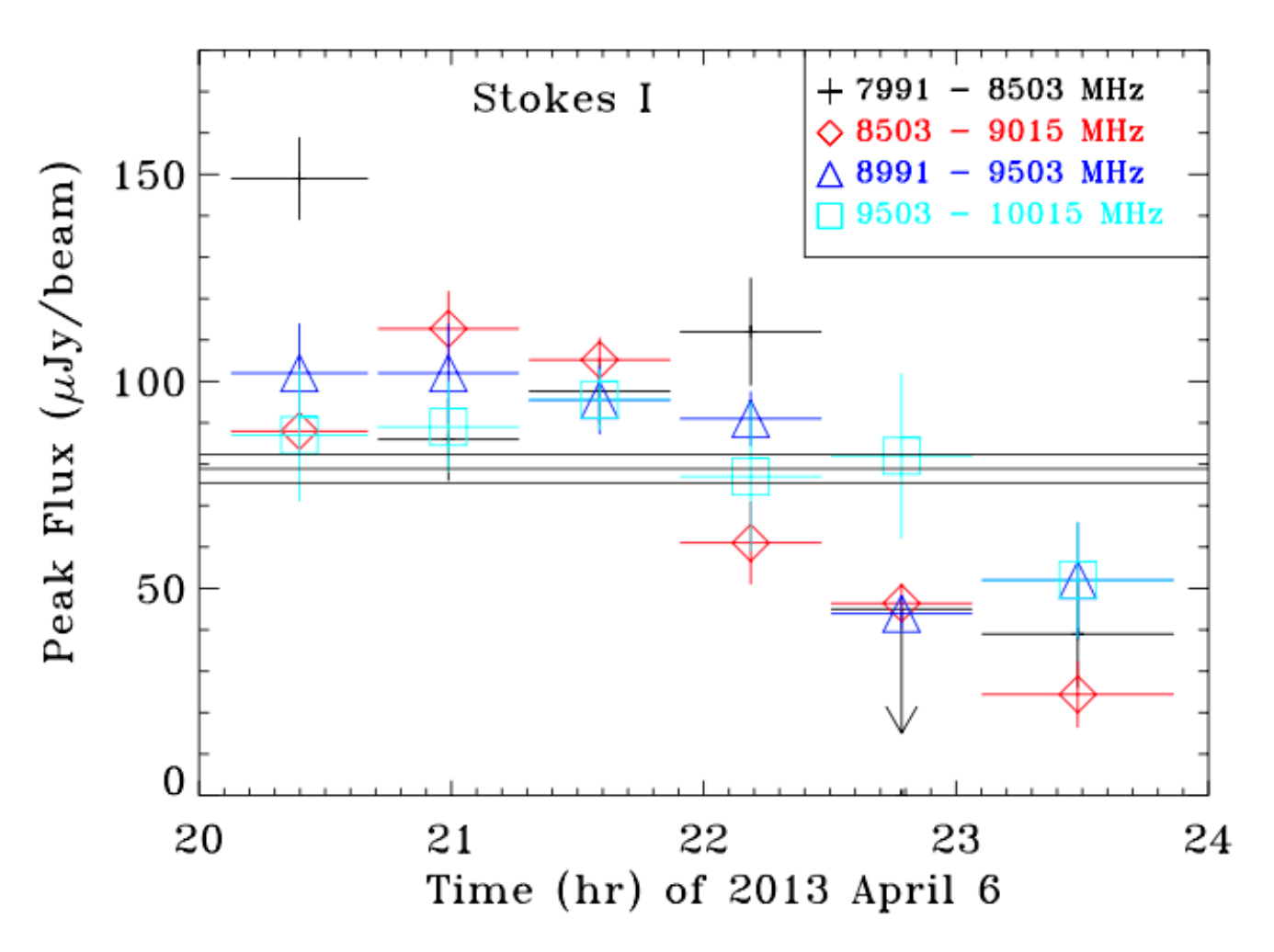}
}
}
\parbox{18cm}{
\parbox{9cm}{
\includegraphics[scale=0.25]{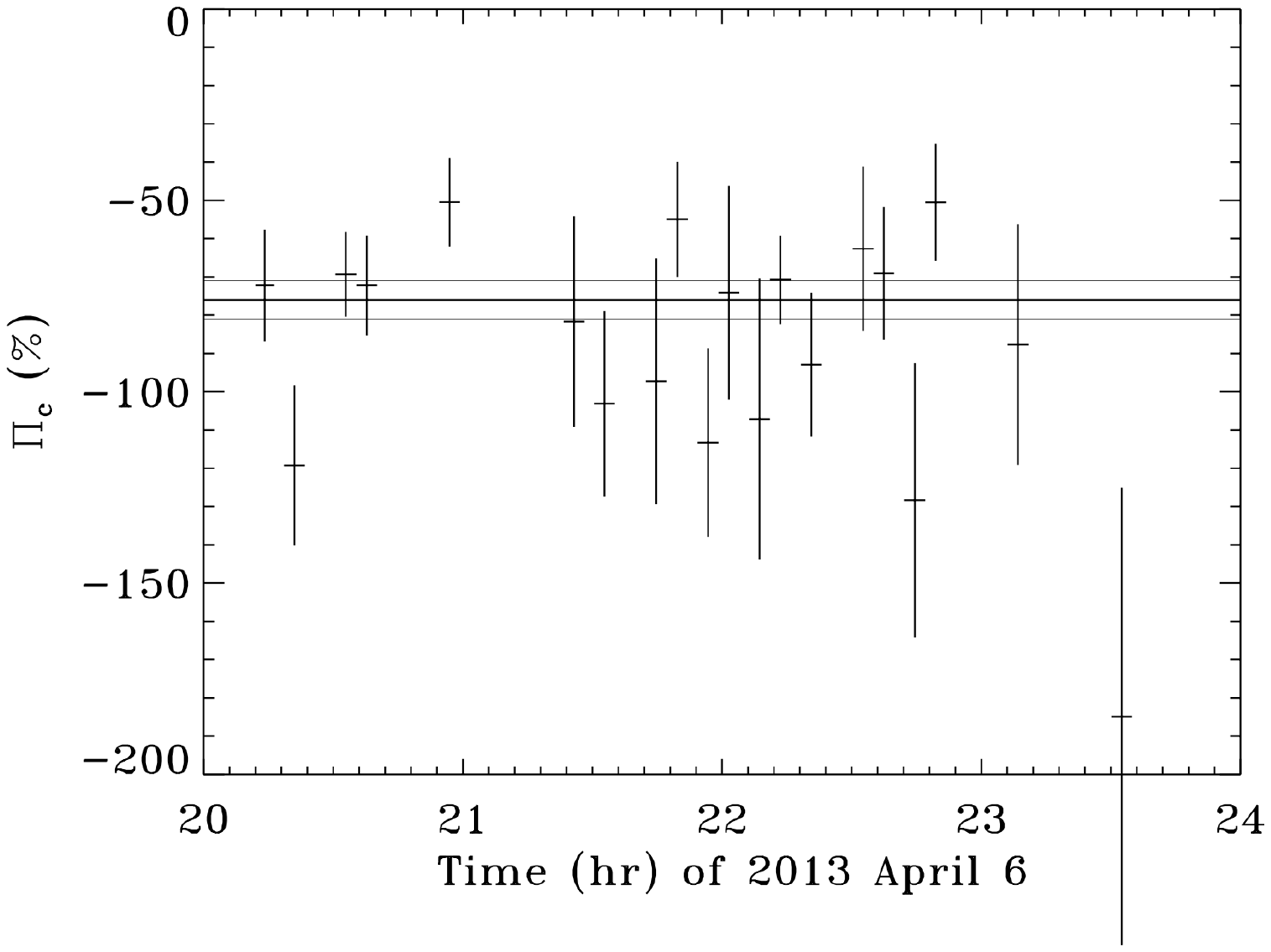}
}
\parbox{9cm}{
\includegraphics[scale=0.3]{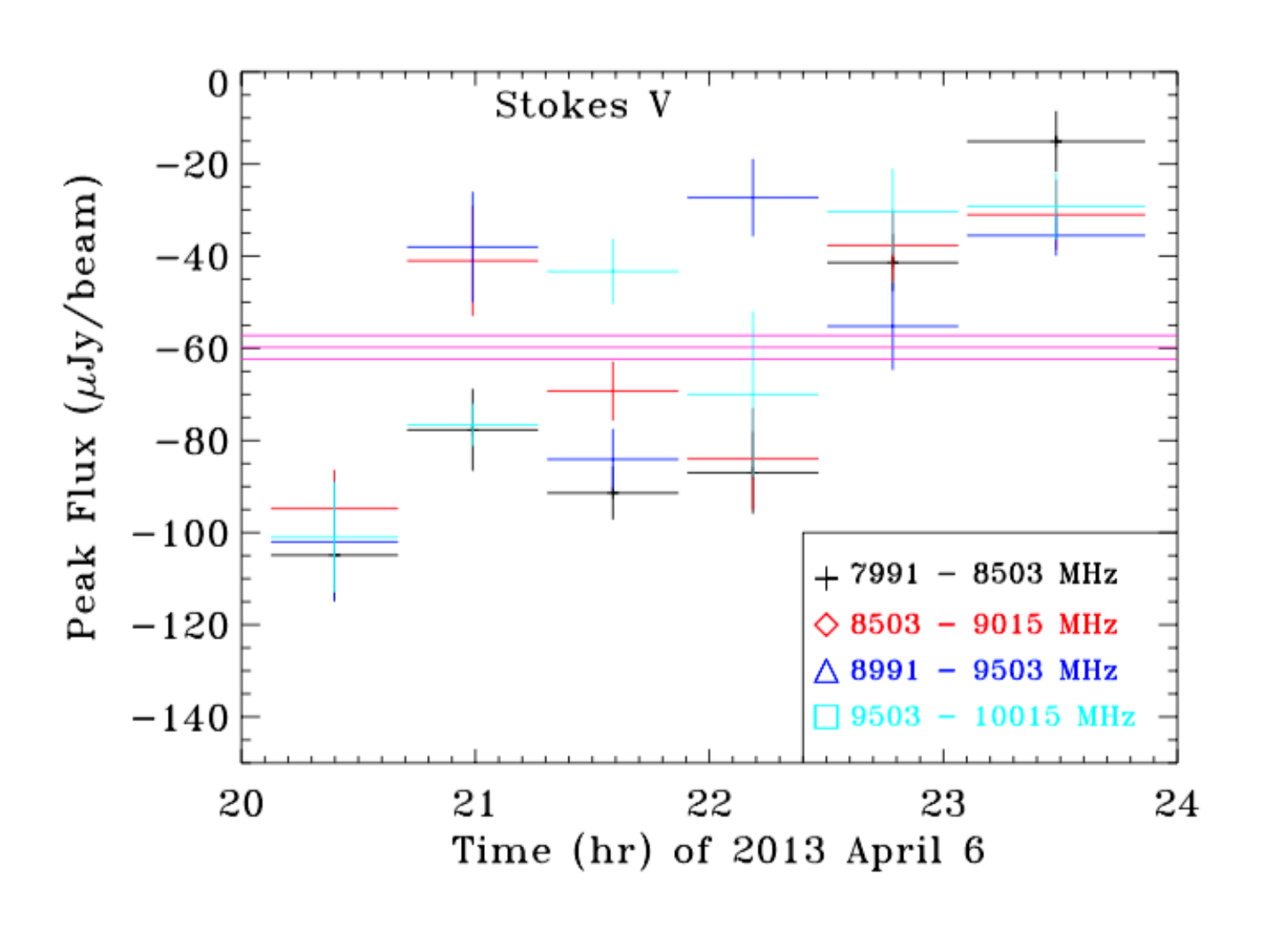}
}
}
\caption{Time-series of JVLA data for 2M0752. 
{\it top left} - Variation of total intensity and circularly polarized flux with time 
integrated over all frequencies
The black and the red line denote, respectively, the time-averaged total intensity and 
the time-averaged amount of Stokes $V$ flux. 
{\it bottom left} - Variation of the percent circular polariation integrated over all 
frequencies and time-average value shown as black line. 
{\it top right} - Time and frequency variations of the Stokes $I$ flux density, 
broken up into six time bins and four frequency spans.
{\it bottom right} - Variations of circularly polarized flux with the same time and 
frequency bins as for total flux density.
\label{fig:2m0752_radio}}
\end{center}
\end{figure*}

\subsection{2MJ0838\label{subsect:results_scr0838}}


The source 2MJ0838 was detected at both C- and X-bands. In addition to a detection in total intensity (Stokes I), the source was evident in Stokes V images as well. 
In the C-band Stokes I image there was a strong source detected in addition to the target. 
We performed image fits to the sources to determine their positions and average characteristics. The average flux density values for the target, 2MJ0838, are listed in Table~\ref{tbl:fluxdetails}. 
Light curve creation proceeded in Miriad
with the task {\sc uvfit}. 
For each invocation of the task,  all source parameters for the fit to the visibility data for the specified time bin 
were fixed except the flux density of the target. 
This process was repeated individually for both Stokes I and V, and for each frequency band, to create light curves with approximately 7.5 minute time bins (roughly
half of each scan length).
The light curves of 2MJ0838 in both radio bands as well as the time evolution of the polarization are shown in Figs.~\ref{fig:scr0838Catcalc} and ~\ref{fig:scr0838Xatcalc}.

Inspection of the temporal behavior of circularly polarized and total intensity flux densities reveals interesting behavior. The C-band light curve shows
evidence of total intensity variability 
at a factor of six at least, with additional periods where the source is undetected. 
The X-band light curve displays far less evidence for variability during the times when the source was detected.  
The average flux densities over the whole observation, derived from image fits, are overlaid as horizontal lines in Figs.~\ref{fig:scr0838Catcalc} and ~\ref{fig:scr0838Xatcalc}. 
For both bands these averages are lower
than the flux densities in time intervals when the source is detected, which reflects the significant fraction of the observing time when
the source is undetected. 
In general, when there is 
a detection in the Stokes V flux density, there is little variability in these detected flux density levels.
We note though that there are relatively few intervals when there is a Stokes V detection. 
The average values of circularly polarized flux density derived from the entire observation are smaller in an absolute value sense than the time intervals when it is detected; this mirrors the 
behavior noted for the average total intensity. 
Table~\ref{tbl:fluxdetails} delineates these average, minimum, and maximum detected flux densities. 

The periods of time with elevated C-band flux densities correspond in the X-band to either upper limits 
or to no evidence for variability.
The time period 07:20:35-09:52:55 is indicative of the first type, variability at C-band with undetected flux densities at X-band, while the time period 10:59:45-12:58:25 shows no indication of variability at X-band and factors of $\sim$three variability at 
C-band. We term these "C1" and "C2", respectively.

We investigated the frequency dependence of the large and varying C-band flux during the decay phase of C2 by repeating the above calculations in each time bin but splitting the 2.048\,GHz bandpass into four 512\,MHz  spectral intervals.  We determined the flux of the bright object noted above in each spectral interval, and repeated the {\sc uvfit} calculations, allowing only the peak flux density of the target to vary.  In order to investigate spectral variations, we filtered on time bins where the signal to noise ratio of the time bin across the entire C-band bandpass was larger than 8, and the signal-to-noise ratio of each of the four subbands was larger than 3.
This resulted in five time bins. 
These times are indicated in Figure~\ref{fig:scr0838Catcalc} 
and the spectral variations, including the flux density measurement at X-band, are shown in
Figure~\ref{fig:scr_freqtime}.

\begin{figure}
    \centering
       \includegraphics[scale=0.4]{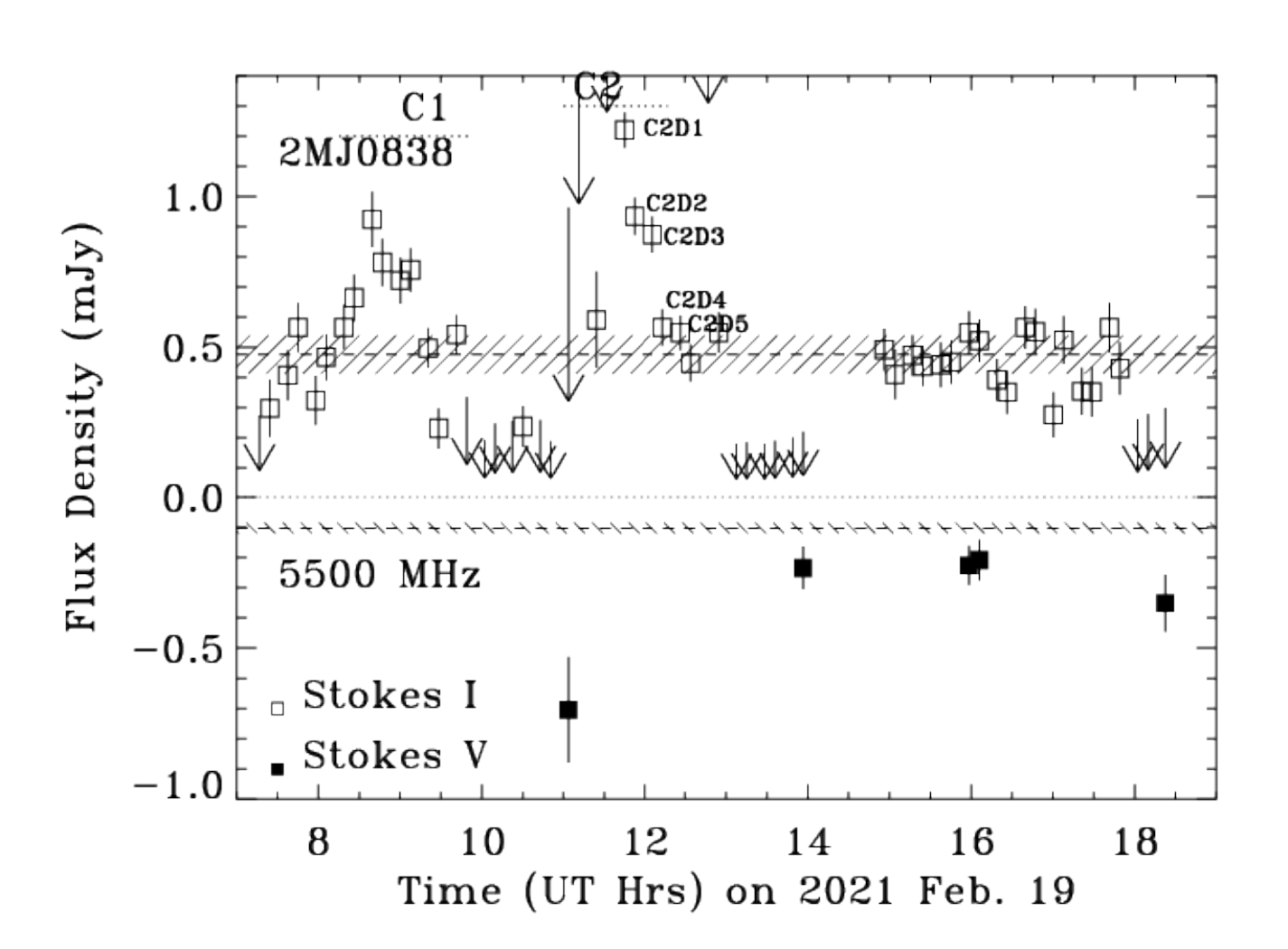}
    \caption{
    Total intensity and circular polarization behavior of 2MJ0838  with ATCA at C band. Light curve bin size is 7.5 minutes. Open squares indicate detections of total intensity, and filled symbols indicate detections of circularly polarized flux density. 
    For upper limits in total intensity, an arrow from 1$\sigma$ to 3$\sigma$ is shown.  
    The dashed lines and forward (backward) hatched regions indicate the average flux density and standard deviation for total intensity (circularly polarized flux) determined from image analysis spanning the entire time. 
    Approximate extent of events 'C1' and 'C2' are indicated. Specific time bins in the decay phase of event C2 are denoted, and their spectral energy distributions are plotted in Figure~\ref{fig:scr_freqtime}.
    }
    \label{fig:scr0838Catcalc}
\end{figure}

\begin{figure}
    \centering
   \includegraphics[scale=0.4,angle=90]{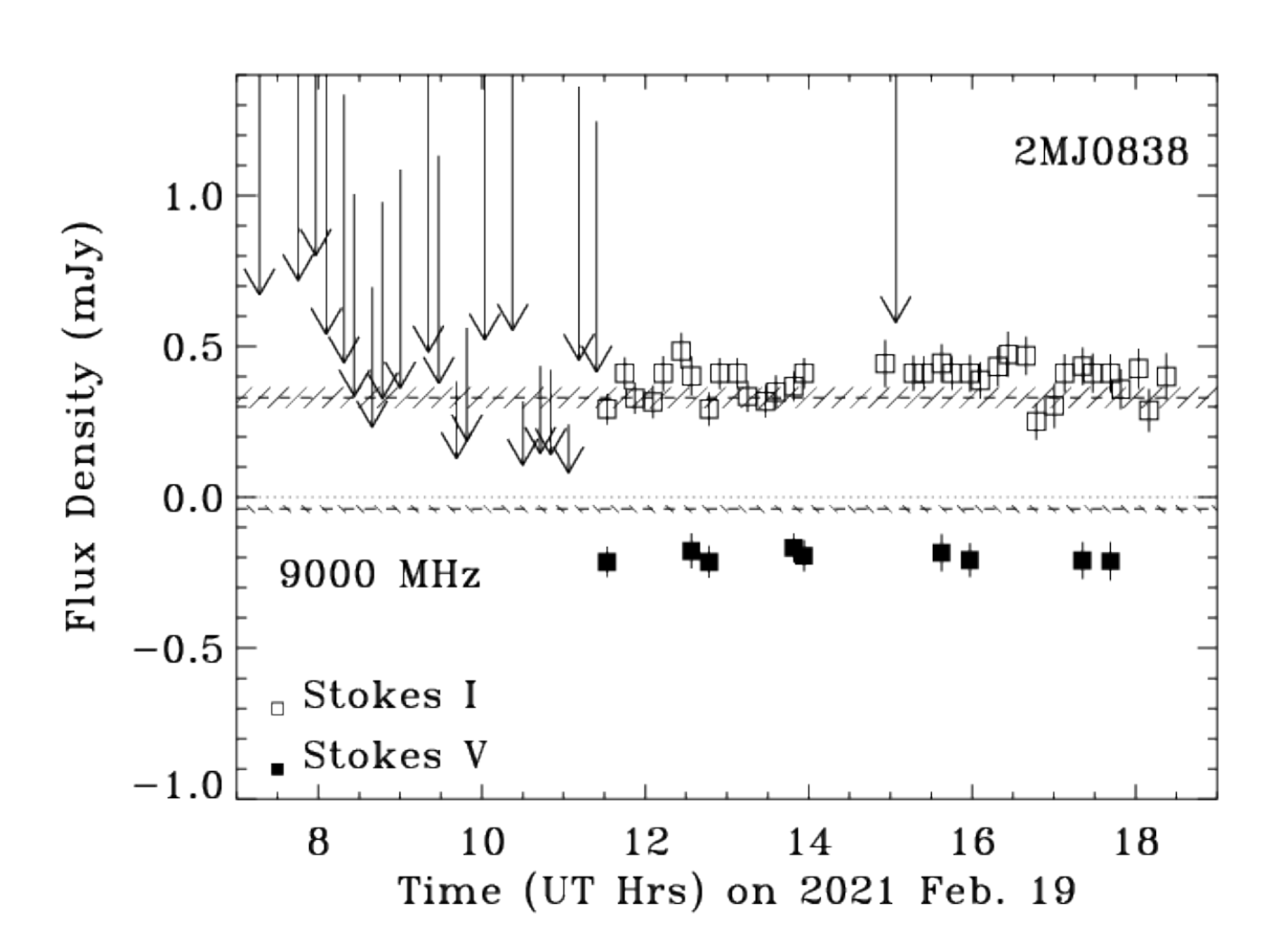}
    \caption{
    Total intensity and circular polarization behavior for 2MJ0838 at X band. Figure labels are the same as for Figure~\ref{fig:scr0838Catcalc}.
    }
    \label{fig:scr0838Xatcalc}
\end{figure}

\begin{figure}
    \includegraphics[scale=0.4,bb=80 80 800 600,clip]{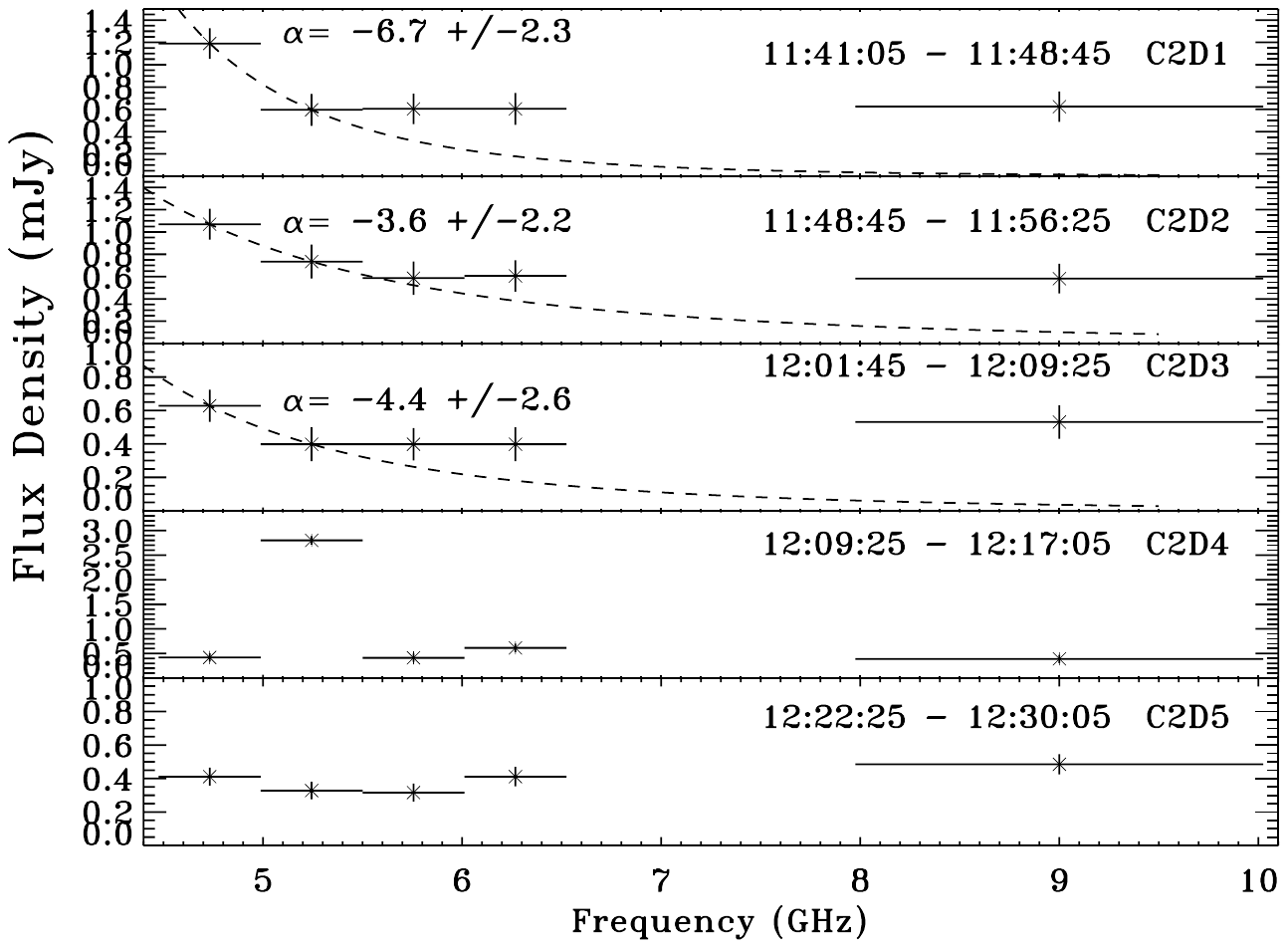}
    \caption{Spectra of individual time bins during the decay phase of event C2 on SCR~0838 noted in Figure~\ref{fig:scr0838Catcalc}. 
    The C-band was divided into four spectral bins for which the flux was computed, while the X-band is represented by a single flux measurement.
    The spectral index calculated from the two lowest  frequency bins is shown for the first three time bins in the decay of C2. 
}
    \label{fig:scr_freqtime}
\end{figure}


\begin{figure*}
\begin{center}
\parbox{18cm}{
\parbox{9cm}{\includegraphics[width=9cm]{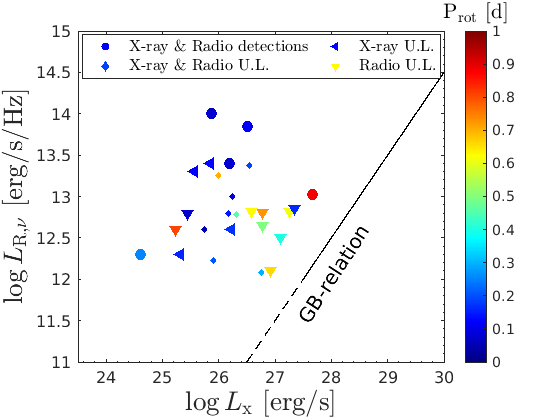}}
\parbox{9cm}{\includegraphics[width=9cm]{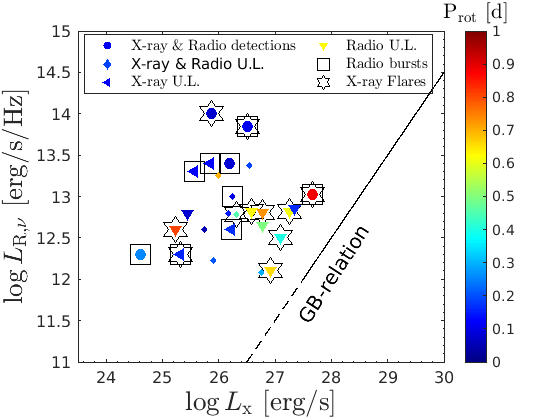}}
}
\caption{Radio vs. X-ray luminosity. {\it Left panel}: the lowest measured X-ray luminosity in the $0.2-2.0$\,keV energy band, including upper limits. 
The black line represents the relation from \cite{Guedel93.2}. {\it Right panel}:  same as left panel with the radio bursts and X-ray flares presented as open squares and stars, respectively. See Sect.~\ref{sect:discussion} for more details.}
\label{fig:radio_vs_xrays}
\end{center}

\end{figure*}

\begin{figure}
\begin{center}
\includegraphics[width=9cm]{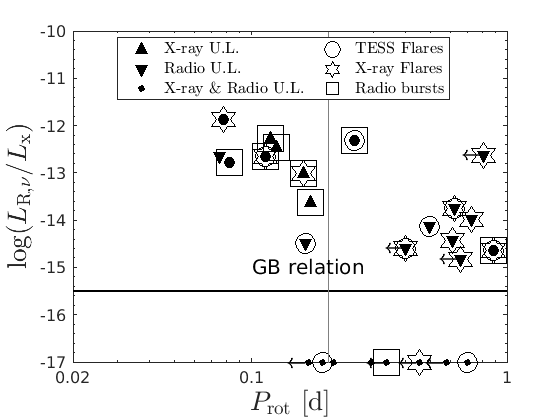}
\caption{Radio to X-ray luminosity ratio vs rotation period. 
TESS \& X-ray flares are shown with open filled circles \& stars, respectively, while radio bursts with open filled squares. UCDs that are placed on the x-axis are undetected with both X-ray \& radio instruments. The solid horizontal line at $\log(L_{\rm R,\nu}/L_{\rm x}) \sim -15.5$ represents the relation of \cite{Guedel93.2} as a function of $P_{\rm rot}$, while the vertical line at $P_{\rm rot}=0.2$\,d is from \citet{Pineda2017} to separate the radio-loud from the radio-quiet regime.}
\label{fig:LrLx_vs_Prot}
\end{center}
\end{figure}

\begin{figure}
\begin{center}
\includegraphics[width=9cm]{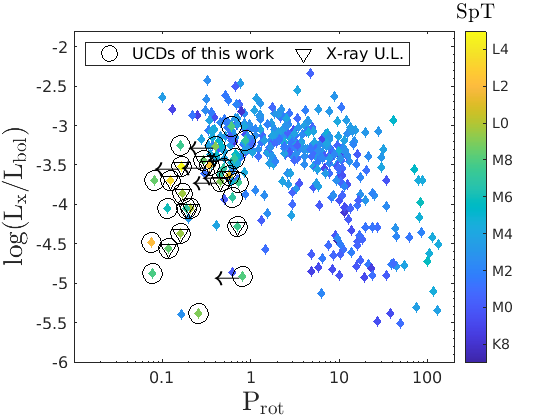}
\caption{The activity-rotation relation in terms of $L_{\rm x}/L_{\rm bol}$ vs $P_{\rm rot}$ and based on SpT-color code. As in Figs.\ref{fig:radio_vs_xrays} \& \ref{fig:LrLx_vs_Prot}, we present the deepest X-ray luminosity, including the upper limits (upside-down open triagles) as a function of rotation periods extracted from TESS light curves or calculated from $v\cdot \sin{i}$ measurements when the rotational modulation do not provide a clear value (left-oriented arrows). The UCD sample of this work (open circles) is compared with the M~dwarfs studied in \citet{Magaudda2020,Magaudda2022}, see Sect.~\ref{sect:discussion}}.
\label{fig:frac_lum_vs_Prot}
\end{center}
\end{figure}


We speculate that  events C1 and C2 on 2MJ0838 are gyrosynchrotron emission from flares, since we have not detected circular polarization associated with these events as is typical for coherent bursty emission.
The initial stages of the decay of C2 shown in Fig.~\ref{fig:scr_freqtime} reveal a
very steep frequency dependence of the flux in the first two frequency bins. 
We constrain the spectral index from these two bins for the first three time intervals, and overplot the fit results. 
This reveals that the
emission from 4.5-5.5 GHz at least in these three time bins during the decay phase of C2 is consistent with a falling spectrum, possibly indicating optically thin emission. 
The flux densities at higher frequencies, 5.5 -10 GHz, show very little variation with frequency.
This constrains the peak frequency of the presumed gyrosynchrotron emission to be at a value below $4.5$\,GHz.  
We note that this value is lower than what is typically seen from solar and stellar gyrosynchrotron tlfares, which have
peak frequencies 5-10 GHz or higher \citep{Shaik2021,Dulk1985}.

Using equation 39 in \citet{Dulk1985},
we explore for the C2 event of 2MJ0838 the dependence of  the peak frequency on the magnetic field strength in the radio-emitting source  as well as the product of the total number density of non-thermal electrons times the length scale of the radio-emitting region, denoted $NL$ by \citet{Dulk1985}. 
This relationship also depends on the power-law index of the accelerated particle distribution determined from radio observations ($\delta_{\rm r}$) 
as well as an assumption about the angle of
propagation with respect to the line of sight, $\theta$ (here assumed to be 45$^{\circ}$. 
We use the values of spectral index $\alpha$ constrained from 
Fig.~\ref{fig:scr_freqtime}
to infer the index of the accelerated
electron distribution $\delta_{r}$, where $\alpha=1.22-0.9\delta_{r}$ under optically thin
conditions for gyrosynchrotron emission \citep[Equation 35 in][]{Dulk1985}. 
Time bins C2D1 and C2D2
in Fig.~\ref{fig:scr_freqtime} provide bounds for the values of $\alpha$ and hence $\delta_{r}$. 
For these  time bins we find
$\delta_{r}=8.8\pm2.6$
and 
$\delta_{r}=5.2\pm2.4$, respectively. 
Figure~\ref{fig:nupk_constraints} shows the relationships among these parameters for a wide range of B and NL; we have assumed that the angle of propagation $\theta$ is 45$^{\circ}$ for these parameter space calculations.
Based on recent studies of solar 
flares the nonthermal electron density appears to 
range between 10$^{5}$ and 10$^{11}$ cm$^{-3}$ \citep{Fleishman2022}, and loop sizes from solar to stellar flares can range from $\leq$ 10$^{8}$ cm \citep{Cargill2006} to the size of the stellar diameter \citep{Stelzer2006}. 
We use a wide range of parameter space for $N$ and $L$ to investigate trends.
Inspection of Figure~\ref{fig:nupk_constraints} 
shows that in combination with the peak frequency constraints, the lower value of $\delta_{r}$ is more
constraining on allowed regions of magnetic field strength.  In particular, magnetic field strengths of only a few tens of Gauss are compatible with  a nonthermal electron density near the maximum found in \citet{Fleishman2022} (10$^{11}$ cm$^{-3}$) and a length scale of order the radius of 2MJ0838 (10$^{10}$cm).
This is similar to the equipartition field strength derived
from X-ray observations in Sect.~\ref{subsubsect:analysis_xrays_xspec}, but whereas the X-ray constraints are determined from the average coronal properties, the current constraints derive from a specific event corresponding to elevated flux levels and are not necessarily characteristic. 
Additional modelling will be needed to constrain these parameters further.

The circular polarization properties of 2MJ0838 are intriguing. They do not appear to be periodic; the detections 
at both X and C bands are just slightly above 3$\sigma$. Additionally, they indicate moderate degrees of circular polarization (40-50\%) when there is a detection, which is different from the highly circularly polarized bursts seen in 2MJ0752 and other UCDs which approach 100\% circular polarization.
This is also a much larger value of circular polarization fraction than is typically seen from quiescent gyrosynchrotron emission, generally $\sim$20\% or less
\citep{White1989}. 
If we identify the frequencies at which the circular polarization is detected
with the electron gyro-frequency, then we can place a constraint on the magnetic field strength in the steady radio-emitting region, with $\nu_{\rm obs}=s \cdot \nu_{B}$, where $s$ is the harmonic number of the emission being observed, and $\nu_{B}$ is the gyrofrequency, $\nu_{B}\approx 2.8\times10^{6} B[G]$ Hz. The large frequencies at which we detected the circular polarization, namely 9000 MHz, imply field strengths of one to several kilogauss. 

We use the method described by \citet{Osten2016,Smith2005}
 to convert the radio light curve into a constraint on the non-thermal energy as a function of magnetic field strength in the radio-emitting source and the distribution of accelerated particles $\delta_{r}$ as derived above. This contour plot is displayed in 
Fig.~\ref{fig:nonthermal_contours}.
While we do not have simultaneous measures of radiated flare energy, we note that this range spans 
10$^{30}$-10$^{34}$ erg for values of $\delta_{r}$
consistent with the observational constraints; these are flare energy ranges noted from a recent analysis of TESS white light flares on ultracool dwarfs
\citep{Petrucci23.0}.
For realistic flare energies (Fig.~\ref{fig:nonthermal_contours}) the magnetic field in the flaring source is on the order of a few $100$\,G, or less.

The equipartition field, derived from average X-ray measurements of this object in Section~\ref{subsubsect:analysis_xrays_xspec}, is considerably smaller than what is derived from the flare analysis. 
On its face, this implies that these coronae could be extremely compact. 
An apt comparison is that of \citet{Ness2004}, whose
analysis of X-ray grating data of late-type stars 
 combined measurements of electron density and volume emission measures, to suggest ``pathologically small'' coronal scale heights of 10$^{7}$ cm or less  ($H/R_{\star}<0.001$ for the M dwarfs considered in their sample). 
 They invoke a filling factor less than unity to overcome this situation.
Constraints on the sizes of coronae of other late-type
stars indicate larger extents: X-ray eclipses of the
G dwarf component in the $\alpha$~CrB system by its X-ray dark A star companion reveal a coronal height above the limb less than a solar radius \citep{Schmitt1994}.
Similarly, imaging observations of the Sun reveal a uniform typical height of the bright portion of the solar corona to $\sim$0.3R$_{\odot}$ \citep{DeForest2007}.

\begin{figure}
    \includegraphics[scale=0.4,angle=90]{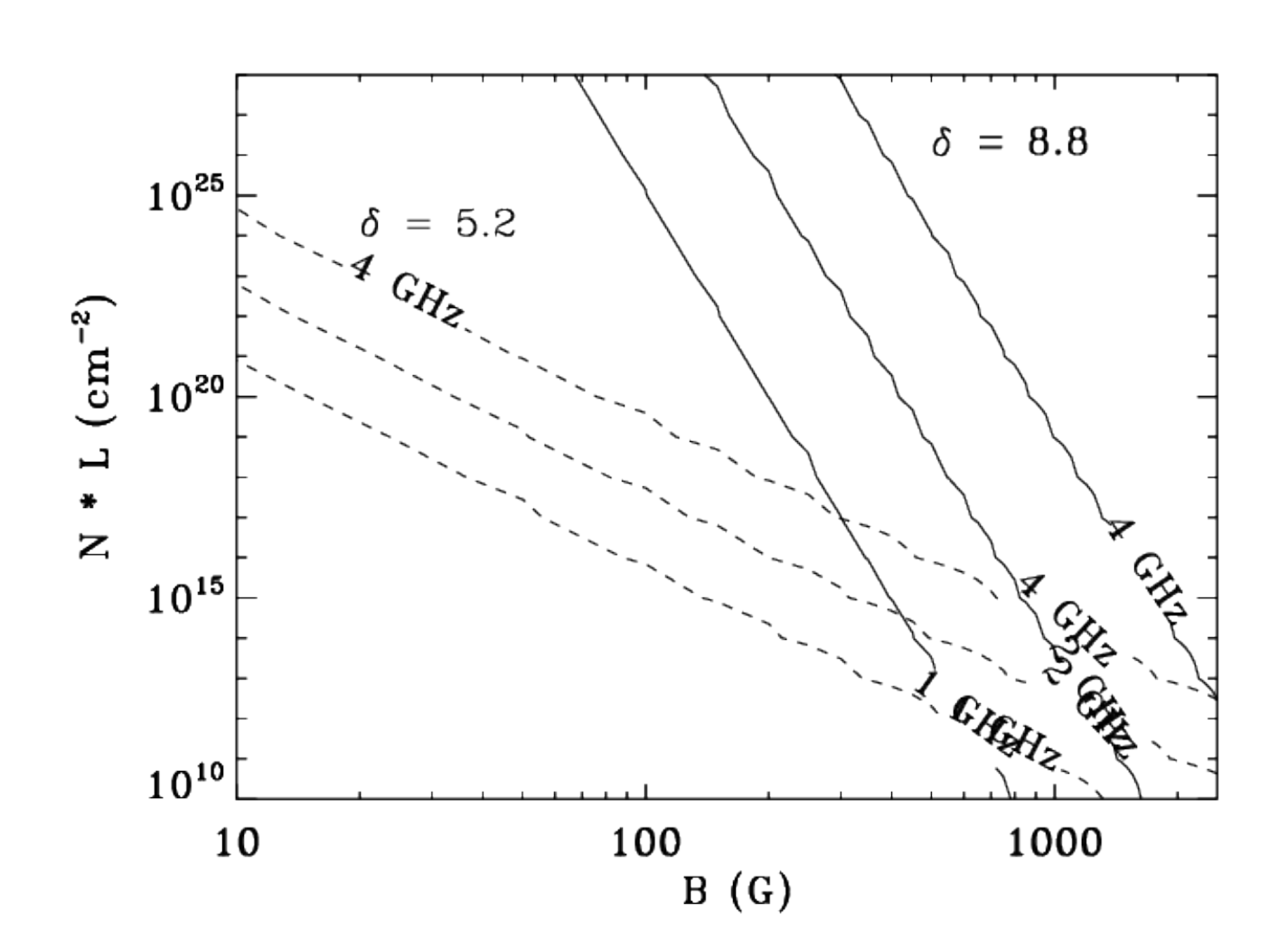}
    \caption{Contour plot of dependence of the peak frequency of gyrosynchrotron emission as a function of magnetic field strength in the radio-emitting source and the product of total number density of accelerated electrons and size scale of the emission $NL$.  Two different sets of contours are shown, which bracket the inferred values of $\delta_{r}$ seen in the frequency-dependent decay of event C2 in SCR0838 shown in Figure~\ref{fig:scr_freqtime}. 
    The larger values of $NL$ provide  more constraints on the field strength in the radio-emitting region, in order to be consistent with the low peak frequency and the smaller value of $\delta_{r}$.
    \label{fig:nupk_constraints}}
\end{figure}

\begin{figure}
    \includegraphics[scale=0.4,angle=90]{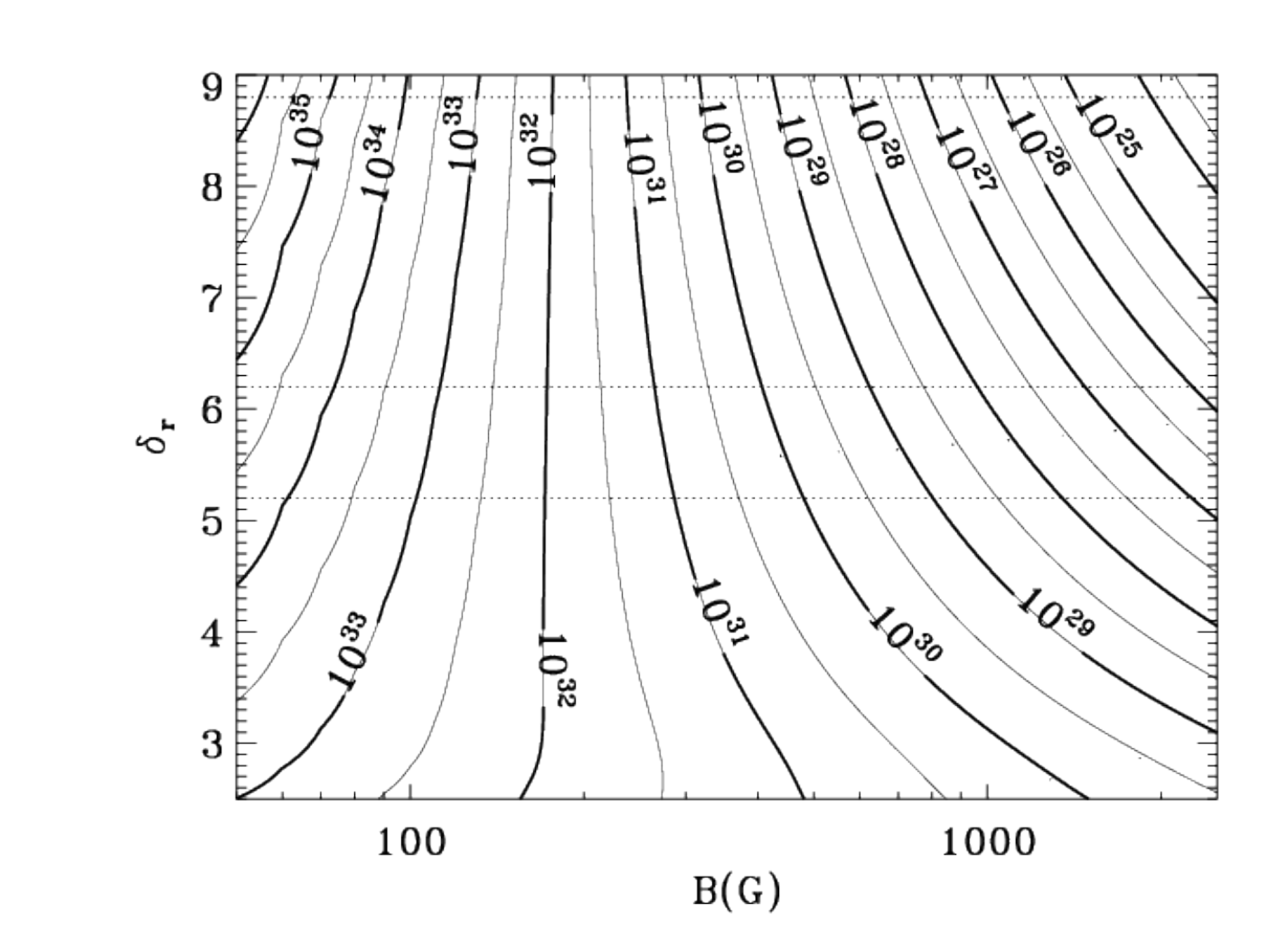}
    \caption{Contour plot of estimated non-thermal energy in radio flare C2 seen on SCR0838, as a function of the magnetic field strength ($B$) in the radio emitting region and the power-law index of the distribution of accelerated particles, $\delta_{\rm r}$. Horizontal lines indicate 
    the range of 
    $\delta_{\rm r}$ from the frequency dependence in Fig.~\ref{fig:scr_freqtime}.
    \label{fig:nonthermal_contours}}
\end{figure}

\section{Discussion}\label{sect:discussion}
\label{sect:results_multil}

With our increased database of UCDs with sensitive X-ray and radio data we present an updated view of the $L_{\rm R,\nu}-L_{\rm x}$ relation (Fig.~\ref{fig:radio_vs_xrays}). 
If multiple observations were obtained in either waveband we show the most sensitive one (left panel), including upper limits in case of non-detections. 
Radio and X-ray upper limits are displayed with filled upside-down and left-oriented triangles, respectively, while upper limits in both bands are presented with smaller filled diamonds and full detections with filled circles. In the right panel of Fig.~\ref{fig:radio_vs_xrays}, we highlight objects for which X-ray flares and radio bursts have been observed in at least one instance.
Among all objects of this work, $26$ UCDs including the literature sample, we detected optical flares for $9$ objects, among which $4$ show X-ray flaring activity, $3$ emitted radio bursts, and only $2$ present activity along all three wavelength bands. 
For the UCDs from our dedicated campaigns whether there is X-ray flaring activity or not is inferred from the variability in the {\it XMM-Newton} light curves (Fig.~\ref{fig:lc}). This choice is justified by the fact that the time resolution of {\it XMM-Newton} is higher than the one of eROSITA, that we recall being only $4$\,h and, thus, longer than typical X-ray flares on UCDs \citep[see e.g.][]{Stelzer2006}. Since we have one radio observation for each target, we account for variability simply if there is a burst in their emission. For the literature sample, instead, we adopted the notion of flaring/bursting versus quiescent X-ray and radio activity as presented in the references. In both panels of Fig.~\ref{fig:radio_vs_xrays} we also show the GB-relation (see Sect.~\ref{sect:intro} for more details). 

We confirm previous findings that UCDs are drastically offset from the GB-relation. With our enhanced  sample statistics we overall confirm  the picture of two groups of UCDs with different activity characteristics outlaid by \citet{Stelzer2012}, with X-ray flaring objects being closer to the GB-relation and radio bursting ones often associated with faint or no measurable X-ray emission.
The majority of the joint X-ray/radio sample is detected in X-rays and at levels of $L_{\rm x}$ fainter than the stars on which the GB-relation was defined (the dashed part of the line in Fig.~\ref{fig:LrLx_vs_Prot}). This (apparently) leftwards displacement with respect to the GB-relation could be explained by the lack of sensitivity at radio wavelengths.
The color code of this figure shows that radio bursts are mostly detected from fast-rotating UCDs in agreement with the results of \citet{Pineda2017}. 
Their two regimes of fast-rotating ``radio-loud'' and slow-rotating ``radio-quiet'' UCDs were the basis of our target selection (see Sect.~\ref{sect:intro}). Since we work with photometric period while \citet{Pineda2017} used $v\sin{i}$ measurements, we had translated their boundary between the two regimes ($v\sin{i}\approx38\,\rm km/s$) to a value of $P_{\rm rot} \approx 0.2$\,d (assuming a typical late-M dwarf radius of $R_{\star}\sim 0.15\,R_{\odot}$ and $i \approx 90^{\circ}$.
In contrast, X-ray flares are found on UCDs that are distributed throughout the range of rotation periods considered in Fig.~\ref{fig:radio_vs_xrays}, 
implying no evident relation for UCDs between their X-ray activity and rotation. 

The X-ray scale height/equipartition magnetic field analysis in Sect.~\ref{subsubsect:analysis_xrays_xspec} reveals either a very compact corona for values of magnetic field strength of a few tens of Gauss, or extremely low confining magnetic field strength for a more extended corona. This could originate from the  assumptions of homogeneity and spherical symmetry.

One UCD of our sample (2MJ0752) is simultaneously detected along the two wavebands adopted for this work and its X-ray emission is one of the brightest. We do not infer X-ray variability from the {\it XMM-Newton} light curve (see Fig.~\ref{fig:lc}), although we clearly see variable X-ray emission from eROSITA detections (see Fig.~\ref{fig:er_LCs}). In Sect.~\ref{subsect:results_2m0752} and  Fig.~\ref{fig:2m0752_radio} we presented the radio light curve of this UCD that shows two apparent double peaks. This brightness modulation suggests the presence of a periodic burst corresponding to a shorter rotation period ($< 2$\,h) than the one of 2MJ0752 (listed in Table~\ref{tab:tess}). We linked this radio emission to a possible auroral electron precipitation, as the one typical of giant planets of the solar system. The X-ray brightest and radio simultaneous detections of this UCD are showing signatures expected from higher mass M~dwarfs along
with emerging evidence of radio auroral emission, reinforcing the hypothesis of a bimodal dynamo mechanism ruling the X-ray/radio activity along the ultracool dwarf regime.

For a direct investigation of the dependence of radio and X-ray activity on rotation we show in Fig.~\ref{fig:LrLx_vs_Prot} the relation between  $L_{\rm R,\nu}/L_{\rm x}$ and $P_{\rm rot}$, emphasizing objects for which an X-ray flare (open star), optical flare (open circle) or a radio burst (open square) is present in our dataset. 
UCDs that are upper limits in both X-ray and radio bands are placed on the bottom of the plotting window, and we indicate with left-oriented arrows the rotation periods that are upper limits, calculated from $v\sin{i}$ measurements. In this $L_{\rm R,\nu}/L_{\rm x}-P_{\rm rot}$ space the GB-relation is a horizontal line, and we placed a vertical solid line at $P_{\rm rot}= 0.2$\,d to separate the radio-loud and radio-quiet regimes from \citet{Pineda2017}.

Since only $2$ of the $26$ UCDs experienced activity along all three wavelenghts, searching for a relation between the mechanism that generates a radio bursts rather than an optical or X-ray flare remains a difficult task. Thus, a larger sample and $B$ measurements are required and in order to investigate such dichotomy that might be caused by the structure of the magnetic field.

Given the fast rotation of UCDs and their peculiar X-ray/radio behavior it is interesting to study their position in the rotation-activity relation compared to earlier M dwarfs. We use the representation in terms of fractional luminosity, $L_{\rm x}/L_{\rm bol}$ versus $P_{\rm rot}$ (see Fig.~\ref{fig:frac_lum_vs_Prot}). The $L_{\rm x}/L_{\rm bol}$ ratio has the advantage of removing most of the mass dependence of X-ray brightness in the saturated regime. In most studies of the activity-rotation relation this parameter is combined with the Rossby number ($R_0 = \tau_{\rm conv}/P_{\rm rot}$, where $\tau_{\rm conv}$ is the convective timescale). For late-type stars activity-rotation relations that make use of the Rossby number display a smaller horizontal spread than those involving only $P_{\rm rot}$ \citep{Wright2011,Magaudda2020}. However, since the convective timescale is notoriously difficult to constrain and no parametrizations exist for UCDs later than SpT = M9 we work with the rotation period. 
In Fig.~\ref{fig:frac_lum_vs_Prot} the SpT-color code applies to both the UCDs from this article and the comparison sample of early- to mid-M dwarfs from \citet{Magaudda2020,Magaudda2022}, but for clarity we additionally highlight the UCDs with open black circles. As in the other figures, we display for the UCDs the most sensitive X-ray measurement, including the upper limits that are plotted as upside-down open triangles. The left-oriented arrows indicate for which object we calculated the rotation period from $v\sin{i}$ measurements. As stated in many previous works \citep{Wright2011,Wright2016,Stelzer2016,Wright2018,GonzalezAlvarez2019,Magaudda2020,Magaudda2022}, the activity rotation relation of early- to mid- M~dwarfs splits into two regimes: for $P_{\rm rot}\lessapprox10$\,d the X-ray activity does not depend on the rotation (saturated regime) and at longer periods the X-ray activity declines with increasing rotational period (unsaturated or `correlated' regime). 


For our analysis of the $L_{\rm x}/L_{\rm bol}-P_{\rm rot}$ relation we only consider the quiescent state of our sample. Our sample of late-M dwarfs spans the entire range of $L_{\rm x}/L_{\rm bol}$ that covers the typical activity level of both the saturated \& unsaturated regime of the earlier SpTs from \citet{Magaudda2022}. This spread and consequent low activity of UCDs evidently show a lack of correlation between $L_{\rm x}$ and $P_{\rm rot}$, and it is usually ascribed to their mostly neutral atmospheres, that $-$ even if the magnetic field is present $-$ do not allow coupling with matter and thus outer magnetic heated atmospheres are not prominent \citep{Mohanty02.1}. 
Previous works already found a significant drop in X-ray activity beyond spectral type $\sim$M7 \citep{Cook14.0,Berger2010}. Specifically, \citet{Berger2010} showed $L_{\rm x}/L_{\rm bol}\approx 10^{-4}$ for M7–M9 and $L_{\rm x}/L_{\rm bol}\approx 10^{-5}$ as deepest limits for L~dwarfs, while \citet{Cook14.0} found a $L_{\rm x}/L_{\rm bol}-$spread of $2$\,dex for UCDs located in the saturated regime, i.e. $P_{\rm rot}\leq10$\,d. 
We found a $L_{\rm x}/L_{\rm bol}-$spread for UCDs with $P_{\rm rot}\leq 1$\,d larger than the one presented in \citet{Cook14.0} only for quiescent X-ray emissions (see Fig.~\ref{fig:frac_lum_vs_Prot}).
In fact, we observe a declining $L_{\rm x}/L_{\rm bol}$ for later SpT than M7 down to $L_{\rm x}/L_{\rm bol}\approx 3\times10^{-6}$. This detection refers to a M9 dwarf from the literature, DENIS\,J1048, for which the deepest $L_{\rm x}$ is from {\it XMM-Newton} observation (see Tables~\ref{tab:lit_sample_xray_radio}~\&~\ref{tab:obslog_er}).
Moreover, with our larger and updated sample of UCDs we confirm the results of \citet{Cook14.0}, for which the authors inferred the decrease of the X-ray activity of fast rotating UCDs caused by a decrease in the effectiveness of the magnetic dynamo.  
Furthermore, with our larger and updated sample of UCDs we confirm the results of \citet{Cook14.0}, whereby the authors coupled the decrease in X-ray activity of rapidly rotating UCDs with a decrease in the effectiveness of the magnetic dynamo.  
\citet{Cook14.0} suggested that the X-ray activity decrease for UCDs is due to the presence of a bimodal dynamo across late-type dwarfs that produces distinct magnetic field topologies. In other words, UCDs with large-scale fields may spin down more efficiently than those with weaker and tangled fields.

\section{Conclusions}
\label{sect:conclusion}

We present a collection of simultaneous X-ray/radio observations of a sample of $10$ UCDs that we enlarge with archival X-ray and radio data from the literature. The final sample counts $26$ late-M dwarfs. We added two more radio-detected dwarfs to the list of known emitters: 2MJ0752 and 2MJ0838. The first object is a slow rotating ($v\sin{i}=9$\,km/s) and one of the X-ray brightest radio-detected UCDs. It exhibits a possible loss cone along the line of sight, suggested by the presence of two double peaks in its radio light curve that correspond to a rotation period shorter that the one extracted from TESS data ($P_{\rm rot}=0.88$\,d). Given that there is a bias in radio detection fraction against the slowest rotating UCDs \citep{McLean12.0,Stelzer2012}, it becomes an even more intriguing object to follow up. The detection of this target during what looks like a long-lasting radio burst, after a previous epoch in which a lower upper limit was obtained, suggests that the actual fraction of radio-emitting objects amongst UCDs is much higher than one would estimate based on single epochs alone. 
The other radio-detected object (2MJ0838) displays evidence for X-ray emission and incoherent radio flares, along with a moderate  circularly polarized flux which does not vary appreciably. We speculate that this object is experiencing a transition of its magnetic behavior, producing signatures typical of higher mass M~dwarfs along with emerging evidence of auroral emission. 

We updated the $L_{\rm R,\nu}-L_{\rm x}$ relation for UCDs with rotation periods lower than $1$\,d. While comparing our results with those from \citet{Guedel93.2} for earlier-type stars we confirmed that rapid rotators lie the furthest away from the GB-relation, and are more likely to exhibit radio bursts indicative of loss cone-auroral emission. 
\citet{Paudel2018ApJ...858...55P} noted the ubiquity of white-light flaring in UCDs. Our results support and extend this conclusion to shorter wavebands by noting that X-ray flaring occurs throughout the region of radio and X-ray luminosities in Fig.~\ref{fig:radio_vs_xrays}, while objects with radio bursts tend to be experienced more by fast rotating UCDs. Based on our examination of radio, X-ray and rotation period results we observed and confirmed the presence of a radio-loud regime for UCDs with $P_{\rm rot}\leq0.2$\,d, already proposed by \citet{Pineda2017}, and a radio-quiet regime where most of the X-ray flares take place for larger $P_{\rm rot}$. 

Finally, we examined $L_{\rm x}/L_{\rm bol}$ as a function of $P_{\rm rot}$ for our sample of UCDs in comparison with the results of \citet{Magaudda2022} for earlier-type  M dwarfs. We observed that despite their fast rotation the X-ray emission of UCDs covers the whole $L_{\rm x}/L_{\rm bol}$ range ($\approx -3.0 ... -5.5$) that comprises the saturated and the unsaturated regimes of earlier M~dwarfs. This shows that no evident relation seems to emerge between the X-ray emission of UCDs with respect to their rotation. The low activity detected for our sample of UCDs can be ascribed to their mostly neutral atmospheres, that do not allow coupling with matter \citep{Mohanty02.1} and to a decrease of dynamo efficiency that suggests the presence of a bimodal dynamo across late-type dwarfs that changes the magnetic field topology \citep{Cook14.0}.

\begin{acknowledgements}

EM is supported by Deutsche Forschungsgemeinschaft under grant STE 1068/8-1. 

This work is based on data from eROSITA, the soft X-ray instrument aboard SRG, a joint Russian-German science mission supported by the Russian Space Agency (Roskosmos), in the interests of the Russian Academy of Sciences represented by its Space Research Institute (IKI), and the Deutsches Zentrum für Luft- und Raumfahrt (DLR). The SRG spacecraft was built by Lavochkin Association (NPOL) and its subcontractors, and is operated by NPOL with support from the Max Planck Institute for Extraterrestrial Physics (MPE).

The development and construction of the eROSITA X-ray instrument was led by MPE, with contributions from the Dr. Karl Remeis Observatory Bamberg \& ECAP (FAU Erlangen-Nuernberg), the University of Hamburg Observatory, the Leibniz Institute for Astrophysics Potsdam (AIP), and the Institute for Astronomy and Astrophysics of the University of Tübingen, with the support of DLR and the Max Planck Society. The Argelander Institute for Astronomy of the University of Bonn and the Ludwig Maximilians Universität Munich also participated in the science preparation for eROSITA.

The eROSITA data shown here were processed using the eSASS/NRTA software system developed by the German eROSITA consortium.

This paper includes data collected with the TESS mission, obtained from the MAST data archive at the Space Telescope Science Institute (STScI). Funding for the TESS mission is provided by the NASA Explorer Program. STScI is operated by the Association of Universities for Research in Astronomy, Inc., under NASA contract NAS 5–26555.


\end{acknowledgements}

\bibliographystyle{aa} 
\bibliography{ucds_XR_Magaudda}

\begin{appendix}
\section{}   
The Appendix displays radio maps of the field around objects which did not produce a detection at radio wavelengths.
\begin{figure*}
\begin{center}
    \parbox{18cm}
    {
    \parbox{6cm}{\includegraphics[width=0.35\textwidth]{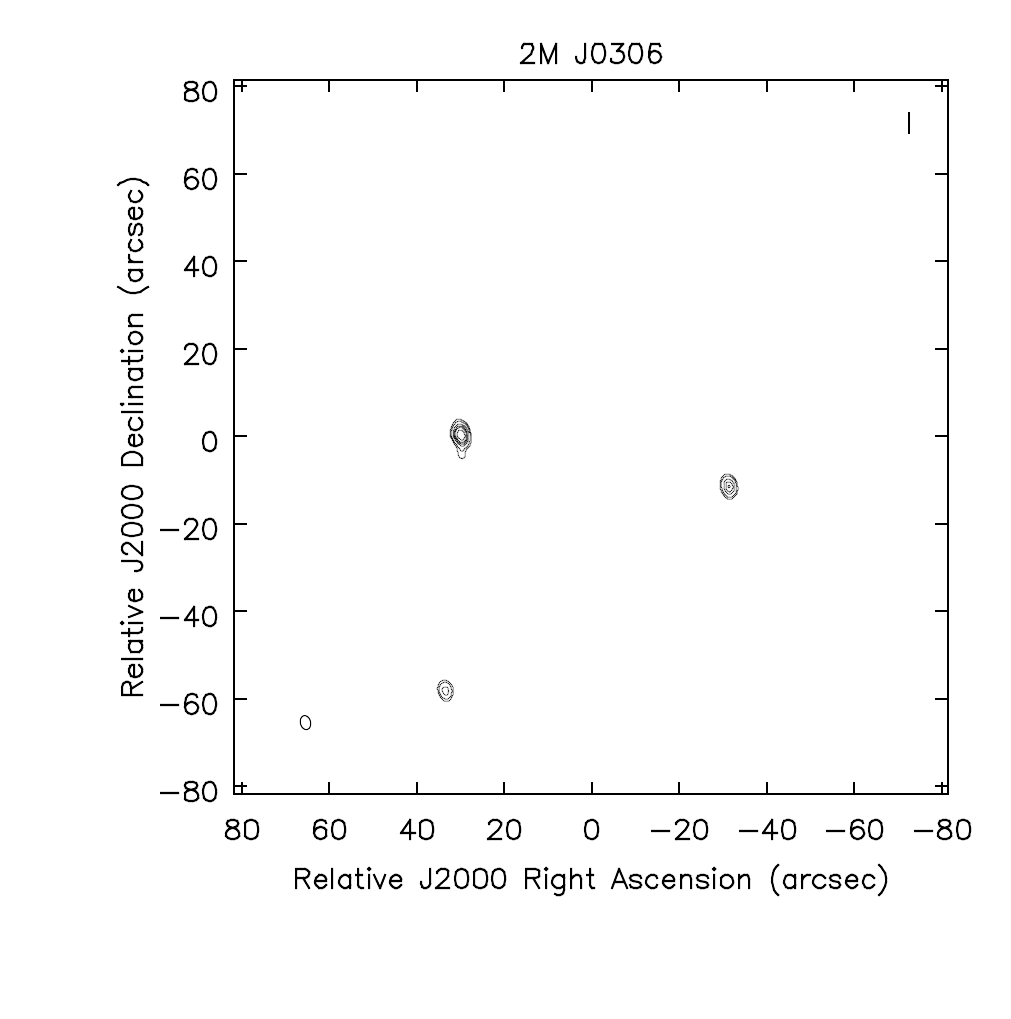}}
    \parbox{6cm}{\includegraphics[width=0.35\textwidth]{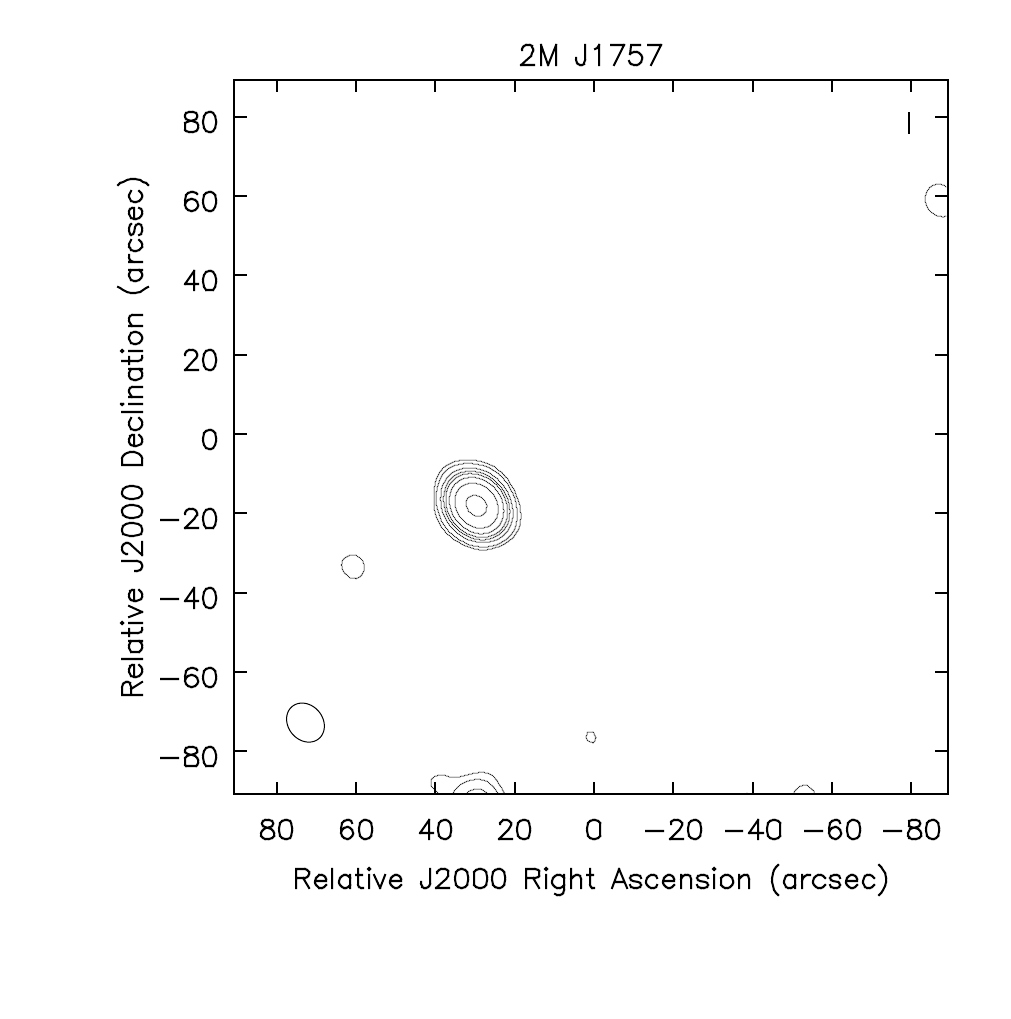}}
    \parbox{6cm}{\includegraphics[width=0.35\textwidth]{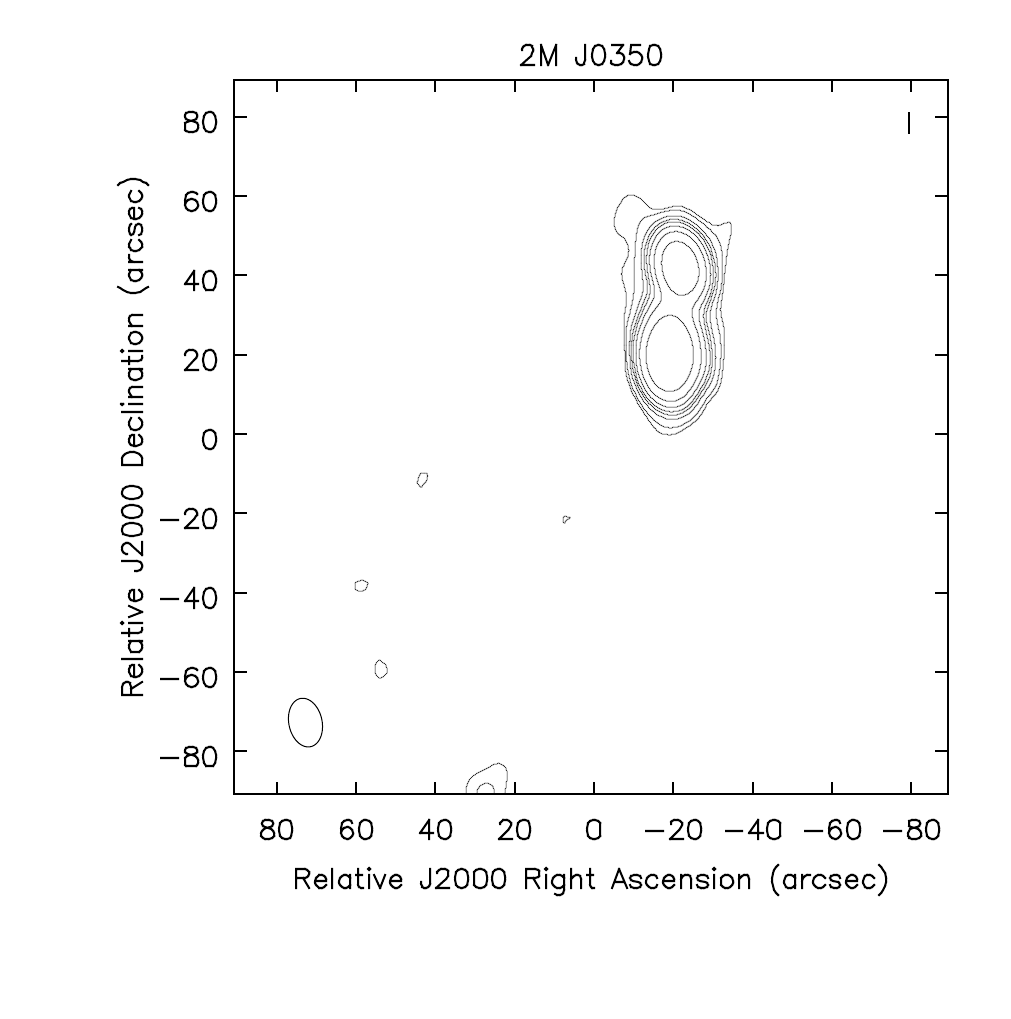}}
    }

\end{center}
\caption{Contour images of the JVLA field around the three JVLA targets in D and CnB configuration as observed at X band which were not detected. The size is $3^{\prime} \times 3^{\prime}$ for the two first images taken in D configuration and $2.6^{\prime} \times 2.6^{\prime}$ for the last one observed in CnB configuration. Contour levels are $3$, $5$, $10$, $15$, $20$, $30$, $50$, and $100$ times the image rms, which is listed in Table~\ref{tab:radio_limits}. The right ascension and declination of the phase center for the observation for each target is listed in Table~\ref{tab:obslog_radio}.}
\label{fig:appendix_JVLA}
\end{figure*}

\begin{figure}
    \begin{center}
    \includegraphics[width=0.4\textwidth]{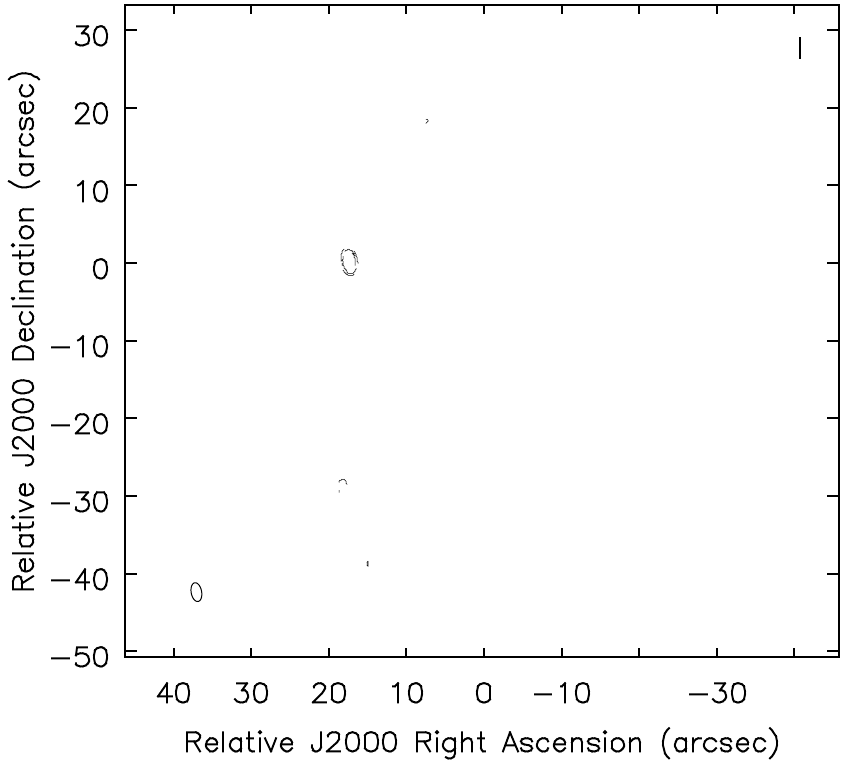}
    \end{center}
    \caption{Zoom in of region around 2MJ0652 with the JVLA as observed at C band. Contours are 5, 8, and 10 times the rms of 1.8$\mu$Jy determined from a blank region of the field.}
    \label{fig:appendix_2MJ0652}
\end{figure}

\begin{figure}
   \includegraphics[width=0.4\textwidth]{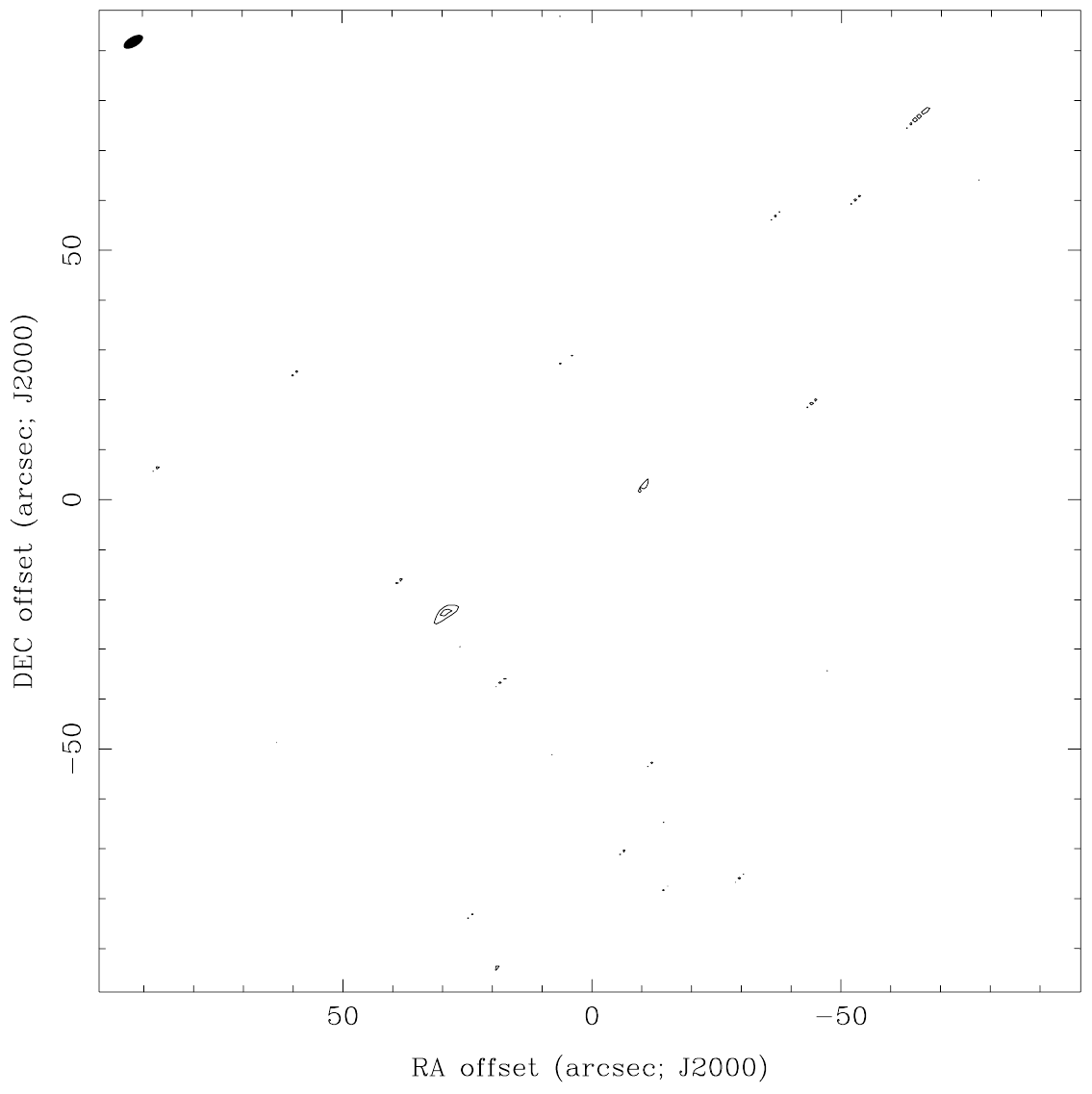}
   \caption{Maps of the three-arcminute region around the lower frequency of the targets 2MJ1055 observed with ATCA, which was not detected at either frequency. Contours are shown at 3, 5, and 10 times the rms values listed in Table~\ref{tab:radio_limits}. 
   \label{fig:appendix_ATCA}}
\end{figure}

\section{}
We present an example of our TESS rotation analysis showing the results we obtained for the UCD TIC\,29890705. 

\begin{figure*}
    \centering
    \includegraphics[width=0.6\textwidth,angle=90]{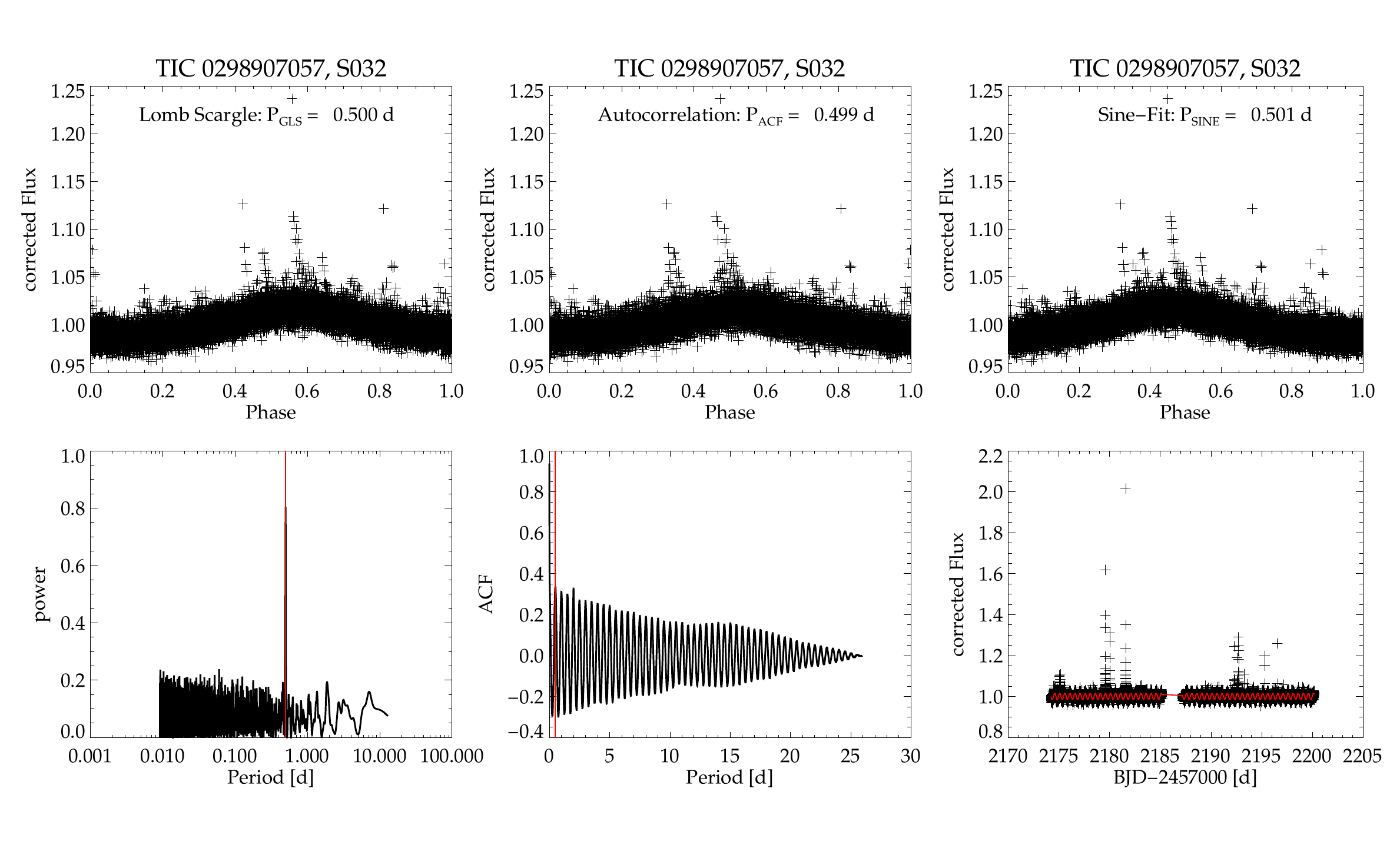}
    \caption{Result of the period search for TIC\,298907057 observed with TESS in sector 32: from left to right results refer to Lomb-Scargle periodogram, auto-correlation and sine-fitting. {\it top panels} -  light curves phase-folded with the periods obtained with each of the different methods; {\it bottom} - periodogram, autocorrelation function  and the original light curve with the sine fit.}
    \label{fig:tessexample}
\end{figure*}
\end{appendix}

\end{document}